\documentclass[preprint]{aastex}
\usepackage{txfonts}
\usepackage{graphicx}
\usepackage{color}
\usepackage{multirow}
\usepackage{natbib}
\usepackage{amssymb}
\usepackage{epstopdf}
\usepackage{placeins}
\bibpunct{(}{)}{;}{a}{}{,}

\shortauthors{T.\ Van Reeth et al.}
\shorttitle{Gravity-mode period spacings for a sample of $\gamma\,$Dor stars}

\received{receipt date}
\revised{revision date}
\accepted{acceptance date}
\begin{document}
\title{Gravity-mode period spacings as seismic diagnostic for a sample of $\gamma\,$
  Doradus stars from {\it Kepler\/} space photometry and high-resolution
  ground-based spectroscopy\footnotemark[1]}

\author{T.~Van~Reeth, A.~Tkachenko\altaffilmark{2}}\affil{Instituut voor Sterrenkunde, KU Leuven, Celestijnenlaan 200D, 3001 Leuven, Belgium}\email{timothy.vanreeth@ster.kuleuven.be}
% \and \author{ }\affil{Instituut voor Sterrenkunde, KU Leuven, Celestijnenlaan 200D, 3001 Leuven, Belgium}\thanks{Postdoctoral Fellow of the Fund for Scientific Research (FWO), Flanders, Belgium.}
\author{C. Aerts}\affil{Instituut voor Sterrenkunde, KU Leuven, Celestijnenlaan  200D, 3001 Leuven, Belgium}
\affil{Department of Astrophysics, IMAPP, University of Nijmegen, PO Box 9010, 6500 GL Nijmegen, The Netherlands}
\affil{Kavli Institute for Theoretical Physics, University of California, Santa Barbara, CA 93106, USA}
\author{P.~I.~P\'apics\altaffilmark{3}}\affil{Instituut voor Sterrenkunde, KU
  Leuven, Celestijnenlaan 200D, 3001 Leuven, Belgium}
\affil{Kavli Institute for Theoretical Physics, University of California, Santa Barbara, CA 93106, USA}
\author{S.~A.~Triana}\affil{Instituut voor Sterrenkunde, KU Leuven, Celestijnenlaan 200D, 3001 Leuven, Belgium}

\author{K.~Zwintz}\affil{Instituut voor Sterrenkunde, KU Leuven, Celestijnenlaan 200D, 3001 Leuven, Belgium}\affil{Institute for Astro- and Particle Physics, University of Innsbruck, Technikerstrasse 25/8, A-6020 Innsbruck, Austria}
\author{P.~Degroote, J.~Debosscher}\affil{Instituut voor Sterrenkunde, KU Leuven, Celestijnenlaan 200D, 3001 Leuven, Belgium}
\author{S. Bloemen}\affil{Department of Astrophysics, IMAPP, University of
  Nijmegen, PO Box 9010, 6500 GL Nijmegen, The Netherlands}
\affil{Kavli Institute for Theoretical Physics, University of California, Santa Barbara, CA 93106, USA}
\author{V.~S.~Schmid\altaffilmark{4}}\affil{Instituut voor Sterrenkunde, KU Leuven, Celestijnenlaan 200D, 3001 Leuven, Belgium}\affil{Kavli Institute for Theoretical Physics, University of California, Santa Barbara, CA 93106, USA}
\author{K.~De~Smedt}\affil{Instituut voor Sterrenkunde, KU Leuven, Celestijnenlaan 200D, 3001 Leuven, Belgium}
\author{ Y. Fremat}\affil{Royal Observatory of Belgium, Ringlaan 3, 1180 Brussels, Belgium}
\author{A.~S.~Fuentes, W.~Homan}\affil{Instituut voor Sterrenkunde, KU Leuven, Celestijnenlaan 200D, 3001 Leuven, Belgium}
\author{ M. Hrudkova, R. Karjalainen}\affil{Isaac Newton Group of Telescopes, Apartado de Correos 321, E-38700 Santa Cruz de la Palma, Canary Islands, Spain}
\author{R.~Lombaert}\affil{Instituut voor Sterrenkunde, KU Leuven, Celestijnenlaan 200D, 3001 Leuven, Belgium}
\author{ P. Nemeth}\affil{Dr. Karl Remeis-Observatory \& ECAP, Astronomisches Inst., FAU Erlangen-Nuremberg, 96049 Bamberg, Germany}
\author{R.~\O{}stensen}\affil{Instituut voor Sterrenkunde, KU Leuven, Celestijnenlaan 200D, 3001 Leuven, Belgium}
\author{G. Van De Steene}\affil{Royal Observatory of Belgium, Ringlaan 3, 1180 Brussels, Belgium}
\author{J.~Vos, G.~Raskin, H.~Van~Winckel}\affil{Instituut voor Sterrenkunde, KU Leuven, Celestijnenlaan 200D, 3001 Leuven, Belgium}
\footnotetext[1]{Based on data gathered with the NASA Discovery mission
  \emph{Kepler\/} and spectra obtained with the HERMES spectrograph, which is
  installed at the Mercator Telescope, operated on the island of La Palma by the
  Flemish Community at the Spanish Observatorio del Roque de los Muchachos of
  the Instituto de Astrof\'isica de Canarias, and supported by the Fund for
  Scientific Research of Flanders (FWO), Belgium, the Research Council of
  KU\,Leuven, Belgium, the Fonds National de la Recherche Scientifique
  (F.R.S.-FNRS), Belgium, the Royal Observatory of Belgium, the Observatoire de
  Gen\`eve, Switzerland, and the Th\"uringer Landessternwarte Tautenburg,
  Germany. All data used in this study are downloadable from
{\tt \scriptsize https://fys.kuleuven.be/ster/Projects/gamma-doradus}}
\altaffiltext{2}{Postdoctoral Fellow of the Fund for Scientific
  Research (FWO), Flanders, Belgium.}  \altaffiltext{3}{Postdoctoral Fellow of
  the Fund for Scientific Research (FWO), Flanders, Belgium.} 
  \altaffiltext{4}{Ph.D. Fellow of the Fund for Scientific Research (FWO), Flanders, Belgium.}

\begin{abstract}
  Gamma Doradus stars (hereafter $\gamma\,$Dor stars) 
are gravity-mode pulsators of spectral type A or F. Such
  modes probe the deep stellar interior, offering a detailed fingerprint of their
  structure. Four-year high-precision space-based {\it Kepler\/} photometry of
  $\gamma$\,Dor stars has become available, allowing us to study these stars
  with unprecedented detail.  We selected, analysed, and characterized a sample
  of 67 $\gamma$\,Dor stars for which we have \emph{Kepler\/} observations 
  available.  For all the targets in the sample we assembled high-resolution 
  spectroscopy to confirm their F-type nature.  We found fourteen binaries, 
  among which four single-lined binaries, five double-lined binaries, two triple 
  systems and three binaries with no detected radial velocity variations.
  We estimated the orbital parameters whenever
  possible. For the single stars and the single-lined binaries, fundamental
  parameter values were determined from spectroscopy. We searched for period
  spacing patterns in the photometric data and identified this diagnostic for 50
  of the stars in the sample, 46 of which are single stars or
  single-lined binaries. We found a strong correlation between the spectroscopic
  $v\sin i$ and the period spacing values, confirming the influence of rotation
  on $\gamma$\,Dor-type pulsations as predicted by theory. We also found
  relations between the dominant g-mode frequency, the longest 
pulsation period detected in series of prograde modes, 
  $v\sin i$, and $\log\,T_{\rm eff}$.
\end{abstract}

\section{Introduction}
\label{sec:intro}

One of the most important elements of modern astronomy and astrophysics is the
theory of stellar structure and evolution. It has a direct impact on research
domains such as the interstellar medium, exoplanets, galactic evolution
etc. While the general outline of this theory is nowadays well understood and
globally accepted, various aspects such as the interior stellar rotation and the
chemical mixing processes inside stars are largely unknown.

A key element of stellar structure and evolution which may vary between
different theoretical models is the evolution of the convective core in
intermediate-mass ($1\,M_\odot \lesssim M_* \lesssim 2\,M_\odot$) main-sequence
stars. This is the transition range from low-mass stars with a radiative core
and a convective envelope to high-mass stars with a convective core and a
radiative envelope. For these stars the evolution of the convective core
throughout the star's core-hydrogen burning phase is dominated by the changes in
the local opacity and the energy produced by the CNO hydrogen burning cycle with
respect to the $pp$ chains. Due to the interplay of these physical processes,
the convective core can either grow or shrink during the core hydrogen burning
phase, depending on the star's initial mass
\citep[e.g.][]{Mitalas1972,Crowe1982,Silva2011}. The precise transition point
between these two regimes in theoretical models not only depends on the
star's chemical composition and the internal mixing processes, but also on their
numerical treatment \citep{Miglio2008a}.

A good way to calibrate and improve upon the existing stellar structure and
evolution theories in this mass range, is a detailed asteroseismic study of
$\gamma\,$Dor pulsators. Gamma Dor stars are slightly more massive (1.4
to 2.5 $M_\odot$) and more luminous than the Sun and exhibit non-radial
gravity-mode pulsations \citep[e.g.,][]{Kaye1999}. The pulsations, which have 
a typical period between 0.3 and 3 days and are excited by the convective 
flux blocking mechanism at the bottom of the convective envelope, probe the
internal stellar structure of the stars up to the edge of the convective core
\citep{Guzik2000,Dupret2005}. It was shown by \citet{Tassoul1980} that for a
non-rotating star consisting of a convective core surrounded by a chemically
homogeneous radiative envelope, pulsations with the same spherical degree $l$
and different radial orders $n$ are equidistantly spaced in the period domain if
$l \ll n$. This model was strongly improved upon by \citet{Miglio2008a}, who
showed that a chemical gradient at the edge of the convective core modifies the
g-mode resonance cavity, which in turn results in periodic dips in the period
spacing pattern, where the depth and the regularity of the dips indicate the
steepness and the location of the gradient, respectively. In the case of a
growing convective core, there is a chemical discontinuity at the edge of the
core, whereas a shrinking core leaves behind a chemical gradient in the near
core region. \citet{Bouabid2013} studied the influence of extra mixing processes
and stellar rotation. Chemical mixing washes out the $\mu$-gradient, which 
strongly reduces the prominence of the dips in the period spacing pattern.

Rotation introduces frequency splitting, which in the case of fast rotation
results in separated period spacing patterns depending on the value of the
azimuthal order $m$. Retrograde mode patterns will typically have an upward
slope, while both prograde and zonal mode patterns have a downward slope in the
graphical representation of $\Delta\,P$ versus $P$ \citep{Bouabid2013}.  For
modes excited with equal amplitude, it is expected that prograde sectoral modes
(with $0 < m = l$) are the easiest to detect. It has been shown that for all
other pulsation modes the Coriolis force will trap the pulsation within an
equatorial waveguide, which becomes more narrow as the rotational velocity
increases. As a result, the geometrical cancellation effects will be more
pronounced for these oscillation modes, making it harder to detect them
\citep{Townsend2003}.

Thanks to the space-mission \emph{Kepler\/} \citep{Koch2010}, we now have
photometric observations of four year duration, which is essential for
gravity-mode pulsators to enable successful seismic modelling
\citep[e.g.,][]{Papics2014}.  The advantage of space-based observations is the
large number of frequencies which can be resolved, as opposed to only a handful of
modes detected in ground-based data of $\gamma\,$Dor stars 
\citep[e.g.,][]{Cuypers2009}.  Various
period spacing patterns for AF-type main-sequence pulsators have meanwhile been
reported \citep[e.g.,][]{Chapellier2012, Kurtz2014, Bedding2014, Saio2015}, as
well as sample compilations and studies of such pulsator classes
\citep[e.g.][]{Grigahcene2010,Uytterhoeven2011,Balona2011,Bradley2015}.

Here, we study the sample of 70 $\gamma\,$Dor candidates as composed by
\citet{Tkachenko2013} in full depth. For all the stars in this sample, we
collected the four-year high-precision \emph{Kepler\/} photometry and we
gathered high-resolution spectroscopy to make a full observational
characterization of the sample. This allows us to do a detailed statistical
analysis of the observed properties for this sample of gravity-mode pulsators.
In Section \ref{sec:obs}, we discuss the available data and the applied data
reduction methods. In Section \ref{sec:spec}, we evaluate the $\gamma$
Dor-type nature of the targets and check for binarity. We also list the
fundamental parameter values derived from the spectroscopy. We then look for
period spacing patterns in the photometric data in Section \ref{sec:perspacings}
and check for correlations with the spectroscopic parameter values (Section
\ref{sec:specvsphot}), before coming to our conclusions (Section
\ref{sec:conclusions}).
All our data produces are made publicly available along with this publication.

\section{Observations and data reduction}
\label{sec:obs}
For each star in our sample we obtained high-resolution spectra between May and
August in 2011, 2013, and 2014 with the HERMES spectrograph (R = 85000,
$\lambda$ = 377\,-\,900\,nm, \citet{Raskin2011}) at the 1.2-m Mercator telescope
(Observatorio del Roque de los Muchachos, La Palma, Canary Islands, Spain). The 
total number of spectra and observation times are listed per star in the appendix
in Table \ref{tab:obs}.

The obtained spectra were reduced with the most recent version of the HERMES
pipeline (release 5) and the individual spectra were subsequently normalised in
two steps by fitting a cubic and linear spline through the flux continuum
respectively, as described by \citet{Papics2012}. Outliers in the spectra
resulting from cosmic hits were identified by comparing the local flux with the
median value in the surrounding spectral region (of 40 pixels). If the
difference exceeded 5$\sigma$, where $\sigma$ is determined as the local 
signal-to-noise (S/N) ratio in the spectrum, the flux point was replaced by 
the median value.

In addition to the spectroscopic observations, all the stars in our sample have
been observed by the \emph{Kepler\/} mission with long cadence (i.e. with an
integration time of 29.4 minutes). For the majority of stars, data are
available for all eighteen quarters (Q0-Q17), while the shortest obtained light
curves cover eight quarters. The available quarters are also listed for each
star in Table\,\ref{tab:obs}.

Because the standard light curve extraction algorithm, provided by the Mikulski
Archive for Space Telescopes (MAST), is known to result in residual instrumental
signal in the low frequency region \citep{Debosscher2013,Tkachenko2013}, we
apply the light curve extraction algorithm developed by one of us (S.B.). The
code is based on the use of customised masks, which allows us to include the
signal from more pixels, thereby reducing the influence of instrumental
trends. The light curves of each individual quarter were subsequently detrended
by changing the scale of the extracted light curves to magnitude and subtracting
a $1^{st}$- or $2^{nd}$-order polynomial, and concatenated. The outliers of the
light curves were removed manually. We then extracted the frequencies from the
reduced light curves by an iterative prewhitening method based on the
Lomb-Scargle periodogram, as described in detail by \citet{VanReeth2015}.
Observed oscillation frequencies above the Nyquist frequency were evaluated by
comparing their amplitudes with those of possible aliases. Following
\citet{Murphy2013}, the frequency with the highest amplitude was retained. The
errors of the frequency values were determined using the method by
\citet{SchwarzenbergCzerny2003}, which is based on the statistical errors
resulting from a non-linear least-squares fit corrected for the correlated
nature of the data.

\section{Spectroscopic analysis}
\label{sec:spec}
Visual inspection of the available spectra revealed that two of the targets,
KIC\,5772452 and KIC\,7516703, are in fact rotationally modulated K-type
stars. They have therefore been discarded for the rest of this paper.
In addition, one of the stars in the sample, KIC\,10080943, had
already been identified as a double-lined binary by \citet{Tkachenko2013}. 
This star has been monitored extensively in spectroscopy meanwhile and 
reveals tidally-excited as well as free oscillations. Given that some of 
its modes are of different origin than those of the other sample stars, it will 
be discussed separately (Schmid et al, in preparation). This leaves us 
with 67 stars for the current study.

\subsection{Binarity}
\label{subsec:binarity}
% TO PROVIDE: electronic table with radial velocities + place on website
In a first step in our data analysis we checked the targets in our sample for
binarity.  We applied the improved least-squares deconvolution (LSD) algorithm,
as described by \citet{Tkachenko2013c}, to the individual spectra of each star
using a line mask with typical parameter values for a $\gamma\,$Dor star
($T_{\rm eff} = 7000\,$K; $\log\,g = 4.0$; $[M/H] = 0.0$; $\zeta = 2\,$km\,s$^{-1}$).
The line mask was computed with the spectrum
synthesis synthV code \citep{Tsymbal1996}, which makes use of a spectral line
list (obtained from the VALD database, \citet{Kupka1999}) and a predefined grid
of atmospheric models (computed with the LLmodels code, \citet{Shulyak2004}). 
This resulted in high S/N average line profiles which were
checked for the presence of a signal from a possible companion. This resulted in the
detection of four single-lined binaries 
(KIC\,6468146, KIC\,7867348, KIC\,10467146, KIC\,11754232), four double-lined 
binaries (KIC\,3222854, KIC\,3952623, KIC\,6367159, KIC\,8693972), two 
triple systems (KIC\,6467639, KIC\,6778063) and three binaries for which we have
spectral contributions from both components, but no detected radial-velocity 
variations. This is also listed in Table\,\ref{tab:characterisation}.

\begin{deluxetable}{lccccccccc}
% \rotate
\tabcolsep=2pt
\tablecolumns{10}
\tablewidth{0pc}
\tabletypesize{\footnotesize}
\tablecaption{\label{tab:bin}The obtained parameter values and their respective error margins for several of the binary systems in our sample. For KIC\,3952623, KIC\,7867348 and KIC\,11754232 we were able to obtain an orbital solution.}
\tablehead{
\colhead{} & KIC\,3222854 & \hspace{0.2cm} & \multicolumn{3}{c}{KIC\,3952623} & \hspace{0.2cm} & KIC\,7836348 & KIC\,8693972 & KIC\,11754232}
\startdata
method & spectroscopy & & spectroscopy & FM method & PM method & & spectroscopy & spectroscopy & spectroscopy\\
$P$ [days] & $\gtrsim 100$ & & 19.515 & 19.5 & 19.526 & & 10.0132 & $\sim$ 60, 90, 120 or 180 & 214.9\\
$\sigma_P$ & \nodata & & 0.006 & 0.2 & 0.001 & & 0.0006 & \nodata & 0.6\\
$e$ & \nodata & & 0.17 & 0 & 0.191 & & 0.23 & \nodata & 0.55\\
$\sigma_e$ & \nodata & & 0.04 & \nodata & 0.001 & & 0.03 & \nodata & 0.01\\
$\omega$ [2$\pi$] & \nodata & & 1.65 & \nodata & 1.5481 & & 1.05 & \nodata & 4.52\\
$\sigma_\omega$ & \nodata & & 0.1 & \nodata & 0.0006 & & 0.11 & \nodata & 0.03\\
$T_0$ [BJD] & \nodata & & 2456442.8 & \nodata & \nodata & & 2455685.96 & \nodata & 2456417.23\\
$\sigma_{T_0}$ & \nodata & & 0.3 & \nodata & \nodata & & 0.15 & \nodata & 1.38\\
$K_1$ [$km\,s^{-1}$] & \nodata & & 30 & 30.08 & 19.36 & & 9.3 & \nodata & 17.8\\
$\sigma_{K_1}$ & \nodata & & 0.8 & \nodata & 0.04 & & 0.2 & \nodata & 0.2\\
$K_2$ [$km\,s^{-1}$] & \nodata & & 52.5 & \nodata & \nodata & & \nodata& \nodata  & \nodata\\
$\sigma_{K_2}$ & \nodata & & 1.4 & \nodata & \nodata & & \nodata & \nodata & \nodata\\
$a_1\sin i$ [AU] & \nodata & & 0.053 & 0.054 & \nodata & & 0.0083 & \nodata & 0.294\\
$\sigma_{a_1\sin i}$ & \nodata & & 0.001 & \nodata & \nodata & & 0.0002 & \nodata & 0.004\\
$a_2\sin i$ [AU] & \nodata & & 0.093 & \nodata & \nodata & & \nodata & \nodata & \nodata\\
$\sigma_{a_2\sin i}$ & \nodata & & 0.003 & \nodata & \nodata & & \nodata & \nodata & \nodata\\
$\gamma$ [$km\,s^{-1}$] & -39.5 & & -15.7 & \nodata & \nodata & & -26 & -5.9 & 13.5\\
$\sigma_{\gamma}$ & 1.7 & & 2.5 & \nodata & \nodata & & 0.2 & 3.9 & 0.4\\
\enddata
\end{deluxetable}

The radial velocities of the different components of a binary were obtained by
taking the center of gravity of their LSD profile.\footnote{These are available
  from\\{\tt \small https://fys.kuleuven.be/ster/Projects/gamma-doradus}}  
For some of the detected single- and double-lined
binaries we have obtained a sufficient number of observations to cover the
binary orbit, allowing us to obtain an (order of magnitude) estimate of the
orbital period $P$, as shown in Table\,\ref{tab:bin}. For KIC\,3952623, KIC\,7867348,
and KIC\,11754232 these values were sufficiently accurate to allow us to
determine values for the remaining spectroscopic orbital parameters (eccentricity 
$e$, the angle $\omega$ between the ascending node and the periapsis, the time 
$T_0$ of periastron passage, the system velocity $\gamma$, velocity amplitudes 
$K_j$ and the projected semi-major axes $a_j\sin i$) as well,
which are again listed in Table\,\ref{tab:bin}. The radial velocity variations of
the two triple systems proved to be complicated to unravel. Here a more detailed
analysis is required,  which is outside the scope of this paper.

We also applied the frequency modulation (FM) method \citep{Shibahashi2012} and
the phase modulation (PM) method \citep{Murphy2014} to the pulsation frequencies
extracted from the \emph{Kepler\/} light curves.  As \citet{Murphy2014} pointed
out, the FM method is  most sensitive for the detection of relatively
short-period binaries, the PM method is more likely to be successful for the
detection of long-period binaries and lower-frequency pulsations.  In the case
of KIC\,3952623, we were able to confirm the spectroscopic results with both the
FM and PM methods, as shown in Table\,\ref{tab:bin}, though the PM method leads 
to an underestimate of the velocity amplitude due to the relatively short orbital 
period.  Neither method provided a satisfactory result for the other observed 
binaries.  There are several possible explanations. Firstly, an important requirement 
for these methods is that $P_{\rm puls} \ll P_{\rm orbit}$. The typical $\gamma\,$Dor
pulsation period has a value between 0.3 and 3 days, so that often the difference 
between $P_{\rm puls}$ and $P_{\rm orbit}$ is too small. Secondly, as we will see in 
Section \ref{sec:perspacings}, we often observe period spacing patterns with a value
between 0.001 and 0.035\,d, depending on the rotation velocity and the spherical
degree $l$ of the mode. The corresponding spacings in the frequency domain are
close to the ones we would obtain for a binary orbit above 20\,d.  Aside from
KIC\,3952623, there are indications of the presence of the expected signal for
KIC\,11754232, but unfortunately we have only eight quarters of data available
for this star, which prevented a reliable analysis.

Out of the fourteen targets that we found to be binaries, seven are identified as
$\gamma\,$Dor/$\delta\,$Sct hybrid pulsators (out of eleven hybrids in the studied
sample of 69 targets). While this could mean that several of these binary
systems consist of a $\gamma\,$Dor and a $\delta\,$Sct star, we cannot exclude
the possibility that one (or both) of the components are hybrid
pulsators. Indeed, three hybrids are identified as single-lined binaries, and from
the application of the FM method to KIC\,3952623 we conclude that the
studied $\delta\,$Sct pulsations belong to the brightest component, which seems
to indicate that the primary is a hybrid, considering the relatively high
amplitudes of the $\gamma\,$Dor pulsations.

\subsection{Fundamental parameters}
\label{subsec:parameters}
In the second part of the spectroscopic analysis, we focused on the stars which
have been classified as either single stars or single-lined binaries. The
spectra of these stars were corrected for any radial velocity variations (in the
case of the single-lined binaries) and merged. This resulted in a single
spectrum for each star, which we subsequently used to determine the fundamental
spectroscopic parameter values. The advantage is that such a combined spectrum 
has a higher S/N ratio and that pulsation line profile variations,
which might otherwise skew the spectral analysis, are largely averaged out.

We use the Grid Search in Stellar Parameters (GSSP) code for the spectral
analysis \citep{Lehmann2011,Tkachenko2012}.  The values of the temperature 
$T_{\rm eff}$, surface gravity $\log\,g$, metallicity $[M/H]$, microturbulence 
$\zeta$ and rotational velocity $v\sin i$ are determined from a
$\chi^2$-minimization for the observed spectra with respect to synthetic spectra
within a relatively large wavelength range. The synthetic spectra are computed
for different values of the listed parameters using a combination of the synthV
code \citep{Tsymbal1996}, the LLmodels code \citep{Shulyak2004} and information
obtained from the VALD database \citep{Kupka1999}, i.e. similar to how the LSD line
mask was computed. In this study we took the spectral range between 4700\,\AA\
and 5800\,\AA, which includes the $H_\beta$-line. For a typical $\gamma\,$Dor star, 
the Balmer lines are mostly sensitive to $T_{\rm eff}$ \citep[e.g.][]{Gray1992}. 
By including the $H_\beta$-line into the spectral fits, we were able to put 
strong constraints on the values of $T_{\rm eff}$.

\begin{figure*}
\centering
 \includegraphics[width=0.49\textwidth]{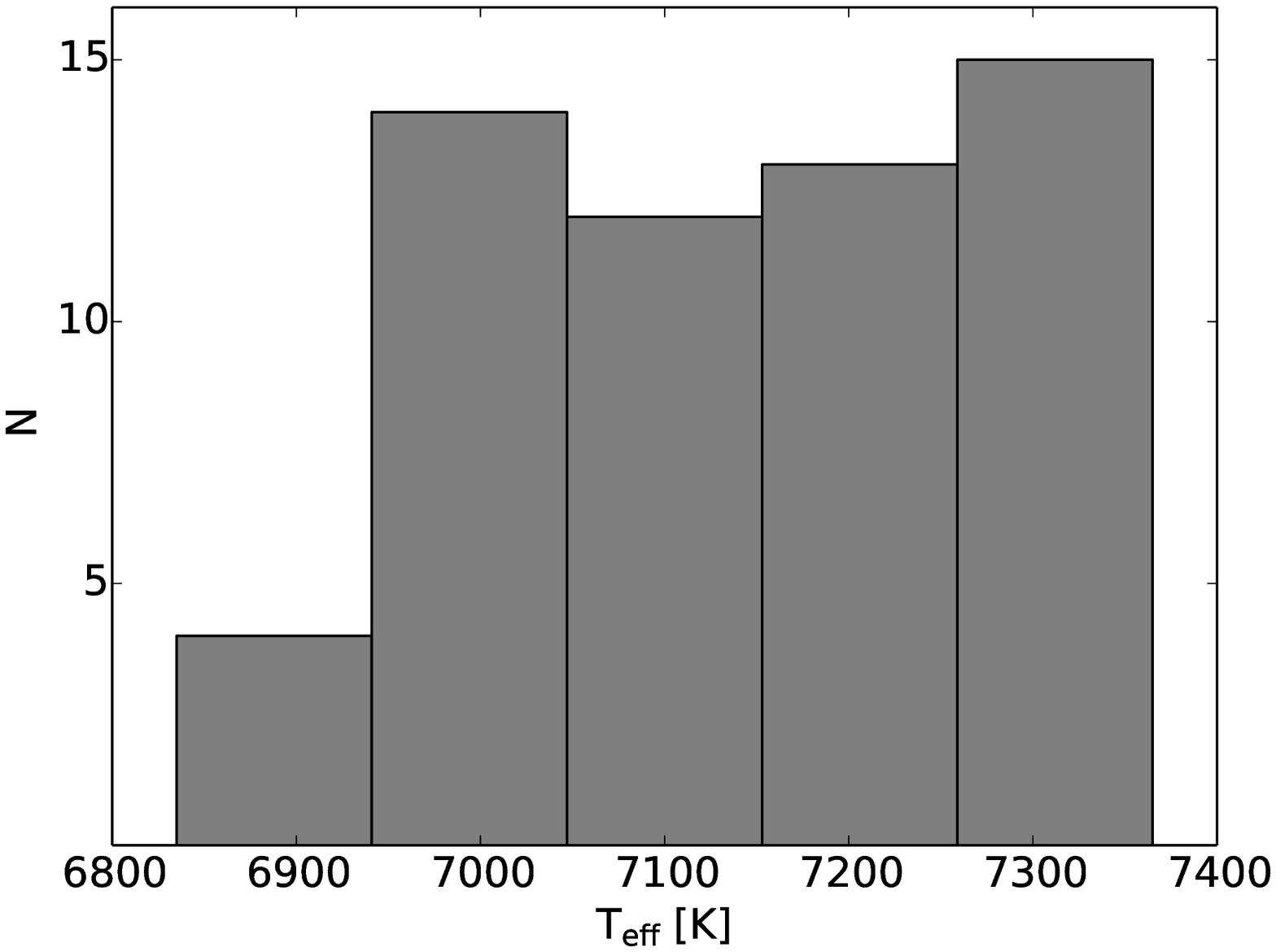}
 \includegraphics[width=0.49\textwidth]{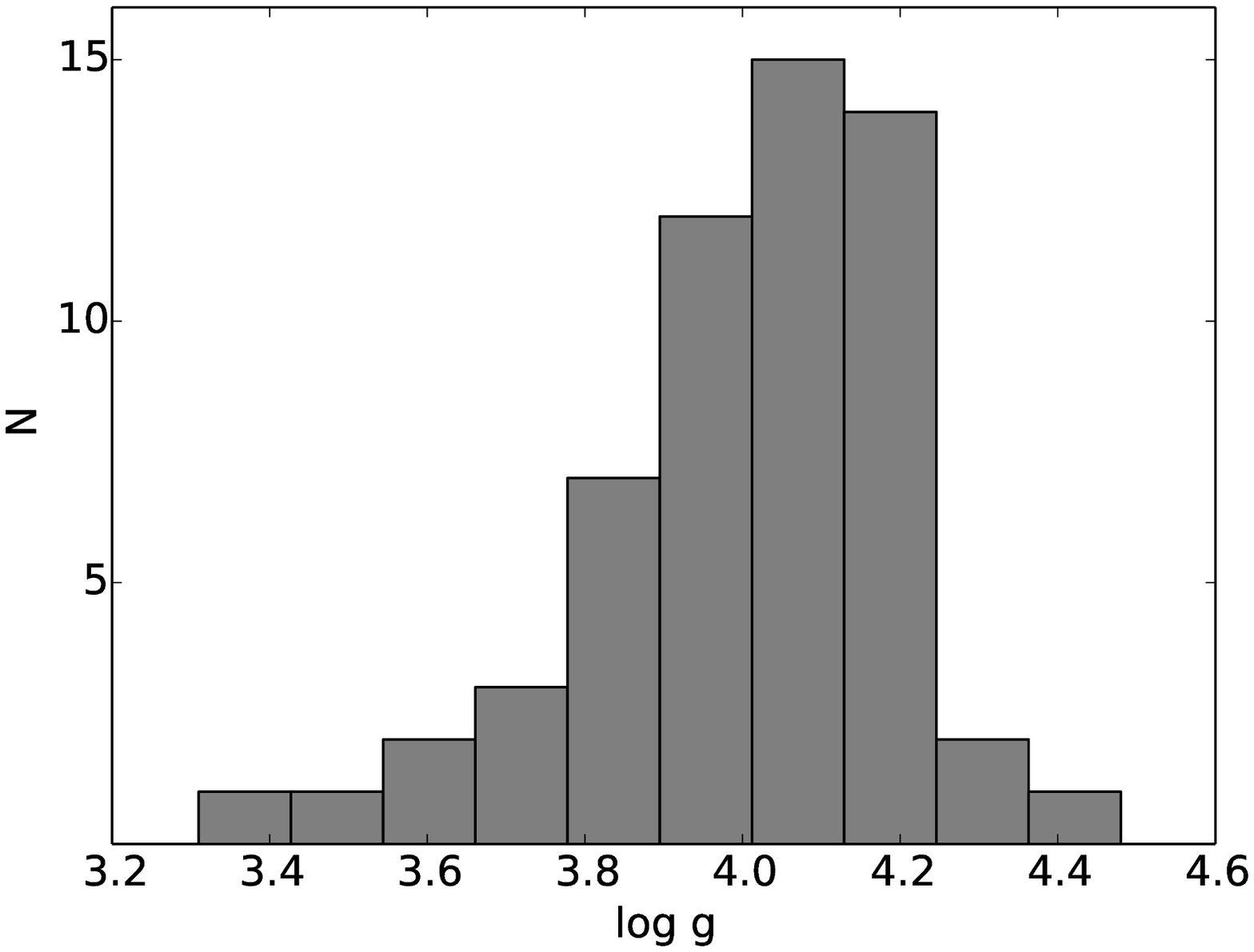}
 \includegraphics[width=0.49\textwidth]{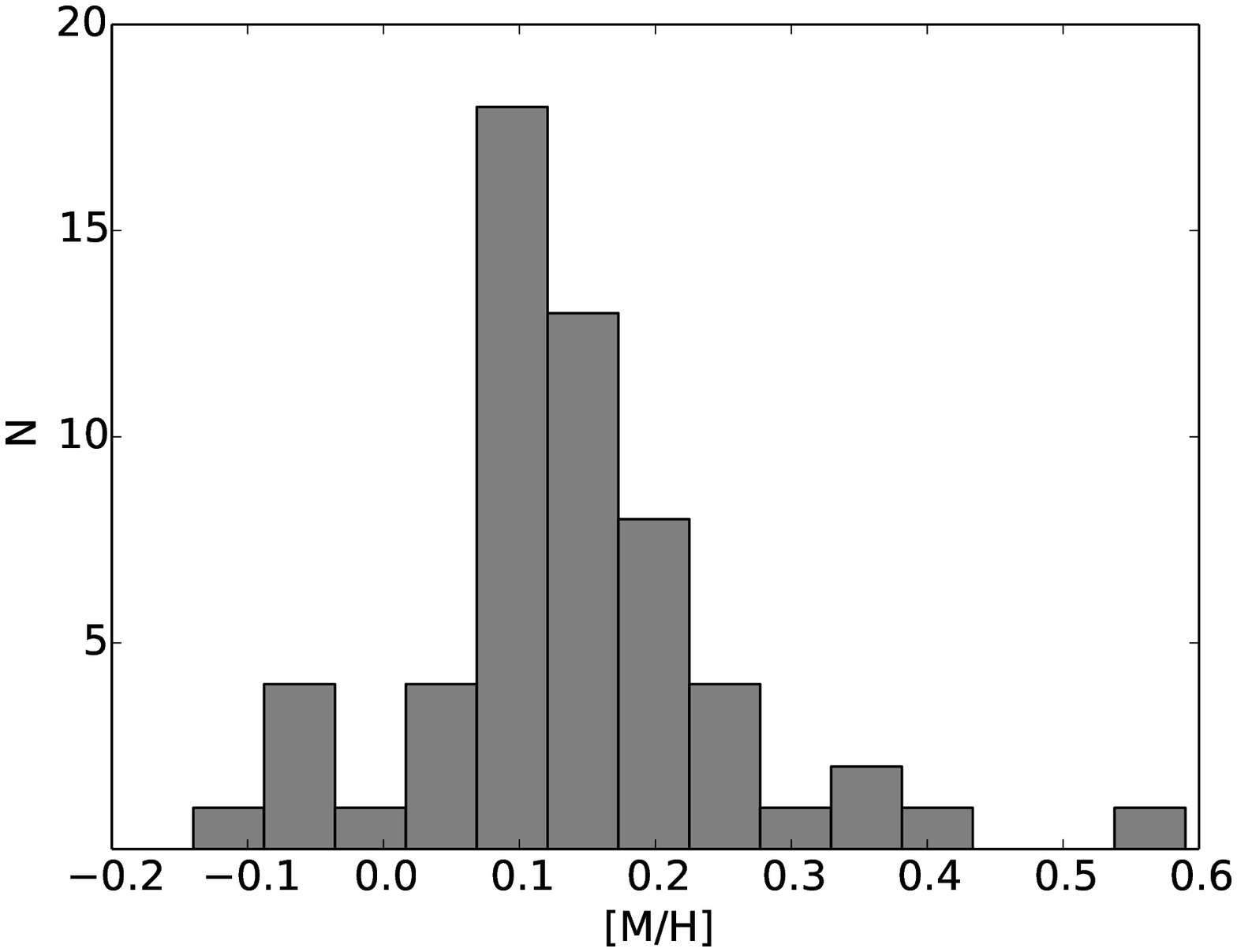}
 \includegraphics[width=0.49\textwidth]{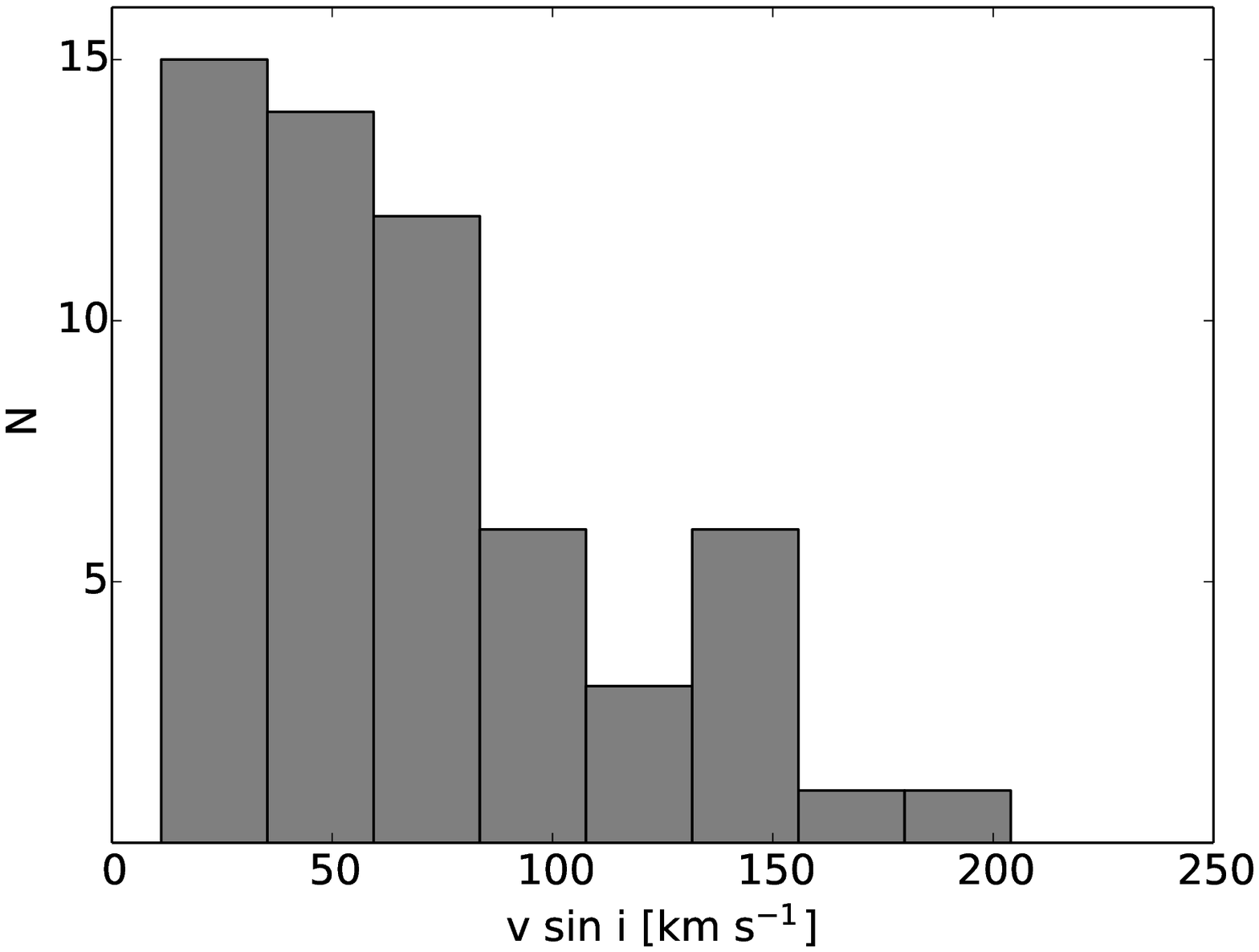}
 \includegraphics[width=0.49\textwidth]{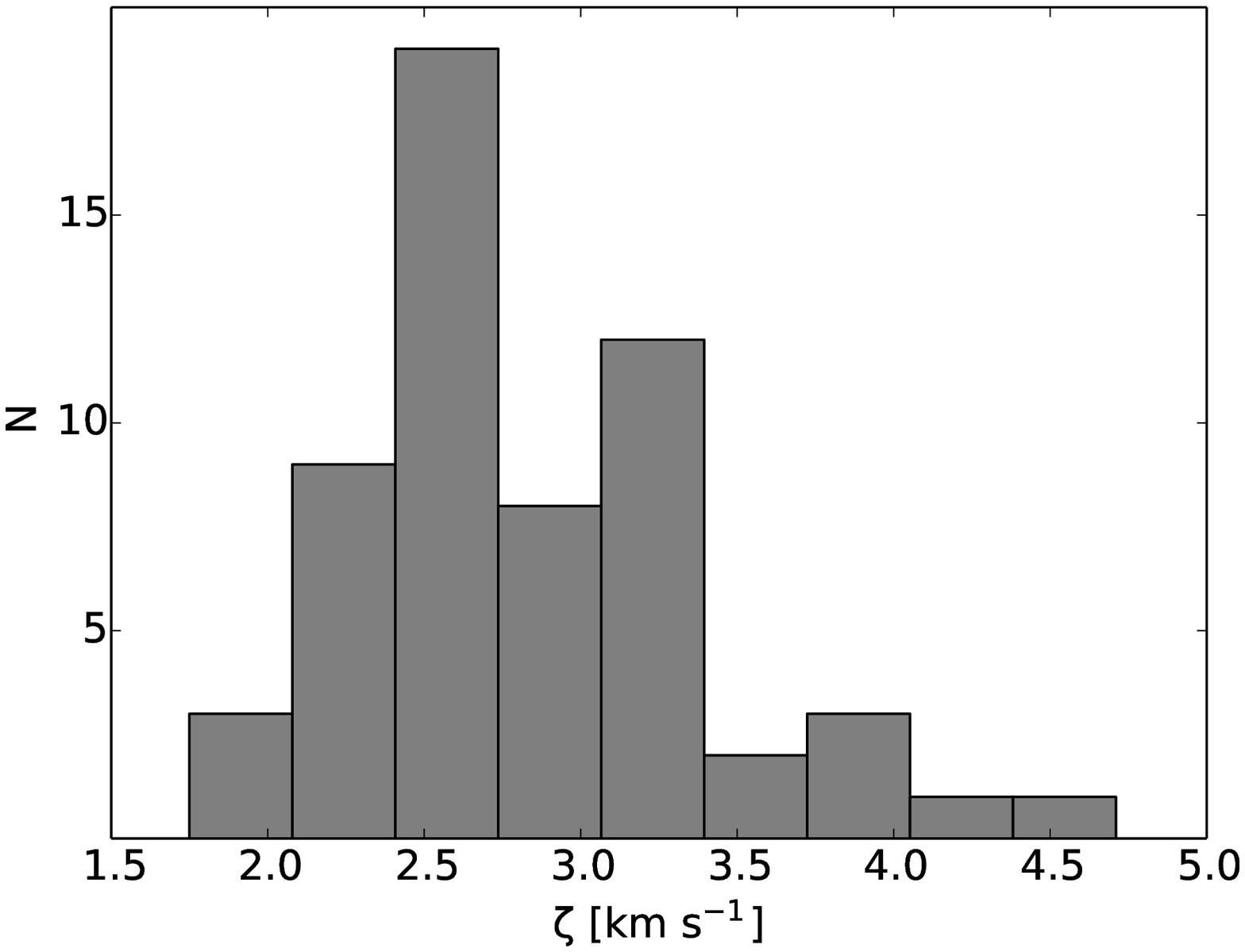}
 \caption{\label{fig:param_hist}Histograms representing the distribution of the
   atmospheric parameter values obtained for the single stars and the
   single-lined binaries in our sample.}
\end{figure*}

The obtained parameter values are listed in Table\,\ref{tab:param} and some
sample distributions are illustrated in Fig.\,\ref{fig:param_hist}. As we can see,
$T_{\rm eff}$ typically has a value near 7100\,K, while $\log\,g$ is close to
4.1. The metallicity, $[M/H]$, is on average slightly higher than the solar
metallicity, while $\zeta$ typically has a value between 2 and
3.5\,km\,s$^{-1}$. These are in close agreement with the parameters we had
chosen for the LSD line mask in Section\,\ref{subsec:binarity}.  The metallicity values
previously computed by \citet{Tkachenko2013} are on average slightly below the
solar value, because $\zeta$ was kept fixed at 2\,km\,s$^{-1}$ in that study,
whereas we included it as a free parameter.  
  We also find that most $\gamma\,$Dor stars in our sample are slow to moderate
  rotators compared to their break-up velocity $v_c =
  \sqrt{2GM_*/3R_*}$, where $M_*$ and $R_*$ are the stellar mass and
  radius. For a $1.6\,M_\odot$ star with a $1.9\,R_\odot$ radius we find $v_c =
  327\,$km\,s$^{-1}$.  Typical values for $\gamma\,$Dor stars are all on the order of
  300\,km\,$s^{-1}$.

\begin{figure}
 \centering
 \includegraphics[width=0.49\textwidth]{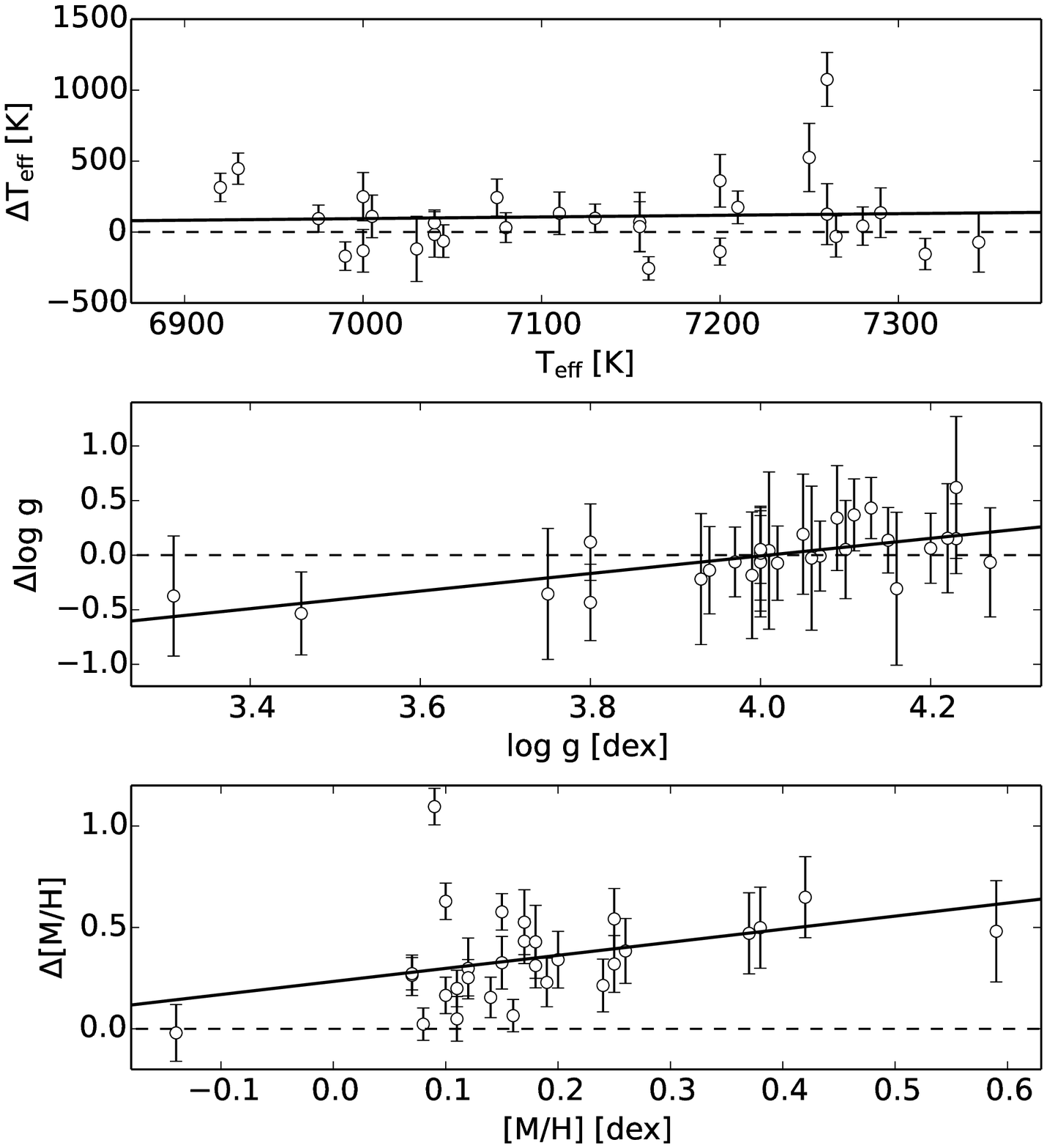}
 \caption{\label{fig:diff_GSSP} The differences between the spectroscopically derived
   values and the KIC values of the fundamental parameters $T_{\rm eff}$ (top),
   $\log\,g$ (middle) and $[M/H]$ (bottom), as a function of the spectroscopic values. 
   Linear fits of the relations are represented by the solid black lines.}
\end{figure}

A comparison of the spectroscopically derived parameters values of $T_{\rm eff}$, 
$\log\,g$ and $[M/H]$ with the values found in the KIC catalogue, which are 
derived from SED fitting, is shown in Fig. \ref{fig:diff_GSSP}. It is found that, 
apart from a few outliers, the $T_{\rm eff}$ values in the KIC catalogue are 
reasonably reliable, while the deviations of the surface gravity and metallicity 
values seem to be correlated to the actual parameter values. This relation is similar
to the one found by \citet{Tkachenko2013b}.

\section{Period spacing patterns in the \emph{Kepler\/} photometry} 
\label{sec:perspacings}
\begin{figure*}
\includegraphics[width=0.49\textwidth]{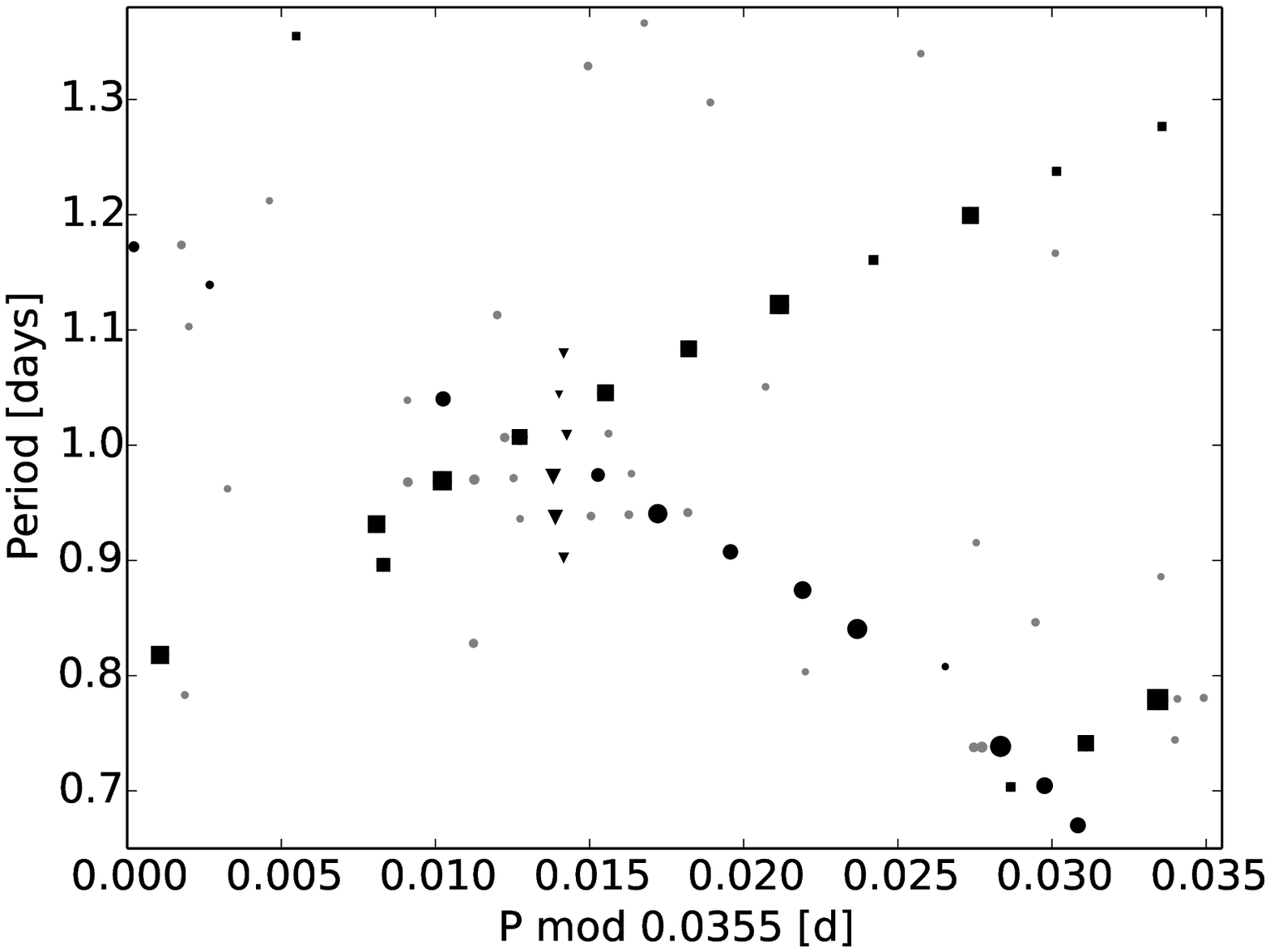}
 \includegraphics[width=0.49\textwidth]{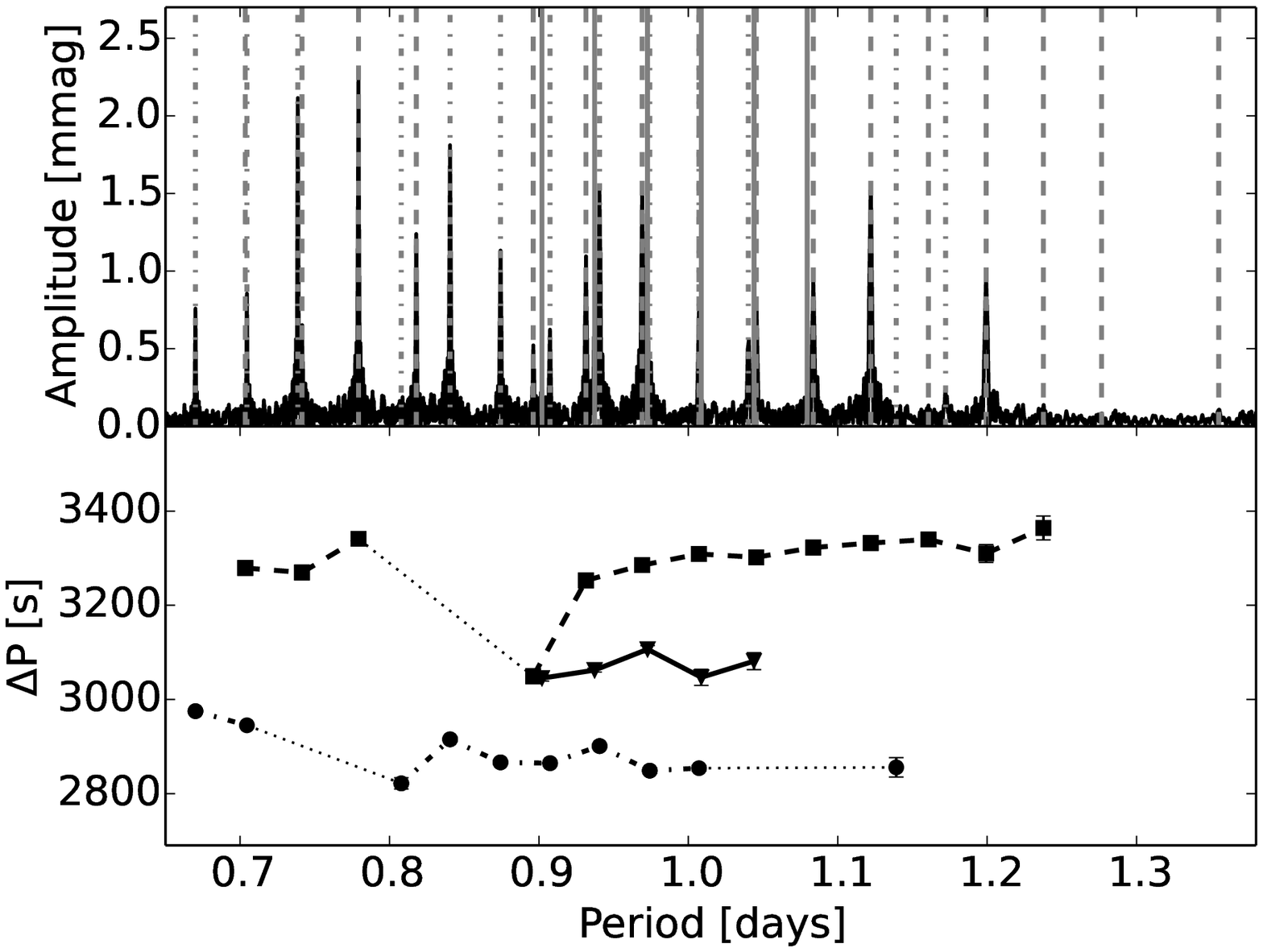}
%  \plottwo{KIC9751996_1.eps}{KIC9751996_2.eps}
 \caption{\label{fig:kic9751996}The period spacing patterns of the slowly 
   rotating star KIC\,9751996. \emph{Left:} the pulsation series in a period \'echelle
   diagram. The prograde (black circles), zonal (black triangles) and retrograde
   (black squares) dipole modes are marked separately, while the other
   pulsation modes are shown in grey. \emph{Top right:} the prograde (dash-dot lines),
   zonal (full lines) and retrograde (dashed lines) modes in the Fourier
   spectrum. \emph{Bottom right:} the period spacing patterns, using the same symbols
   as before. The dotted lines indicate missing frequencies.}
\end{figure*}

\begin{figure*}
\centering
 \includegraphics[width=\textwidth]{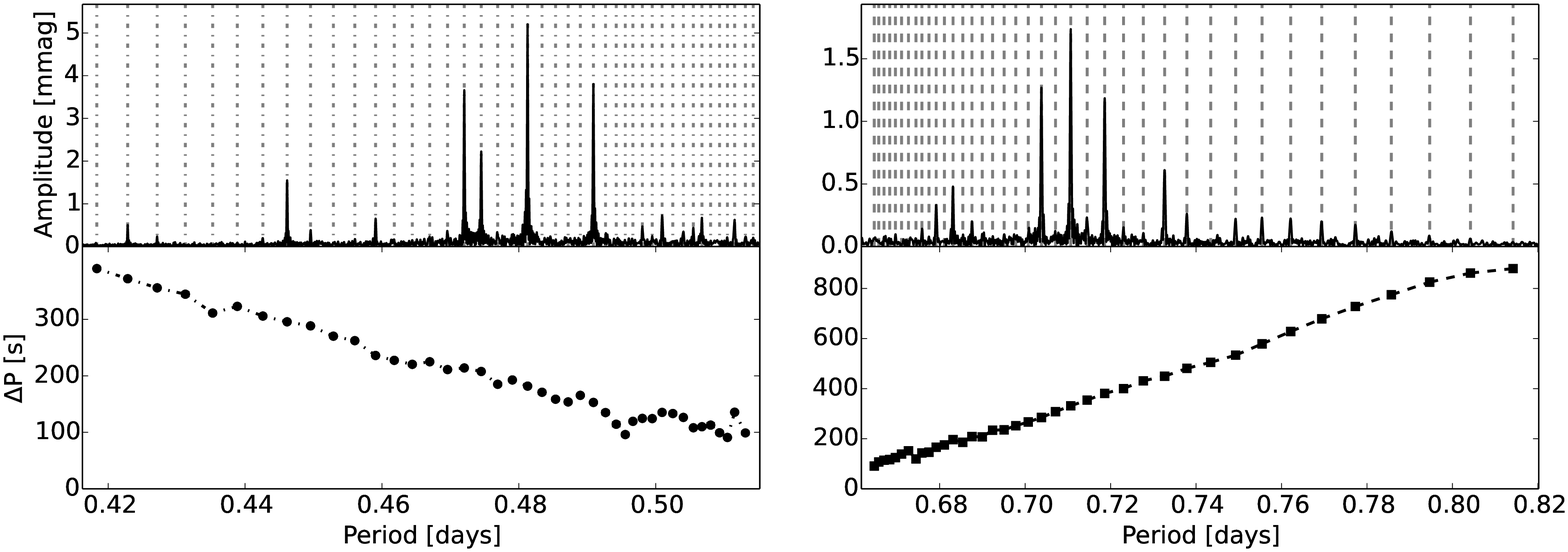}
 \caption{\label{fig:kic8375138}The prograde (left) and retrograde (right)
   period spacing patterns of KIC\,8375138. The used symbols are the same as in
   Fig.\,\ref{fig:kic9751996}.}
\end{figure*}

\begin{figure*}
\centering
 \includegraphics[width=\textwidth]{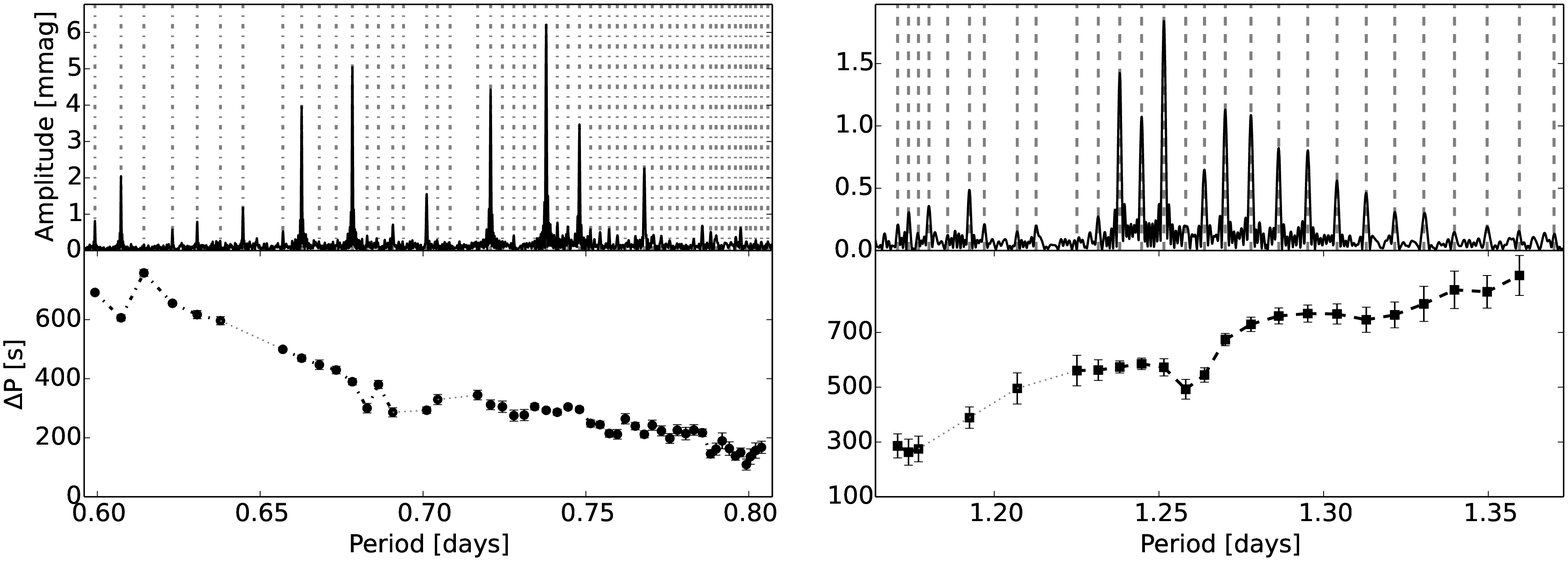}
 \caption{\label{fig:kic2710594}The prograde (left) and retrograde (right)
   period spacing patterns of KIC\,2710594. The used symbols are the same as in
   Fig.\,\ref{fig:kic9751996}.}
\end{figure*}

\begin{figure}
\centering
 \includegraphics[width=0.49\textwidth]{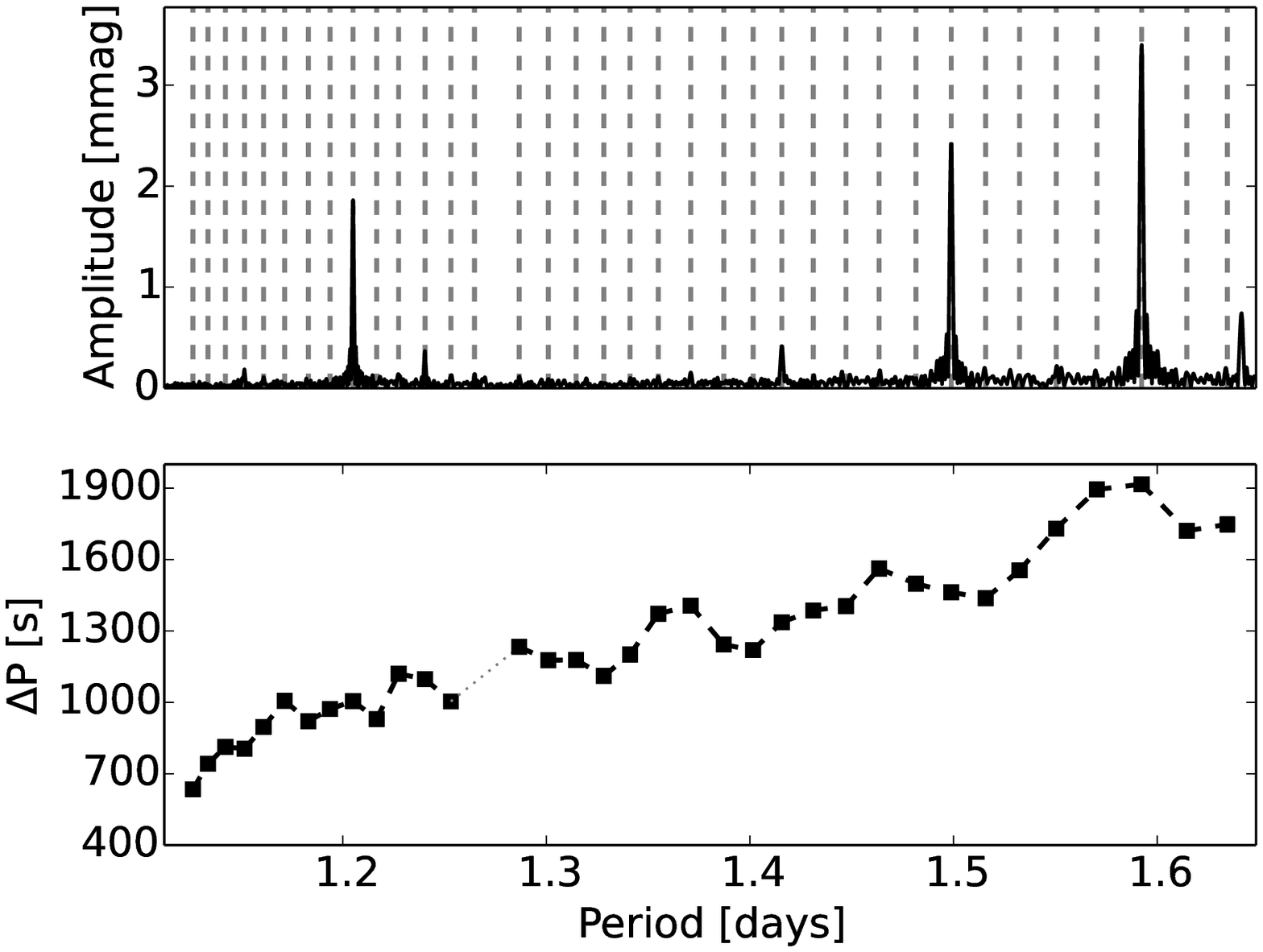}
 \caption{\label{fig:kic11668783b}The retrograde period spacing patterns of
   KIC\,11668783. The used symbols are the same as in
   Fig.\,\ref{fig:kic9751996}.}
\end{figure}

Having confirmed the $\gamma\,$Dor nature for the  vast majority of the targets in our
sample and having identified the binaries, we used the extracted pulsation
frequencies to
look for period spacings by means of the methodology described in
\citet{VanReeth2015}. This resulted in the {\it detection of period spacings for
  45 of the 67 studied targets}. These are listed in
Tables\,\ref{tab:characterisation} and \ref{tab:persp}. The found spacing
patterns are all shown in Appendix\,\ref{app:persp}. A selection can also be found
in Figs.\,\ref{fig:kic9751996} to \ref{fig:kic11668783b}.

Most of the period spacing patterns have a clear slope. This has been observed
before \citep{Bedding2014,VanReeth2015} and is a consequence of frequency shifts
due to rotation. Interestingly, most of the detected period spacing patterns
seem to be complementary to the ones presented by \citet{Bedding2014}.  These
authors considered slowly rotating stars without spectroscopic confirmation that
it actually concerns F-type stars, while our method is also suitable to detect
the period spacing patterns of the moderate to fast rotating stars in our
sample. While this difference is clearly visible in the increasingly strong
downward slopes of many of the detected patterns, there are also several
detected spacing series with an upward slope.  

As discussed in Section\,\ref{sec:intro}, the period spacing patterns with a
downward slope are either prograde or zonal modes. Given that zonal modes are
expected to be more susceptible to trapping in an equatorial waveguide under the
influence of the Coriolis force \citep{Townsend2003}, we expect most, if not
all, of the detected modes to be prograde. According to theory, spacing 
series with an upward slope correspond to retrograde modes that are shifted 
in the frequency domain because of the stellar rotation. It is expected that 
the average period spacing value of retrograde modes increases for higher 
rotation rates \citep{Bouabid2013}, as found for KIC\,11145123 \citep{Kurtz2014}, 
KIC\,9244992 \citep{Saio2015} and KIC\,9751996 (see
Fig.\,\ref{fig:kic9751996}). Nevertheless, for several stars, such as
e.g. KIC\,8375138 (see Fig.\,\ref{fig:kic8375138}), the observed mean period
spacing value for the series with the upward slope is smaller than it would be
if the star did not rotate. The reason is that for these stars, the rotation
frequency is so high the retrograde modes are actually shifted below zero in the
inertial reference frame. When observed, pulsation frequencies are assumed to be
positive, so that when the retrograde modes are observed, we end up with a
spacing pattern such as the one seen in Fig.\,\ref{fig:kic8375138}. While
\citet{Bouabid2013} already mentioned that this might occur, we find this to be
fairly common since no less than ten stars with a detected period spacing
pattern are fast rotators exhibiting both prograde and retrograde modes.

For six stars in our sample, such as KIC\,11668783 illustrated in
Fig.\,\ref{fig:kic11668783b}, we find a period spacing pattern connected with
retrograde modes only, all of which having similar characteristic amplitude
patterns in the sense that the mode amplitude grows as the pulsation period
increases and drops sharply (see Figs.\,\ref{fig:kic8375138} and
\ref{fig:kic2710594}).  The amplitude comb of KIC\,8375138 as seen in
Fig.\,\ref{fig:kic8375138} reveals dips in the amplitudes in an otherwise
monotonically growing period series.

Even though many of the observed periods spacing patterns are reasonably smooth,
there are also several patterns which show regular dips.  This is particularly
visible for the pulsation spacing pattern of KIC\,2710594, shown in
Fig.\,\ref{fig:kic2710594}.  Finally, several of the stars for which we find
period spacing patterns are hybrid $\gamma\,$Dor/$\delta\,$Sct stars, most
notably KIC\,8645874, KIC\,9751996 and KIC\,11754232 (listed in
Table\,\ref{tab:characterisation}). Not only are these single stars or
single-lined binaries (KIC\,11754232), they are also slow rotators and have a
relatively rich pulsation spectrum in the $\delta\,$Sct regime, making them
interesting targets to probe the interior stellar rotation profile, as done 
in \citet{Kurtz2014} and \citet{Saio2015}.

\section{Statistical analysis of the sample properties}
\label{sec:specvsphot}
Among the 45 stars for which we found period spacing patterns, 38 are single
stars and three are single-lined binaries. We have therefore 41 stars for which
we have both spectroscopic and pulsational parameters based on period spacings,
allowing us to look for possible correlations between their properties and
compare those qualitatively with the relations predicted by theory.

\begin{figure}
\centering
 \includegraphics[width=0.49\textwidth]{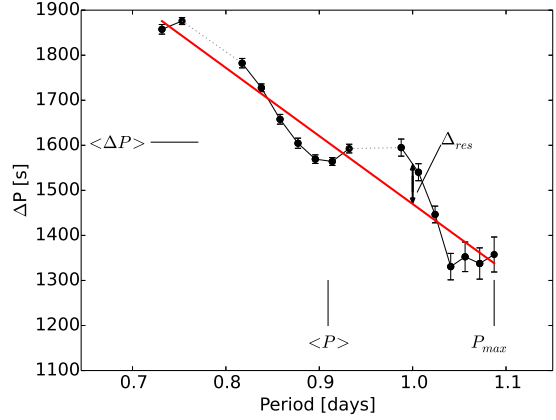}
 \caption{\label{fig:pattern_par}Illustration of some of the photometric parameters
   from the period spacing patterns (shown here for KIC\,10224094) for the
   statistical analysis: $\langle P \rangle$ is the mean pulsation period, $P_{max}$ 
   is the longest pulsation period, $\langle \Delta P \rangle$ is the mean spacing, 
   and $\langle\frac{\mathrm{d}\Delta P}{\mathrm{d}P}\rangle$ is the mean derivative
   (computed from the polynomial fit, here shown in red). Finally,
   $\Delta_{res}$ is a measure of the difference between the polynomial fit and
   the period spacing pattern.}
\end{figure}

In order to properly take into account the characteristics of the period spacing
patterns in the analysis, we limited ourselves to the patterns with a downward
slope, because these are by far the most frequently detected. We derived various 
quantities from them, as illustrated in Fig.\,\ref{fig:pattern_par}.  The mean 
pulsation period of the g-modes that  are part of the found series is denoted as 
$\langle P \rangle$, while $\langle \Delta P \rangle$ is the mean spacing value 
of the pattern. In addition, we have the mean derivative 
$\mathrm{d}(\Delta P)/\mathrm{d}P$ of the pattern, which is actually computed 
from a fitted $2^{nd}$ order polynomial (shown in red in 
Fig.\,\ref{fig:pattern_par}) as this allows us to avoid the influence of any
irregularities or dips in the spacing series.

We also took into account the parameter $\Delta_{res}$, which is a measure of
the residuals of the polynomial fit to the period spacing series as a proxy of
the smoothness of the found pattern. The parameter $P_{\rm max}$, on the other hand,
is the longest detected pulsation period of the downward spacing pattern. We here
consider it to be 
   indicative of the cut-off frequency for which the radiative damping of the
  pulsations in the g-mode cavity becomes dominant over the excitation
  \citep{Dupret2005,Bouabid2013}.  In general, it is the interplay between
  the non-adiabatic excitation and damping, taking into account 
  advective damping due to wave leakage caused by the Coriolis force, that
  determines whether a mode will be excited to detectable amplitude or not
  \citep{Townsend2000}.  The observed quantity $P_{\rm max}$ is therefore not
  only of interest to deduce the boundaries of the $\gamma\,$Dor instability
  strip, but also to derive the physical conditions that determine the delicate
  balance between damping and excitation of the modes.

The other included parameters in our analysis are the frequency $f_g$ and the
amplitude $A_g$ of the dominant g-mode as well as the obtained spectroscopic stellar
parameters. While it would have been valuable to also include the frequency
$f_p$ and the amplitude $A_p$ of the $\delta\,$Sct p-mode with the highest
amplitude or a proxy for the acoustic cutoff frequency for the hybrid
$\gamma\,$Dor/$\delta\,$Sct stars, there are not enough hybrids in the sample
for a meaningful statistical analysis based on these quantities. Similarly, the 
number of sample stars with retrograde modes is also too small for such a study.

We looked for linear relations between the different established
parameters, where we took the logarithm for those quantities whose range is two
or more orders of magnitude.  These are the twelve parameters listed in
Tables\,\ref{tab:param} 
and \ref{tab:persp} for the 41 stars with a downward period spacing pattern.
We performed several multivariate analyses, each
time choosing one of the available parameters as the dependent variable $Y$ and
taking the others as covariates $X_i$. In order to make sure that there was no
offset in the parameter relation, we subtracted the mean values of all the
parameters prior to the regression analysis.  The latter was based on forward
selection and backward elimination \citep[e.g.,][for a similar
application]{Aerts2014}.  With the forward selection method, we add covariates
one at a time as long as their addition implies a significantly better model
fit, starting with the one showing the strongest correlation with the chosen
dependent variable. In the case of the backward elimination, we start with all
covariates under study and remove those showing insignificant correlation with
the dependent variable. The significance tests for each individual covariate were
done using a $t$-test, while the evaluation of each step of the fit was based on
the coefficient of determination $R^2$, i.e., the fraction of the variance of
the dependent variable explained by the model. A summary of the multiple
regression results is provided in Table\,\ref{tab:multivar}.

\begin{figure}
\centering
 \includegraphics[width=0.49\textwidth]{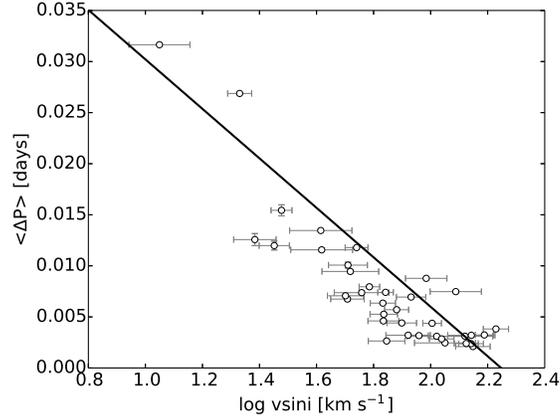}
 \caption{\label{fig:multivar_sp}The average period spacing values of the
   different stars with respect to the logarithm of their $v\sin i$ values.}
\end{figure}

The clearest result from these analyses is the strong correlation between the
stellar rotation and the found period spacing, as predicted by theory
\citep{Bouabid2013} and already discussed in Sections\,\ref{sec:intro} and
\ref{sec:perspacings}. The faster a star is rotating, the larger the rotational
shifts of the pulsation frequencies, which for prograde modes translates into a
period spacing pattern with a steeper downward slope and consisting of more
closely spaced, shorter pulsation periods. The bivariate relation between the
mean period spacing and $\log(v\sin i)$ is shown in Fig.\,\ref{fig:multivar_sp}.

\begin{figure}
\centering
 \includegraphics[width=0.49\textwidth]{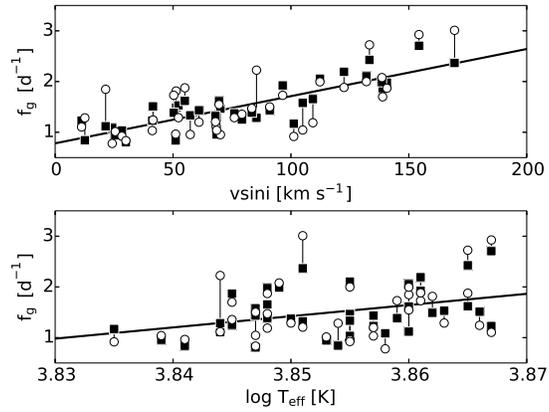}
 \caption{\label{fig:multivar_fg}Observed values (white dots) for the dominant frequency 
 $f_g$, with lines connected to the values predicted by the multivariate models (black 
 squares) described in Table \ref{tab:multivar} for the rotation velocity $v\sin i$ 
 (\emph{top}) and the temperature $\log\,T_{\rm eff}$ (\emph{bottom}). The bivariate 
 model fits listed in the left part of Table\,\ref{tab:multivar} are shown by the full 
 lines.}
\end{figure}

Secondly, we found that the value of dominant mode frequency, $f_g$, 
is strongly correlated with the values of $v\sin i$
and $\log\,T_{\rm eff}$, as shown in Fig.\,\ref{fig:multivar_fg}. As listed in
Table\,\ref{tab:multivar}, the $R^2$-value we obtain for the bivariate model 
based on $f_g$ and $v\sin i$ is 0.468. When we include $\log\,T_{\rm eff}$ as
well, the value of $R^2$ increases to 0.665 for the trivariate model fit, given
by
$$f_g\ =\ 30.9(6.6)\cdot\log\,T_{\rm eff} + 0.0103(0.0013)\cdot v\sin i - 118.23(0.32),$$
where the frequency $f_g$ is expressed in $d^{-1}$, $T_{\rm eff}$ in Kelvin and 
$v\sin i$ in $\rm km\,s^{-1}$.
The relation was found for both the forward and the backward modelling and is clearly 
appropriate. However, there are two unknowns causing
variance in this trivariate model that we cannot take into account at present: the
inclination angle $i$ of the star and the spherical degree $l$ of the pulsation
modes. This may explain some of the scatter in Fig.\,\ref{fig:multivar_fg}.

It is noteworthy that the trend between $f_g$ and $\log\,T_{\rm eff}$ is
opposite to the one found for B-type g-mode pulsators, which do not have an
outer convective envelope and are, in general, slower rotators than the sample
we studied here \citep{DecatAerts2002,Aerts2014}. However, the two correlations
are not directly comparable because the instability strip of SPB stars is much
more extended along the main sequence than the $\gamma\,Dor$ instability strip.
The latter is confined to a much more narrow $\log\,T_{\rm eff}$ range than the
SPB instability region. Therefore the main contributor to the above mentioned
trend for $\gamma\,Dor$ stars, is the stellar radius, whereas this is the
initial mass for SPB pulsators.

\begin{deluxetable}{lccccccc}
% \rotate
\tabcolsep=2pt
\tabletypesize{\footnotesize}
\tablecolumns{8}
\tablewidth{0pc}
\tablecaption{\label{tab:multivar}Results of the multiple linear regression.
We list the coefficients of the covariates for the different model fits as well as the
$p$-values (obtained from a $t$-test) and the $R^2$ values.}
\tablehead{
Explanatory variables & \multicolumn{3}{c}{Separate models}& & \multicolumn{3}{c}{Multivariate models}\\
\cline{2-4} \cline{6-8} \\
\colhead{} & Estimate ($\sigma$) & $p$-value & $R^2$ & & Estimate ($\sigma$) & $p$-value & $R^2$}
\startdata
& \multicolumn{7}{c}{Models for the dependent variable $\langle\frac{\mathrm{d}\Delta P}{\mathrm{d}P}\rangle$}\\
\cline{1-8} \\
$\log (v\sin i)$ & -0.032(0.004) & 0.025 & 0.606 & & \nodata & \nodata & \nodata \\
\cline{1-8} \\
& \multicolumn{7}{c}{Models for the dependent variable $\langle P\rangle$}\\
\cline{1-8} \\
Intercept & \nodata & \nodata & \nodata & & 77.09 & \nodata & \\   
\cline{2-4}\\
$\log (v\sin i)$ & -0.45(0.10) & 0.0001 & 0.350 & & -0.52(0.09) & $<0.0001$ & \\
\cline{2-4}\\
$\log\,T_{\rm eff}$ & -5.8(4.2) & 0.313 & 0.050 & & -10.4(3.2) & 0.0346 & \\
 & & & & & & & 0.501\\
\cline{1-8} \\
& \multicolumn{7}{c}{Models for the dependent variable $\langle\Delta P\rangle$}\\
\cline{1-8} \\
$\log (v\sin i)$ & -0.024(0.002) & $<0.0001$ & 0.796 & & \nodata & \nodata & \nodata\\
\cline{1-8}\\
& \multicolumn{7}{c}{Models for the dependent variable $f_g$}\\
\cline{1-8}\\
Intercept & \nodata & \nodata & \nodata & & -118.23(0.32) & \nodata & \\
\cline{2-4}\\
$v\sin i$ & 0.0093(0.0016) & $<0.0001$ & 0.468 & & 0.0103(0.0013) & $<0.0001$ & \\
\cline{2-4}\\
$\log\,T_{\rm eff}$ & 22.1(10.5) & 0.142 & 0.104 & & 30.9(6.6) & 0.0047 & \\
 & & & & & & & 0.665\\
\cline{2-8}\\
$\zeta$ & 0.61(0.14) & $<0.0001$ & 0.346 & & \nodata & \nodata & \nodata \\
\cline{2-4}\\
$\langle P\rangle$ & -1.93(0.28) & $<0.0001$ & 0.549 & & \nodata & \nodata & \nodata \\
\cline{1-8}\\
& \multicolumn{7}{c}{Models for the dependent variable $P_{\rm max}$}\\
\cline{1-8}\\
Intercept & \nodata & \nodata & \nodata & & 39.50(0.14) & \nodata & \\
$v\sin i$ & -0.0060(0.0006) & $<0.0001$ & 0.703 & & -0.0064(0.0006) & $<0.0001$ & \\
\cline{2-4}\\
$\log\,T_{\rm eff}$ & -4.4(5.9) & 0.550 & 0.015 & & -10.0(2.9) & 0.0292 & \\
 & & & & & & & 0.773\\
\cline{2-8}\\
$\zeta$ & -0.33(0.07) & 0.0079 & 0.358 & & \nodata & \nodata & \nodata \\
\cline{2-4}\\
$\langle P\rangle$ & 1.31(0.08) & $<0.0001$ & 0.876 & & \nodata & \nodata & \nodata \\
\cline{2-4}\\
$\langle\frac{\mathrm{d}\Delta P}{\mathrm{d}P}\rangle$ & 18.4(2.8) & 0.0130 & 0.526 & & \nodata & \nodata & \nodata \\
\cline{2-4}\\
$f_g$ & -0.38(0.06) & $<0.0001$ & 0.505 & & \nodata & \nodata & \nodata \\
\enddata
\end{deluxetable}
\FloatBarrier

\begin{figure}
\centering
 \includegraphics[width=0.49\textwidth]{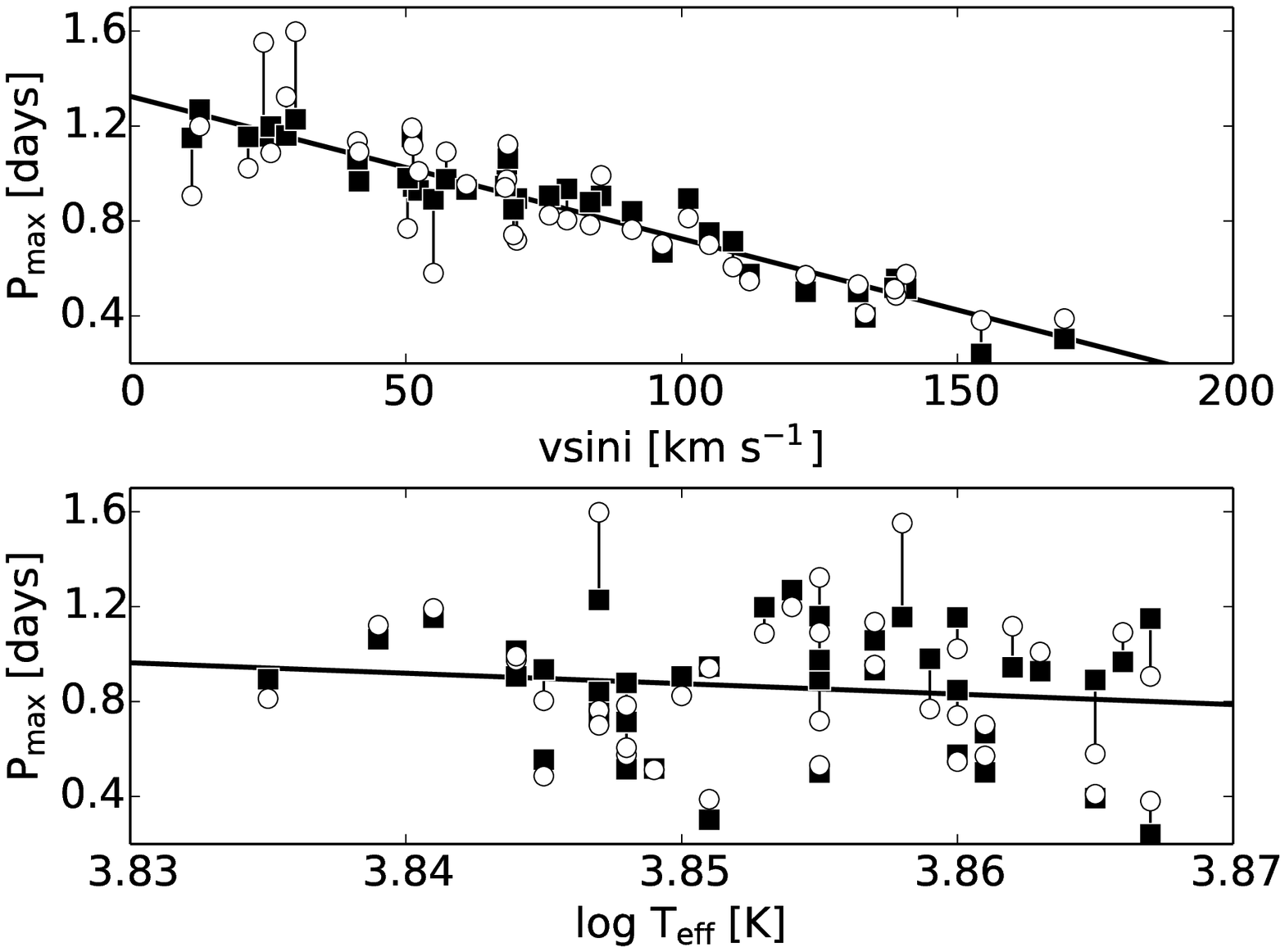}
 \caption{\label{fig:multivar_pmax}Same as Fig.\,\ref{fig:multivar_fg}, but for the
 pulsation period $P_{\rm max}$ instead of the frequency $f_g$.}
\end{figure}

We also found a bivariate relation between $f_g$ and $\zeta$. This is not
surprising because the overall pulsational line broadening, which is dominated
by the mode with frequency $f_g$, is captured by
$\zeta$. Indeed, we could not perform line-profile modelling due to the individual
pulsation modes as it would require time-resolved spectroscopy covering the
phases of each and every g-mode \citep[e.g.,][Chapter 6]{Aerts2010}. Hence, the
overall pulsational broadening is included in the simple broadening function
parametrized with $\zeta$ \citep[Cf.][]{morel2006}.

Finally, we also found a bivariate relation between $f_g$ and $\langle
P\rangle$, which indicates that the period spacing patterns are typically found
in frequency regions around the dominant g-mode frequency.

The correlations between $f_g$ and the other variables also apply to $P_{\rm max}$
to almost the same extent, as can be seen in Table\,\ref{tab:multivar} and as
illustrated in Fig.\,\ref{fig:multivar_pmax}. The main difference is that the
correlation between $P_{\rm max}$ and $v\sin i$ is stronger than between $f_g$ and
$v\sin i$, while the correlation with $\log\,T_{\rm eff}$ is weaker. Thus,
rotation seems to have a larger impact on the period $P_{\rm max}$ than on the frequency
of the dominant mode.

\section{Discussion and conclusions}
\label{sec:conclusions}

\begin{figure}
\centering
 \includegraphics[width=0.49\textwidth]{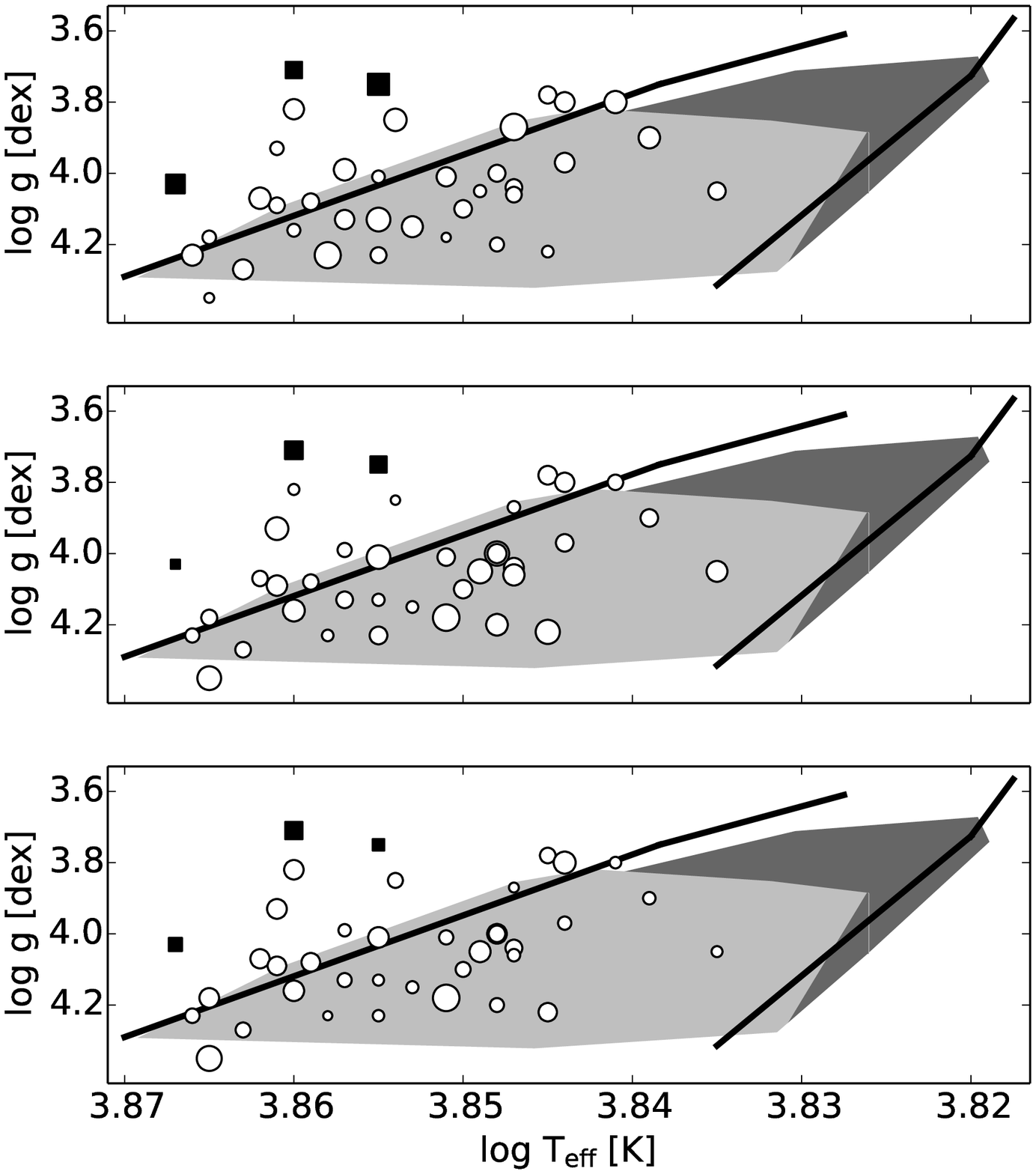}
 \caption{\label{fig:HRD}($\log\,T_{\rm eff}$,$\log\,g$)-diagrams with the sample 
 stars for which period spacing patterns were found. The white dots indicate the single
 stars, whereas the black squares are the single-lined binaries. The full lines 
 mark the instability strip taken from \citet{Dupret2005} (with $\alpha_{\rm MLT} = 2.0$).
 The light grey and dark grey areas are the instability strips for the main-sequence 
 $\gamma$\,Dor stars presented by \citet{Bouabid2011}, with $\alpha_{\rm ov} = 0$ and 
 $\alpha_{\rm ov} = 0.2$ respectively. The marker sizes indicate the longest pulsation
period $P_{\rm max}$ in the spacing patterns (\emph{top}), the rotation velocity $v\sin i$ 
(\emph{middle}) and the g-mode frequency $f_g$ with the highest amplitude (\emph{bottom}).}
\end{figure}

This paper offers an observational asteroseismic analysis of 67 of the
$\gamma\,$Dor stars in the sample presented by \citet{Tkachenko2013}. Thanks to
the available combination of \emph{Kepler\/} photometry and ground-based
high-resolution spectroscopy, we were able to carry out a detailed statistical
analysis.  We found thirteen of the 67 stars to be members of a binary system.  We
computed the stellar parameters for all 67 stars in the sample and found
them to be consistent with the typical values expected for $\gamma$\,Dor
stars. Most of the stars in our sample are slow to moderate rotators with respect
to their breakup velocity.

For 45 stars in the sample we were able to detect period spacings using the
methodology described by \citet{VanReeth2015}. The found spacing patterns are in
line with theoretical expectations \citep{Miglio2008a,Bouabid2013}.  For several
of the stars we detected periodic dips in the period spacing versus period
relations, as predicted by \citet{Miglio2008a}. These dips are caused by the
presence of a chemical gradient near the edge of the convective core. For many
of the sample stars, the period spacing patterns are smooth, pointing either
to extremely young stars or, more likely, to the presence of mixing processes
inside the star that have washed out the chemical gradient. A priori one could
expect to see a correlation between the smoothness of the period spacing patterns
and the observed rotation velocity. In this paper, such a correlation would be
reflected between the values of $\Delta_{res}$ and $v\sin i$, but we could
not find a significant relation between the two parameters. It is likely that, if 
such a correlation exists, the uncertainties on the values of $\Delta_{res}$ are 
too large.

In order to determine the precise cause of the smooth patterns, detailed forward
seismic modeling is required for each individual star. In addition, each of the
found spacing patterns reveals a clear slope.  This is as predicted
theoretically and points towards frequency shifts of the g-modes due to
rotational effects, resulting in a downward or upward slope for prograde and
retrograde modes, respectively.

Our results based on the {\it Kepler\/} time series photometry were further
confronted with the spectroscopic parameters of the pulsators.
We searched for correlations between the seismic quantities for the stars
exhibiting period spacing patterns with a downward slope, since only these
constitute a large enough sample to perform multivariate linear regressions.
We found a strong correlation between $v\sin i$ and the average period spacing,
confirming the influence of rotation on $\gamma\,$Dor-type pulsations as
predicted by theory. In addition, we found a correlation between the dominant
g-mode frequency and both $v\sin i$ and $\log\,T_{\rm eff}$. Similarly, $P_{\rm
  max}$ correlates with $v\sin i$ and $\log\,T_{\rm eff}$.

While our sample contains only eleven hybrid $\gamma\,$Dor/$\delta\,$Sct
pulsators, these constitute an interesting sub-sample for various reasons.
First, seven of these stars were found to be in a binary.  At least four of them
contain a hybrid pulsator as three were identified as single-lined binaries,
while in the case of KIC\,3952623 the observed $\gamma\,$Dor and $\delta\,$Sct
variability originates from the primary component.

Two of the single stars (KIC\,8645874 and KIC\,9751996) and one of the
single-lined binaries (KIC\,11754232) were found to be slow rotators.
Interestingly, some of the recently studied hybrid $\gamma\,$Dor pulsators are
also very slow rotators \citep{Kurtz2014,Bedding2014,Saio2015}.  On the other
hand, the hybrid star KIC\,4749989 is a fast rotator, similar to KIC\,8054146
\citep[e.g.][]{Breger2012}.

We compared the locations of the sample stars in the Kiel diagram with the
$\gamma\,$Dor instability strips determined by \citet{Dupret2005} and
\citet{Bouabid2011} (as shown in Fig.\,\ref{fig:HRD}).  Given the large
uncertainties on $\log\,g$, the positions are compatible with the strips,
although most of our sample stars clearly cluster around the blue edge.  The
symbol sizes in these plots indicate the longest pulsation period $P_{\rm max}$ in
the spacing patterns (\emph{top}), the rotation velocity $v\sin i$
(\emph{middle}) and the g-mode frequency $f_g$ with the largest amplitude
(\emph{bottom}). The quantity $P_{\rm max}$ is of interest to understand
$\gamma$\,Dor instability, in the sense that modes with even longer period are 
more sensitive to radiative damping and/or wave energy leakage.

While we found a correlation between $f_g$ as well as $P{\rm max}$ and
$\log\,T_{\rm eff}$ in Section \ref{sec:specvsphot}, this is not obvious from
the figure because of the strong influence of $v\sin i$ on these two quantities
and due to the large errors of $\log\,g$.  We do see from the symbol sizes that
a higher rotation velocity typically correlates with a higher frequency $f_g$
and a lower period $P_{\rm max}$, as found from the forward and backward modelling.

Our sample study constitutes a very rich observational data base to persue
forward asteroseismic modeling of $\gamma\,$Dor stars. All the necessary data
to perform such stellar modelling is provided to the community through
{\tt \small https://fys.kuleuven.be/ster/Projects/gamma-doradus}.

\acknowledgments The research leading to these results was based on funding from
the Fund for Scientific Research of Flanders (FWO), Belgium under grant
agreement G.0B69.13, from the European Community’s Seventh Framework Programme
FP7-SPACE-2011-1 project number 312844 (SPACEINN), from the Research Council of
KU\,Leuven under grant GOA/2013/012, and from the National Science Foundation of
the USA under grant No.\,NSF PHY11-25915.  KZ acknowledges support by the Austrian 
Fonds zur F\"orderung der wissenschaftlichen Forschung (FWF, project V431-NBL). 
SB is supported by the Foundation for
Fundamental Research on Matter (FOM), which is part of the Netherlands
Organisation for Scientific Research (NWO). TVR thanks Aaron Greenwood (University
of Groningen, the Netherlands) for the very useful discussions on the
interaction of rotation and pulsations in $\gamma\,$Dor stars. CA, SB, PP and VSS
are grateful to the Kavli Institute of Theoretical Physics at the University of
California, Santa Barbara, for the kind hospitality during the research
programme ``Galactic Archaeology and Precision Stellar Astrophysics'' during
which the current study was finalized.  Funding for the \emph{Kepler\/} mission
is provided by NASA's Science Mission Directorate. We thank the whole team for
the development and operations of the mission. This research made use of the
SIMBAD database, operated at CDS, Strasbourg, France, and the SAO/NASA
Astrophysics Data System. This research has also made use of the VizieR
catalogue access tool, CDS, Strasbourg, France, and of the VALD database,
operated at Uppsala University, the Institute of Astronomy RAS in Moscow, and
the University of Vienna.

\newpage

\appendix
% \section{Observation log}

\begin{deluxetable}{cccrrcrc}
\rotate
\tablenum{A1}
\tabcolsep=2.5pt
\tabletypesize{\small}
\tablecolumns{8}
\tablewidth{0pc}
\tablecaption{\label{tab:obs} Observation log of the stars in our sample. $N_Q$ and $N_S$ indicate the total number of available \emph{Kepler\/} mission quarters and HERMES spectra respectively.}
\tablehead{
 KIC & RA (J2000) & DEC (J2000) & $m_V$ & $N_Q$ & Available quarters & $N_S$ & Spectroscopic observation times}
\startdata
2710594 & 19 23 33.317 & +37 59 00.55 & 11.79 & 18 & 0-17 & 4 & May-August 2011\\
3222854 & 19 11 38.539 & +38 20 43.83 & 12.5 & 18 & 0-17 & 7 & August 2013, July-September 2014\\
3448365 & 19 30 35.779 & +38 35 16.67 & 9.92 & 18 & 0-17 & 4 & May-August 2011\\
3744571 & 19 23 05.597 & +38 48 51.91 & 12.03 & 18 & 0-17 & 4 & June-July 2013\\
3952623 & 19 24 08.225 & +39 00 40.25 & 12.09 & 18 & 0-17 & 14 & May-July 2013, July-August 2014\\
4547348 & 19 08 18.888 & +39 41 36.53 & 11.78 & 18 & 0-17 & 4 & May-August 2011\\
4749989 & 19 31 05.251 & +39 52 24.78 & 9.64 & 15 & 0-5, 7-9, 11-13, 15-17 & 2 & May-June 2011\\
4757184 & 19 38 36.922 & +39 50 26.10 & 11.7 & 8 & 0-1, 11-13, 15-17 & 5 & May-August 2011\\
4846809 & 19 39 19.027 & +39 55 10.70 & 12.36 & 15 & 0-5, 7-9, 11-13, 15-17 & 4 & June-July 2013\\
5000454 & 19 13 53.400 & +40 10 44.76 & 13.56 & 17 & 1-17 & 3 & June-July 2013\\
5114382 & 19 42 43.543 & +40 15 00.66 & 11.55 & 8 & 0-1, 11-13, 15-17 & 4 & July-August 2011\\
5254203 & 18 56 08.364 & +40 27 12.87 & 12.92 & 14 & 0-4, 6-8, 10-12, 14-16 & 4 & August 2013, June-July 2014\\
5350598 & 19 11 54.360 & +40 34 01.78 & 12.1 & 18 & 0-17 & 4 & June-July 2013\\
5522154 & 19 12 23.386 & +40 42 14.00 & 10.44 & 10 & 0-1, 10-17 & 3 & May-June 2011\\
5708550 & 19 31 03.494 & +40 56 13.89 & 11.93 & 18 & 0-17 & 4 & July-August 2011\\
5772452 & 19 00 04.541 & +41 05 17.20 & 12.46 & 14 & 0-4, 6-8, 10-12, 14-16 & 2 & August-September 2013\\
5788623 & 19 24 11.066 & +41 00 41.47 & 12.07 & 18 & 0-17 & 4 & June-July 2013\\
6185513 & 18 57 29.040 & +41 30 58.10 & 11.89 & 8 & 0-1, 10-12, 14-16 & 4 & May 2011\\
6342398 & 18 51 34.322 & +41 46 35.05 & 12.37 & 14 & 0-4, 6-8, 10-12, 14-16 & 4 & May-June 2013\\
6367159 & 19 31 21.144 & +41 44 55.50 & 12.6 & 18 & 0-17 & 8 & August-September 2013, June-August 2014\\
6425437 & 19 00 03.295 & +41 50 16.36 & 11.48 & 14 & 0-4, 6-8, 10-12, 14-16 & 3 & May 2011\\
6467639 & 19 53 37.829 & +41 49 05.02 & 12.02 & 18 & 0-17 & 12 & June-July 2013, July-August 2014\\
6468146 & 19 54 04.241 & +41 48 31.01 & 9.96 & 18 & 0-17 & 4 & July 2011, June-August 2014\\
6468987 & 19 54 49.006 & +41 48 13.54 & 12.67 & 18 & 0-17 & 4 & August 2013, June-August 2014\\
6678174 & 19 10 05.712 & +42 09 54.58 & 11.74 & 10 & 0-1, 10-17 & 4 & May-August 2011\\
6778063 & 19 28 06.622 & +42 15 29.60 & 12.18 & 18 & 0-17 & 9 & June-July 2013, July-August 2014\\
6935014 & 19 08 39.984 & +42 27 11.77 & 10.91 & 18 & 0-17 & 5 & May-June 2011\\
6953103 & 19 32 51.240 & +42 28 46.53 & 12.56 & 18 & 0-17 & 4 & August-September 2013, June-August 2014\\
7023122 & 19 13 42.062 & +42 31 13.02 & 10.84 & 18 & 0-17 & 4 & May-July 2011\\
7365537 & 19 30 35.237 & +42 58 10.69 & 9.19 & 18 & 0-17 & 3 & May 2011\\
7380501 & 19 46 41.587 & +42 59 35.34 & 11.98 & 18 & 0-17 & 4 & July-August 2011\\
7434470 & 19 13 57.941 & +43 04 57.98 & 12.02 & 18 & 0-17 & 4 & June-July 2013\\
7516703 & 19 12 48.742 & +43 11 54.34 & 12.51 & 13 & 1-3, 6-7, 10-17 & 2 & August-September 2013\\
7583663 & 18 46 57.458 & +43 16 15.31 & 12.69 & 18 & 0-17 & 4 & August 2013, June-July 2014\\
7691618 & 19 39 02.666 & +43 19 01.20 & 12.36 & 18 & 0-17 & 4 & June-July 2013\\
7746984 & 19 18 45.175 & +43 24 01.54 & 12.47 & 10 & 0-1, 3, 5, 7, 9, 11, 13, 15, 17 & 2 & August 2013\\
7867348 & 18 40 40.774 & +43 36 17.09 & 10.84 & 18 & 0-17 & 12 & May 2011, June-August 2014\\
7939065 & 18 49 11.976 & +43 47 53.52 & 12.14 & 18 & 0-17 & 4 & May-June 2013\\
8364249 & 19 25 16.824 & +44 23 46.75 & 11.95 & 18 & 0-17 & 4 & July-August 2011\\
8375138 & 19 40 47.623 & +44 20 11.83 & 11.02 & 18 & 0-17 & 4 & May-August 2011\\
8378079 & 19 44 30.607 & +44 22 58.04 & 11.84 & 18 & 0-17 & 3 & August 2011\\
8611423 & 19 03 12.216 & +44 44 35.05 & 11.59 & 10 & 0-1, 10-17 & 4 & May-June 2011\\
8645874 & 19 53 51.715 & +44 43 09.37 & 9.92 & 10 & 0-1, 10-17 & 2 & July 2011\\
8693972 & 19 34 44.590 & +44 52 24.03 & 12.81 & 18 & 0-17 & 7 & August-September 2013, July-August 2014\\
8739181 & 19 00 06.043 & +44 56 43.47 & 12.08 & 18 & 0-17 & 5 & June-July 2013\\
8836473 & 19 51 12.305 & +45 01 34.10 & 12.78 & 18 & 0-17 & 4 & August 2013, June-August 2014\\
9210943 & 19 14 04.512 & +45 39 08.99 & 11.81 & 18 & 0-17 & 5 & May-August 2011, June-August 2014\\
9419694 & 19 49 57.994 & +45 57 07.09 & 12.74 & 15 & 0-6, 8-10, 12-14, 16-17 & 4 & August 2013, June-August 2014\\
9480469 & 19 49 34.555 & +46 04 20.90 & 12.78 & 15 & 0-6, 8-10, 12-14, 16-17 & 5 & August-September 2013, June-August 2014\\
9595743 & 19 39 19.226 & +46 14 57.49 & 12.09 & 11 & 0-1, 5, 10-17 & 4 & June-July 2013\\
9751996 & 18 45 59.486 & +46 30 43.23 & 10.96 & 10 & 0-1, 10-17 & 4 & May 2011\\
10080943 & 19 36 27.305 & +47 05 02.29 & 11.81 & 18 & 0-17 & 27 & August 2011, May-July 2013, July-October 2014\\
10224094 & 19 50 06.643 & +47 17 06.58 & 11.9 & 15 & 0-6, 8-10, 12-14, 16-17 & 4 & July-August 2011\\
10256787 & 18 53 47.189 & +47 23 35.01 & 12.25 & 18 & 0-17 & 14 & May-June 2013, June-August 2014\\
10467146 & 19 22 06.050 & +47 40 49.33 & 12.66 & 18 & 0-17 & 4 & August-September 2013, June-August 2014\\
11080103 & 19 18 50.141 & +48 37 13.90 & 12.94 & 18 & 0-17 & 4 & August-September 2013, June-August 2014\\
11099031 & 19 52 37.514 & +48 38 44.01 & 10.02 & 15 & 0-6, 8-10, 12-14, 16-17 & 4 & July 2011\\
11196370 & 19 38 44.486 & +48 50 55.64 & 12.71 & 18 & 0-17 & 2 & August-September 2013\\
11294808 & 19 22 45.396 & +49 04 47.53 & 11.73 & 18 & 0-17 & 4 & May-August 2011\\
11456474 & 19 30 10.495 & +49 20 28.06 & 12.49 & 18 & 0-17 & 4 & August-September 2013, June-August 2014\\
11668783 & 19 41 27.422 & +49 47 41.24 & 12.58 & 18 & 0-17 & 4 & August-September 2013, June-August 2014\\
11721304 & 19 43 27.847 & +49 51 25.41 & 11.71 & 18 & 0-17 & 4 & July-August 2011\\
11754232 & 19 05 57.926 & +49 56 33.53 & 12.24 & 8 & 0-1, 10-11, 13-15, 17 & 13 & June-July 2013, June-August 2014\\
11826272 & 19 49 50.494 & +50 00 57.36 & 10.21 & 18 & 0-17 & 3 & May-August 2011\\
11907454 & 19 11 29.407 & +50 14 29.72 & 11.38 & 18 & 0-17 & 4 & May-August 2011\\
11917550 & 19 34 22.565 & +50 15 46.29 & 11.12 & 18 & 0-17 & 4 & May-August 2011\\
11920505 & 19 39 57.108 & +50 13 32.09 & 9.88 & 18 & 0-17 & 2 & May 2011\\
12066947 & 19 37 03.209 & +50 30 20.41 & 10.23 & 10 & 0-1, 10-17 & 3 & May-July 2011\\
12458189 & 19 20 49.123 & +51 20 09.25 & 11.46 & 18 & 0-17 & 4 & May-August 2011\\
12643786 & 19 13 49.205 & +51 46 48.32 & 11.52 & 18 & 0-17 & 4 & May-August 2011\\
\enddata
\end{deluxetable}
\clearpage

% \section{Sample characterisation}
\begin{deluxetable}{ccccl}
\tablenum{A2}
\tablecolumns{5}
\tablewidth{0pc}
\tablecaption{\label{tab:characterisation}The global observational characteristics
    of the stars in our sample. We have listed information on the variability
    type, where the label ``ROT'' indicates rotational modulation, and *
    indicates that frequencies are found outside of the traditional range for
    $\gamma\,$Dor stars, but likely correspond to combination frequencies. In
    addition, stars are listed as single (S), single-lined binaries (SB1),
    double-lined binaries (SB2), binaries with no detected radial velocity
    variations (SB) and triple systems (SB3). We also report the detections of
    period spacing patterns and comment on their suspected nature.}
\tablehead{
KIC & Class & binarity & Period spacings & Comments}
\startdata
2710594 & $\gamma\,$Dor & S & yes & Prograde and retrograde\\
3222854 & $\gamma\,$Dor & SB2 & no & -\\
3448365 & $\gamma\,$Dor & S & yes & Prograde and retrograde\\
3744571 & $\gamma\,$Dor & S & no & -\\
3952623 & $\gamma\,$Dor/$\delta\,$Sct hybrid & SB2 & no & -\\
4547348 & $\gamma\,$Dor & S & no & -\\
4749989 & $\gamma\,$Dor/$\delta\,$Sct hybrid & S & no & -\\
4757184 & $\gamma\,$Dor & S & yes & Retrograde (quadrupole)\\
4846809 & $\gamma\,$Dor & S & yes & Prograde (dipole and quadrupole)\\
5000454 & $\gamma\,$Dor & S & no & -\\
5114382 & $\gamma\,$Dor & S & yes & Prograde and retrograde\\
5254203 & $\gamma\,$Dor & S & no & -\\
5350598 & $\gamma\,$Dor & S & yes & Retrograde (quadrupole)\\
5522154 & $\gamma\,$Dor* & S & yes & Prograde\\
5708550 & $\gamma\,$Dor & S & yes & Prograde\\
5772452 & K-type (ROT) & - & - & -\\
5788623 & $\gamma\,$Dor & S & yes & Prograde\\
6185513 & $\gamma\,$Dor* & S & yes & Retrograde (quadrupole)\\
6342398 & $\gamma\,$Dor & S & no & -\\
6367159 & $\gamma\,$Dor/$\delta\,$Sct hybrid & SB2 & no & -\\
6425437 & $\gamma\,$Dor & S & yes & Retrograde (quadrupole)\\
6467639 & $\gamma\,$Dor/$\delta\,$Sct hybrid & SB3 & no & -\\
6468146 & $\gamma\,$Dor/$\delta\,$Sct hybrid & SB1 & yes & Prograde\\
6468987 & $\gamma\,$Dor & S & yes & Prograde and retrograde\\
6678174 & $\gamma\,$Dor & S & yes & Prograde\\
6778063 & $\gamma\,$Dor/$\delta\,$Sct hybrid & SB3 & yes & Prograde\\
6935014 & $\gamma\,$Dor & S & yes & Prograde\\
6953103 & $\gamma\,$Dor & S & yes & Prograde\\
7023122 & $\gamma\,$Dor & S & yes & Prograde\\
7365537 & $\gamma\,$Dor* & S & yes & Prograde\\
7380501 & $\gamma\,$Dor & S & yes & Prograde\\
7434470 & $\gamma\,$Dor* & S & yes & Prograde and strong rotation frequency\\
7516703 & K-type (ROT) & - & - & -\\
7583663 & $\gamma\,$Dor & S & yes & Prograde and retrograde\\
7691618 & $\gamma\,$Dor & S & no & -\\
7746984 & $\gamma\,$Dor & SB & yes & Prograde\\
7867348 & $\gamma\,$Dor/$\delta\,$Sct hybrid & SB1 & yes & Retrograde (quadrupole)\\
7939065 & $\gamma\,$Dor* & S & yes & Prograde\\
8364249 & $\gamma\,$Dor & S & yes & Prograde\\
8375138 & $\gamma\,$Dor* & S & yes & Prograde and retrograde\\
8378079 & $\gamma\,$Dor & S & no & -\\
8611423 & $\gamma\,$Dor & S & no & -\\
8645874 & $\gamma\,$Dor/$\delta\,$Sct hybrid & S & yes & Prograde\\
8693972 & $\gamma\,$Dor & SB2 & no & -\\
8739181 & $\gamma\,$Dor & S & no & -\\
8836473 & $\gamma\,$Dor/$\delta\,$Sct hybrid & S & yes & Prograde\\
9210943 & $\gamma\,$Dor & SB & yes & Prograde and retrograde\\
9419694 & $\gamma\,$Dor & S & yes & Prograde\\
9480469 & $\gamma\,$Dor & S & yes & Prograde and retrograde\\
9595743 & $\gamma\,$Dor & S & yes & Prograde\\
9751996 & $\gamma\,$Dor/$\delta\,$Sct hybrid & S & yes & Prograde, retrograde and zonal\\
10080943 & $\gamma\,$Dor/$\delta\,$Sct hybrid & SB2 & - & -\\
10224094 & $\gamma\,$Dor & S & yes & Prograde\\
10256787 & $\gamma\,$Dor & SB & yes & Prograde\\
10467146 & $\gamma\,$Dor & SB1 & yes & Prograde\\
11080103 & $\gamma\,$Dor & S & yes & Prograde\\
11099031 & $\gamma\,$Dor* & S & yes & Prograde\\
11196370 & $\gamma\,$Dor* & S & no & -\\
11294808 & $\gamma\,$Dor* & S & yes & Prograde (dipole and quadrupole)\\
11456474 & $\gamma\,$Dor & S & yes & Prograde\\
11668783 & $\gamma\,$Dor & S & yes & Retrograde\\
11721304 & $\gamma\,$Dor & S & yes & Prograde\\
11754232 & $\gamma\,$Dor/$\delta\,$Sct hybrid & SB1 & yes & Prograde\\
11826272 & $\gamma\,$Dor & S & yes & Prograde\\
11907454 & $\gamma\,$Dor & S & yes & Prograde and retrograde\\
11917550 & $\gamma\,$Dor & S & yes & Prograde\\
11920505 & $\gamma\,$Dor & S & yes & Prograde\\
12066947 & $\gamma\,$Dor* & S & yes & Prograde and retrograde\\
12458189 & $\gamma\,$Dor* & S & yes & Prograde\\
12643786 & $\gamma\,$Dor & S & no & -\\
 \enddata
\end{deluxetable}

\clearpage
% \section{Atmospheric parameter values}

\begin{deluxetable}{ccccccccccccc}
\rotate
\tablenum{A3}
\tabcolsep=4pt
% \tabletypesize{\footnotesize}
\tablecolumns{13}
\tablewidth{0pc}
\tablecaption{\label{tab:param}Spectroscopic parameter values computed for the
    merged spectra of the single stars and single-lined binaries in our sample. The
provided errors are the 1-$\sigma$ values obtained from the $\chi^2$-statistics.}
\tablehead{
KIC & $\log\,T_{\rm eff}$ [K] &  $\sigma_{\log\,T_{\rm eff}}$ & $T_{\rm eff}$ [K] & $\sigma_{T_{\rm eff}}$ & $\log\,g$ & $\sigma_{\log\,g}$ & $\zeta$ [km\,$\mathrm{s}^{-1}$] & $\sigma_{\zeta}$ & $v\sin i$ [km\,$\mathrm{s}^{-1}$] & $\sigma_{v\sin i}$ & $[M/H]$ & $\sigma_{[M/H]}$}
\startdata
2710594 & 3.845 & 0.006 & 7000 & 100 & 3.78 & 0.33 & 2.93 & 0.42 &  79.2 &  4.1 &  0.11 & 0.09\\
3448365 & 3.847 & 0.004 & 7030 &  60 & 4.04 & 0.22 & 2.52 & 0.26 &  91.0 &  3.0 &  0.05 & 0.07\\
3744571 & 3.850 & 0.008 & 7075 & 130 & 3.94 & 0.40 & 2.47 & 0.58 &  52.5 &  4.0 &  0.11 & 0.11\\
4547348 & 3.850 & 0.006 & 7080 & 105 & 4.02 & 0.34 & 2.54 & 0.44 &  68.5 &  4.0 &  0.10 & 0.09\\
4749989 & 3.865 & 0.003 & 7320 &  55 & 4.24 & 0.21 & 3.33 & 0.22 & 204.0 &  8.0 &  0.13 & 0.06\\
4757184 & 3.862 & 0.008 & 7285 & 135 & 4.16 & 0.43 & 1.90 & 0.63 &  36.0 &  3.0 &  0.02 & 0.13\\
4846809 & 3.862 & 0.008 & 7280 & 135 & 4.07 & 0.32 & 3.00 & 0.59 &  51.3 &  3.5 &  0.18 & 0.11\\
5000454 & 3.840 & 0.006 & 6920 & 100 & 3.46 & 0.38 & 3.91 & 0.50 & 101.0 &  4.4 &  0.07 & 0.10\\
5114382 & 3.855 & 0.007 & 7165 & 120 & 4.23 & 0.37 & 2.45 & 0.45 &  70.1 &  4.5 &  0.19 & 0.10\\
5254203 & 3.866 & 0.014 & 7350 & 235 & 4.09 & 0.72 & 3.11 & 1.10 &  42.7 &  6.0 &  0.21 & 0.21\\
5350598 & 3.864 & 0.007 & 7315 & 110 & 4.11 & 0.33 & 2.64 & 0.42 &  27.8 &  1.7 &  0.10 & 0.09\\
5522154 & 3.851 & 0.004 & 7090 &  70 & 4.18 & 0.26 & 3.15 & 0.36 & 169.4 &  7.5 &  0.08 & 0.08\\
5708550 & 3.844 & 0.006 & 6990 & 100 & 3.97 & 0.32 & 2.44 & 0.42 &  68.3 &  3.7 &  0.15 & 0.09\\
5788623 & 3.858 & 0.007 & 7210 & 115 & 4.23 & 0.32 & 2.20 & 0.40 &  24.2 &  1.8 &  0.09 & 0.09\\
6185513 & 3.867 & 0.007 & 7365 & 120 & 4.10 & 0.40 & 3.91 & 0.72 &  82.9 &  5.5 &  0.16 & 0.11\\
6342398 & 3.855 & 0.010 & 7165 & 165 & 3.95 & 0.49 & 2.61 & 0.60 &  31.6 &  2.9 &  0.17 & 0.13\\
6425437 & 3.848 & 0.007 & 7045 & 112 & 3.88 & 0.35 & 2.64 & 0.43 &  49.7 &  2.5 &  0.21 & 0.09\\
6468146 & 3.860 & 0.004 & 7240 &  65 & 3.71 & 0.22 & 3.56 & 0.33 &  69.5 &  1.9 & -0.05 & 0.06\\
6468987 & 3.855 & 0.013 & 7155 & 210 & 4.01 & 0.72 & 4.08 & 1.10 & 132.0 & 12.8 &  0.38 & 0.20\\
6678174 & 3.852 & 0.006 & 7105 &  90 & 3.77 & 0.33 & 2.59 & 0.35 &  45.5 &  2.3 &  0.02 & 0.08\\
6935014 & 3.851 & 0.004 & 7090 &  66 & 4.01 & 0.22 & 2.66 & 0.32 &  68.0 &  3.0 &  0.14 & 0.07\\
6953103 & 3.863 & 0.010 & 7290 & 175 & 4.27 & 0.50 & 2.50 & 0.81 &  52.3 &  5.2 &  0.25 & 0.15\\
7023122 & 3.865 & 0.005 & 7335 &  82 & 4.18 & 0.30 & 2.81 & 0.43 &  55.0 &  2.2 &  0.04 & 0.08\\
7365537 & 3.867 & 0.004 & 7360 &  65 & 4.48 & 0.23 & 3.23 & 0.26 & 154.3 &  4.9 &  0.11 & 0.05\\
7380501 & 3.841 & 0.007 & 6930 & 110 & 3.80 & 0.35 & 2.33 & 0.47 &  51.1 &  3.0 &  0.14 & 0.10\\
7434470 & 3.845 & 0.009 & 7005 & 150 & 4.22 & 0.50 & 3.33 & 0.90 & 138.9 & 11.5 &  0.25 & 0.14\\
7583663 & 3.847 & 0.014 & 7030 & 230 & 4.06 & 0.66 & 3.38 & 1.20 & 105.0 & 11.0 &  0.42 & 0.20\\
7691618 & 3.852 & 0.009 & 7110 & 150 & 4.10 & 0.45 & 2.33 & 0.62 &  43.3 &  3.5 &  0.15 & 0.13\\
7867348 & 3.847 & 0.005 & 7030 &  75 & 3.56 & 0.22 & 2.94 & 0.20 &  17.2 &  1.0 & -0.06 & 0.06\\
7939065 & 3.861 & 0.009 & 7265 & 145 & 4.09 & 0.48 & 3.38 & 0.78 &  96.5 &  7.0 &  0.24 & 0.13\\
8364249 & 3.848 & 0.007 & 7045 & 115 & 4.00 & 0.46 & 3.36 & 0.55 & 140.7 &  8.5 &  0.17 & 0.11\\
8375138 & 3.849 & 0.005 & 7070 &  80 & 4.05 & 0.34 & 3.14 & 0.40 & 138.6 &  6.0 &  0.08 & 0.08\\
8378079 & 3.845 & 0.011 & 7000 & 170 & 3.31 & 0.55 & 1.75 & 0.47 &  11.7 &  1.2 & -0.14 & 0.14\\
8611423 & 3.853 & 0.006 & 7135 & 105 & 4.21 & 0.30 & 1.86 & 0.34 &  20.7 &  1.5 &  0.10 & 0.08\\
8645874 & 3.860 & 0.005 & 7240 &  90 & 3.82 & 0.27 & 3.38 & 0.31 &  21.4 &  0.9 & -0.05 & 0.06\\
8739181 & 3.848 & 0.007 & 7055 & 120 & 3.59 & 0.40 & 2.24 & 0.44 &  27.4 &  1.8 & -0.02 & 0.10\\
8836473 & 3.861 & 0.011 & 7260 & 190 & 3.93 & 0.60 & 4.71 & 1.00 & 122.5 & 11.0 &  0.37 & 0.20\\
9419694 & 3.857 & 0.011 & 7200 & 185 & 3.99 & 0.58 & 2.50 & 0.77 &  41.2 &  4.5 &  0.17 & 0.16\\
9480469 & 3.860 & 0.014 & 7250 & 240 & 4.16 & 0.70 & 3.74 & 1.25 & 112.3 & 12.0 &  0.59 & 0.25\\
9595743 & 3.859 & 0.008 & 7230 & 125 & 4.08 & 0.40 & 2.44 & 0.59 &  50.3 &  3.2 &  0.14 & 0.10\\
9751996 & 3.854 & 0.006 & 7145 & 100 & 3.85 & 0.24 & 3.07 & 0.27 &  12.6 &  0.9 &  0.28 & 0.07\\
10224094 & 3.853 & 0.006 & 7130 & 100 & 4.15 & 0.30 & 2.25 & 0.39 &  25.5 &  1.5 &  0.07 & 0.08\\
10467146 & 3.855 & 0.011 & 7155 & 175 & 3.75 & 0.60 & 2.85 & 0.80 &  57.3 &  5.5 &  0.12 & 0.15\\
11080103 & 3.866 & 0.012 & 7345 & 210 & 4.23 & 0.65 & 2.80 & 0.95 &  41.5 &  4.5 &  0.18 & 0.18\\
11099031 & 3.835 & 0.004 & 6835 &  70 & 4.05 & 0.20 & 2.60 & 0.20 & 101.2 &  3.3 &  0.20 & 0.05\\
11196370 & 3.861 & 0.013 & 7260 & 215 & 4.05 & 0.55 & 2.75 & 0.68 &  16.0 &  2.0 &  0.26 & 0.16\\
11294808 & 3.844 & 0.006 & 6975 &  95 & 3.80 & 0.35 & 2.95 & 0.44 &  85.4 &  4.4 &  0.12 & 0.09\\
11456474 & 3.848 & 0.010 & 7040 & 160 & 4.00 & 0.50 & 3.21 & 0.79 &  83.4 &  6.5 &  0.20 & 0.14\\
11668783 & 3.845 & 0.009 & 7000 & 150 & 4.00 & 0.43 & 2.31 & 0.52 &  33.5 &  2.7 &  0.19 & 0.12\\
11721304 & 3.855 & 0.005 & 7160 &  82 & 4.13 & 0.28 & 2.15 & 0.33 &  28.3 &  1.5 &  0.08 & 0.08\\
11754232 & 3.867 & 0.007 & 7360 & 120 & 4.03 & 0.33 & 2.40 & 0.38 &  11.2 &  1.2 & -0.05 & 0.09\\
11826272 & 3.847 & 0.004 & 7030 &  70 & 3.87 & 0.21 & 2.37 & 0.21 &  30.0 &  1.1 &  0.09 & 0.06\\
11907454 & 3.848 & 0.006 & 7040 &  90 & 4.20 & 0.32 & 2.45 & 0.27 & 109.3 &  4.7 &  0.16 & 0.08\\
11917550 & 3.850 & 0.005 & 7080 &  75 & 4.10 & 0.24 & 2.61 & 0.32 &  76.0 &  3.2 &  0.14 & 0.07\\
11920505 & 3.857 & 0.004 & 7200 &  60 & 4.13 & 0.19 & 2.55 & 0.19 &  61.0 &  2.2 &  0.10 & 0.05\\
12066947 & 3.865 & 0.004 & 7330 &  70 & 4.35 & 0.30 & 3.40 & 0.33 & 133.3 &  5.6 &  0.09 & 0.07\\
12458189 & 3.839 & 0.006 & 6895 &  90 & 3.90 & 0.31 & 2.62 & 0.39 &  68.5 &  3.3 &  0.16 & 0.09\\
12643786 & 3.857 & 0.006 & 7200 &  95 & 4.00 & 0.31 & 2.63 & 0.45 &  76.7 &  4.5 &  0.11 & 0.09\\
\enddata
\end{deluxetable}
\clearpage

%\section{Pulsation parameter values}
% \tabcolsep=2pt
\begin{deluxetable}{ccccccccccccccc}
\tablenum{A4}
\tabcolsep=4pt
\tabletypesize{\small}
\rotate
\tablecolumns{15}
\tablewidth{0pc}
\tablecaption{\label{tab:persp}The values of pulsation parameters and their error margins derived from the downward period spacing patterns for the multivariate analysis (see Section \ref{sec:specvsphot}).}
\tablehead{
KIC & $\langle P\rangle$ [days] & $\sigma_P$ & $\langle\Delta P\rangle$ [days] & $\sigma_{\Delta P}$ & $\langle\frac{\mathrm{d}\Delta P}{\mathrm{d}P}\rangle$ & $\sigma_{\frac{\mathrm{d}\Delta P}{\mathrm{d}P}}$ & $f_g$ [$d^{-1}$] & $\sigma_{f_g}$ & $A_g$ [mmag] & $\sigma_{A_g}$ & $\Delta_{res}$ & $\sigma_{\Delta_{res}}$ & $P_{\rm max}$ [days] & $\sigma_{P_{\rm max}}$}
\startdata
2710594 & 0.710059 & 0.000106 & 0.004405 & 0.000129 & -0.025 & 0.006 & 1.35536 & 0.00004 & 6.16 & 0.71 & 0.81 & 2.02 & 0.80391 & 0.00017\\
3448365 & 0.708845 & 0.000131 & 0.003169 & 0.000137 & -0.017 & 0.026 & 1.50016 & 0.00004 & 7.57 & 0.75 & 0.45 & 6.89 & 0.76368 & 0.00016\\
4846809 & 0.920973 & 0.000197 & 0.010070 & 0.000287 & -0.010 & 0.006 & 1.81324 & 0.00004 & 2.53 & 0.29 & 1.82 & 1.21 & 1.11765 & 0.00035\\
5114382 & 0.670536 & 0.000096 & 0.002642 & 0.000098 & -0.025 & 0.022 & 0.95265 & 0.00006 & 3.21 & 0.53 & 0.89 & 5.39 & 0.71833 & 0.00014\\
5522154 & 0.352971 & 0.000009 & 0.003819 & 0.000008 & -0.052 & 0.010 & 3.00986 & 0.00002 & 1.58 & 0.10 & 0.31 & 0.15 & 0.38857 & 0.00002\\
5708550 & 0.832343 & 0.000123 & 0.004606 & 0.000158 & -0.022 & 0.003 & 1.11550 & 0.00005 & 2.71 & 0.39 & 1.30 & 2.12 & 0.97390 & 0.00015\\
5788623 & 1.133749 & 0.000558 & 0.012569 & 0.000590 & -0.015 & 0.002 & 0.77895 & 0.00007 & 9.44 & 1.84 & 1.97 & 2.67 & 1.55211 & 0.00104\\
6468146 & 0.608196 & 0.000056 & 0.007406 & 0.000084 & -0.035 & 0.008 & 1.54570 & 0.00001 & 1.15 & 0.04 & 1.18 & 0.85 & 0.74142 & 0.00012\\
6468987 & 0.476347 & 0.000058 & 0.003106 & 0.000105 & -0.038 & 0.017 & 1.99899 & 0.00002 & 4.52 & 0.18 & 0.19 & 0.76 & 0.53144 & 0.00001\\
6678174 & 0.917967 & 0.000103 & 0.013403 & 0.000171 & -0.015 & 0.542 & 1.12777 & 0.00007 & 3.00 & 0.59 & 3.02 & 55.65 & 0.95240 & 0.00010\\
6935014 & 0.832490 & 0.000104 & 0.006348 & 0.000111 & -0.025 & 0.005 & 1.20670 & 0.00005 & 5.73 & 0.79 & 1.88 & 3.21 & 0.94147 & 0.00030\\
6953103 & 0.812258 & 0.000173 & 0.009450 & 0.000148 & -0.031 & 0.004 & 1.28760 & 0.00004 & 33.46 & 3.23 & 1.44 & 1.14 & 1.00870 & 0.00045\\
7023122 & 0.479343 & 0.000024 & 0.011788 & 0.000043 & -0.048 & 0.034 & 1.87611 & 0.00001 & 13.90 & 0.40 & 1.27 & 0.43 & 0.58005 & 0.00006\\
7365537 & 0.344913 & 0.000007 & 0.003234 & 0.000008 & -0.049 & 0.011 & 2.92563 & 0.00001 & 8.10 & 0.23 & 0.00 & 0.33 & 0.38088 & 0.00001\\
7380501 & 0.965164 & 0.000137 & 0.006745 & 0.000164 & -0.017 & 0.003 & 0.96329 & 0.00005 & 2.26 & 0.31 & 1.12 & 4.95 & 1.19243 & 0.00024\\
7434470 & 0.417794 & 0.000013 & 0.003200 & 0.000013 & -0.040 & 0.001 & 1.69873 & 0.00000 & 3.82 & 0.04 & 1.26 & 0.09 & 0.48637 & 0.00002\\
7583663 & 0.633695 & 0.000072 & 0.003101 & 0.000093 & -0.031 & 0.007 & 1.04474 & 0.00003 & 6.28 & 0.54 & 0.90 & 1.44 & 0.70027 & 0.00007\\
7939065 & 0.462545 & 0.000037 & 0.008775 & 0.000037 & -0.036 & 0.002 & 1.72817 & 0.00003 & 10.68 & 0.73 & 0.22 & 0.08 & 0.70123 & 0.00014\\
8364249 & 0.522644 & 0.000017 & 0.002100 & 0.000019 & -0.029 & 0.003 & 1.86938 & 0.00002 & 3.94 & 0.17 & 0.57 & 0.96 & 0.57555 & 0.00003\\
8375138 & 0.469786 & 0.000023 & 0.002377 & 0.000025 & -0.034 & 0.007 & 2.07777 & 0.00002 & 5.16 & 0.31 & 0.52 & 0.91 & 0.51310 & 0.00003\\
8645874 & 0.499508 & 0.000019 & 0.026876 & 0.000029 & -0.016 & 0.005 & 1.84701 & 0.00001 & 5.71 & 0.17 & 1.15 & 0.06 & 1.02248 & 0.00017\\
8836473 & 0.505219 & 0.000035 & 0.007475 & 0.000021 & -0.034 & 0.061 & 1.88341 & 0.00003 & 1.47 & 0.13 & 0.44 & 0.51 & 0.57115 & 0.00001\\
9419694 & 0.937988 & 0.000193 & 0.013458 & 0.000118 & -0.016 & 0.023 & 1.03279 & 0.00004 & 14.74 & 1.51 & 2.02 & 3.78 & 1.13479 & 0.00037\\
9480469 & 0.518274 & 0.000039 & 0.002455 & 0.000033 & -0.028 & 0.035 & 1.99485 & 0.00003 & 7.14 & 0.66 & 0.23 & 3.80 & 0.54748 & 0.00006\\
9595743 & 0.683181 & 0.000099 & 0.006987 & 0.000146 & -0.033 & 0.040 & 1.72911 & 0.00003 & 6.93 & 0.62 & 1.09 & 2.68 & 0.76882 & 0.00014\\
9751996 & 0.858863 & 0.000051 & 0.037537 & 0.000072 & 0.002 & 0.010 & 1.28331 & 0.00006 & 2.34 & 0.36 & 2.68 & 0.50 & 1.19934 & 0.00010\\
10224094 & 0.875860 & 0.000110 & 0.019195 & 0.000134 & -0.017 & 0.010 & 1.01242 & 0.00010 & 2.81 & 0.71 & 0.00 & 1.25 & 1.08730 & 0.00031\\
10467146 & 0.996770 & 0.000141 & 0.007373 & 0.000165 & -0.018 & 0.061 & 0.95498 & 0.00004 & 3.47 & 0.36 & 1.09 & 28.11 & 1.09130 & 0.00018\\
11080103 & 0.911278 & 0.000154 & 0.011062 & 0.000194 & -0.019 & 0.016 & 1.24139 & 0.00002 & 12.13 & 0.75 & 0.97 & 4.74 & 1.09122 & 0.00017\\
11099031 & 0.698765 & 0.000055 & 0.004375 & 0.000067 & -0.026 & 0.003 & 0.91665 & 0.00003 & 1.42 & 0.13 & 0.97 & 0.99 & 0.81286 & 0.00006\\
11294808 & 0.912759 & 0.000098 & 0.006943 & 0.000118 & -0.043 & 0.023 & 2.22475 & 0.00004 & 2.32 & 0.22 & 1.17 & 13.74 & 0.99168 & 0.00013\\
11456474 & 0.690040 & 0.000060 & 0.003209 & 0.000071 & -0.025 & 0.002 & 1.47147 & 0.00002 & 4.17 & 0.26 & 1.19 & 1.38 & 0.78283 & 0.00007\\
11721304 & 1.026382 & 0.000331 & 0.011981 & 0.000430 & -0.020 & 0.002 & 0.92287 & 0.00009 & 4.10 & 1.03 & 2.42 & 0.77 & 1.32327 & 0.00053\\
11754232 & 0.665339 & 0.000034 & 0.031646 & 0.000051 & -0.002 & 0.034 & 1.10338 & 0.00006 & 2.03 & 0.30 & 1.94 & 0.37 & 0.90630 & 0.00005\\
11826272 & 1.074839 & 0.000399 & 0.015434 & 0.000543 & -0.018 & 0.001 & 0.83370 & 0.00006 & 11.61 & 1.94 & 1.87 & 0.50 & 1.59754 & 0.00086\\
11907454 & 0.548784 & 0.000061 & 0.002847 & 0.000084 & -0.033 & 0.006 & 1.18715 & 0.00003 & 4.43 & 0.36 & 0.87 & 0.89 & 0.60608 & 0.00007\\
11917550 & 0.716579 & 0.000084 & 0.005704 & 0.000104 & -0.028 & 0.003 & 1.28768 & 0.00004 & 9.21 & 0.87 & 1.86 & 1.08 & 0.82445 & 0.00018\\
11920505 & 0.787720 & 0.000142 & 0.007948 & 0.000144 & -0.027 & 0.003 & 1.19884 & 0.00004 & 13.26 &1.43 & 1.90 & 1.58 & 0.95371 & 0.00020\\
12066947 & 0.359580 & 0.000016 & 0.002396 & 0.000017 & -0.040 & 0.003 & 2.72379 & 0.00004 & 2.77 & 0.30 & 0.17 & 0.14 & 0.40962 & 0.00003\\
12458189 & 0.917404 & 0.000135 & 0.005263 & 0.000178 & -0.019 & 0.003 & 1.03961 & 0.00003 & 4.44 & 0.30 & 1.09 & 2.24 & 1.12169 & 0.00019
\enddata
\end{deluxetable}

\clearpage

\section{Detected period spacing patterns}
\label{app:persp}
Below we provide figures of all the detected period spacing patterns for the
sample stars. The used symbols are the same as the ones used  in
Fig.\,\ref{fig:kic9751996} of the main text.
\begin{figure*}[h]
 \includegraphics[width=\textwidth]{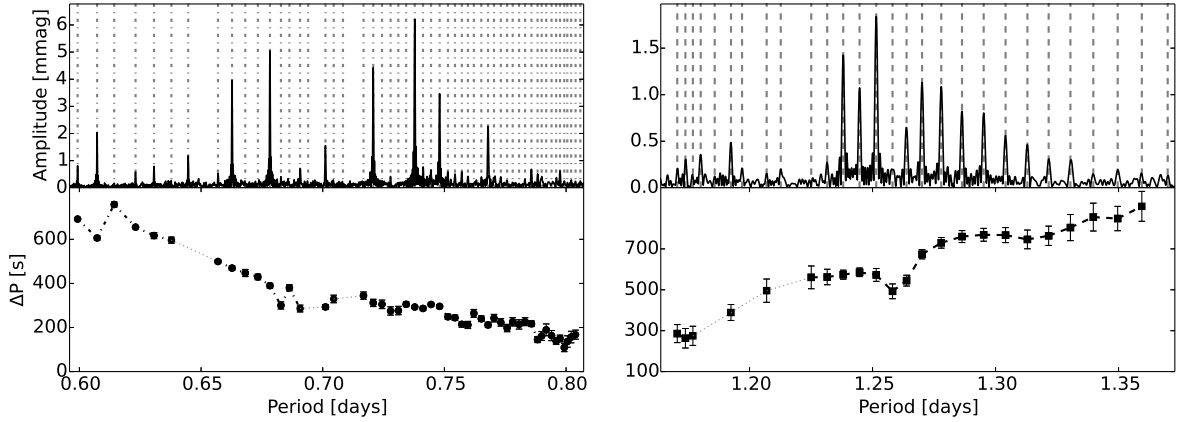}
 \caption{\label{fig:kic2710594b}The period spacing patterns of KIC\,2710594.}
\end{figure*}

\begin{figure*}[h]
 \includegraphics[width=\textwidth]{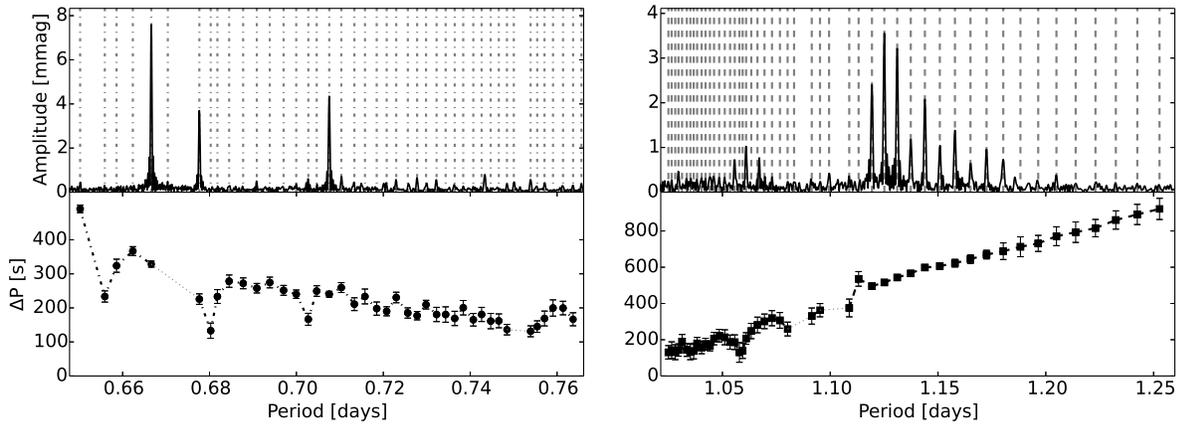}
 \caption{\label{fig:kic3448365}The period spacing patterns of KIC\,3448365.}
\end{figure*}

\begin{figure*}
 \includegraphics[width=\textwidth]{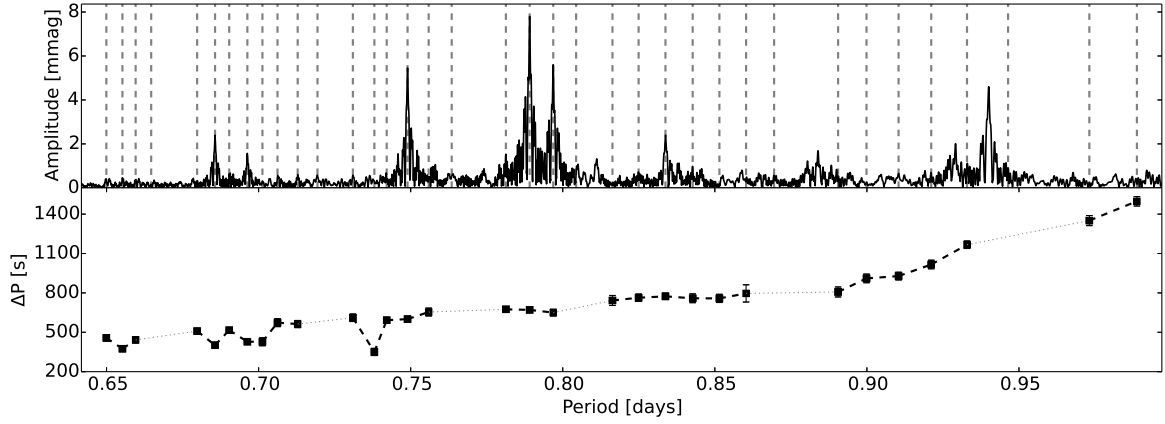}
 \caption{\label{fig:kic4757184}The period spacing patterns of KIC\,4757184.}
\end{figure*}

\begin{figure*}
 \includegraphics[width=\textwidth]{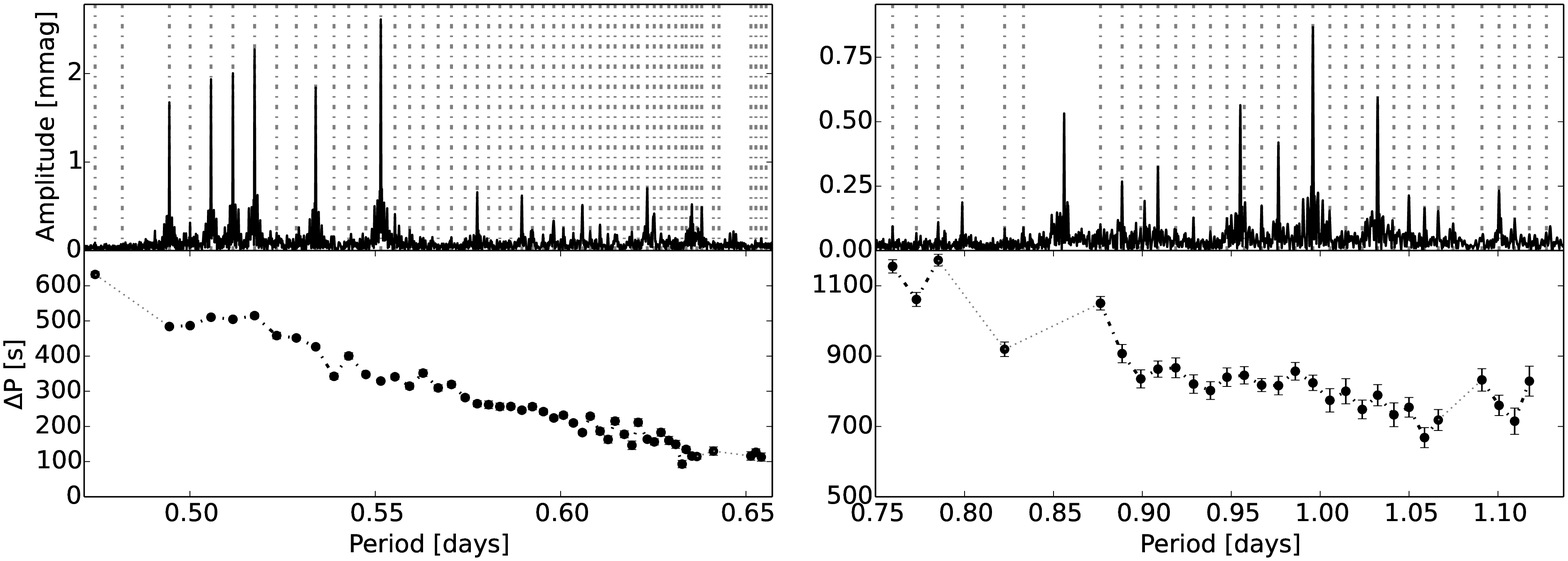}
 \caption{\label{fig:kic4846809}The period spacing patterns of KIC\,4846809.}
\end{figure*}

\begin{figure*}
 \includegraphics[width=\textwidth]{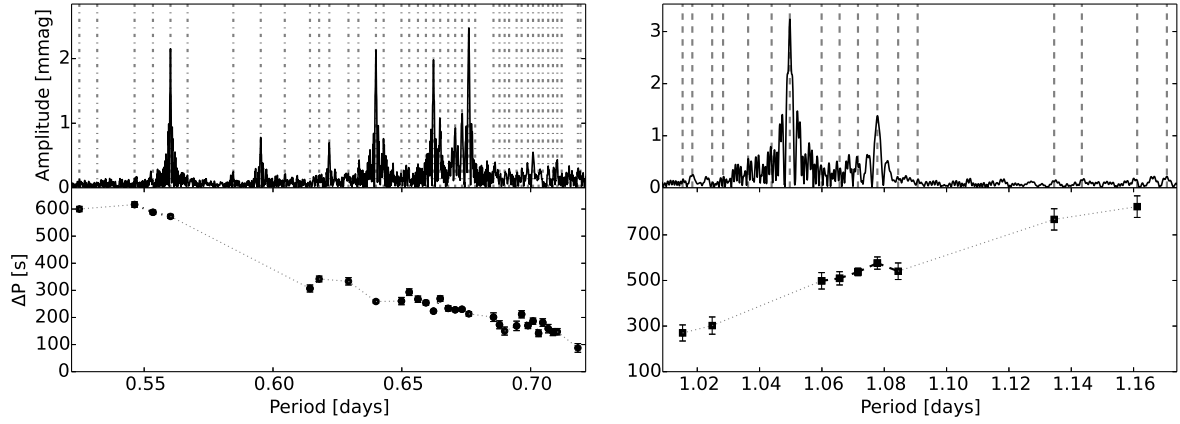}
 \caption{\label{fig:kic5114382}The period spacing patterns of KIC\,5114382.}
\end{figure*}

\begin{figure*}
 \includegraphics[width=\textwidth]{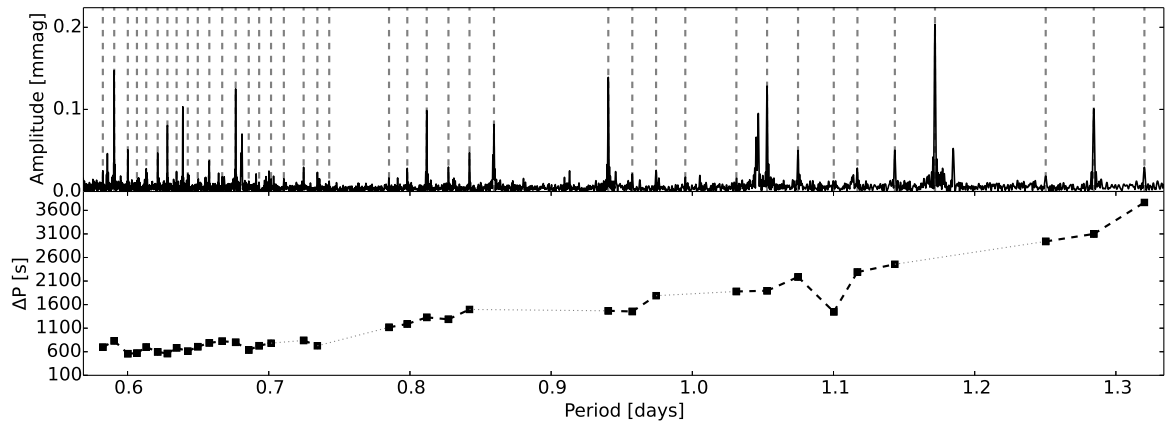}
 \caption{\label{fig:kic5350598}The period spacing patterns of KIC\,5350598.}
\end{figure*}

\begin{figure*}
 \includegraphics[width=\textwidth]{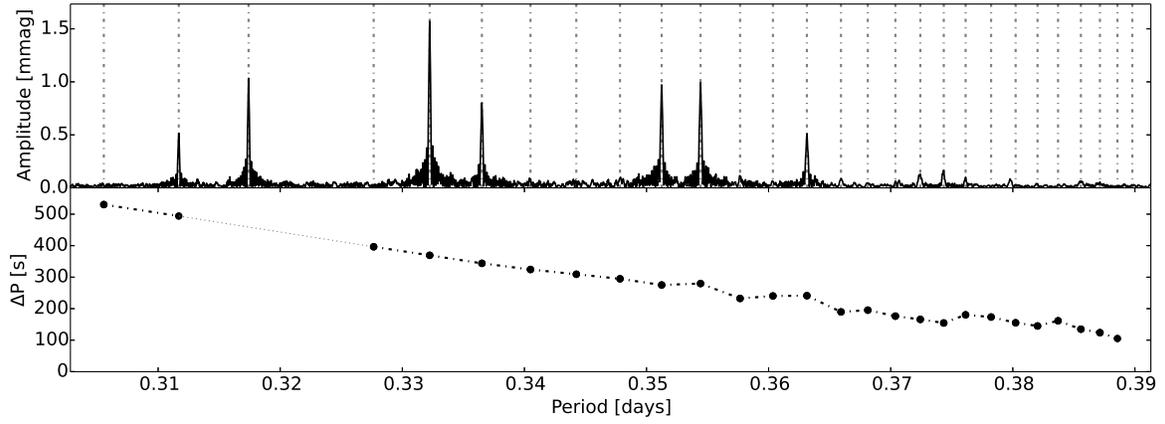}
 \caption{\label{fig:kic5522154}The period spacing patterns of KIC\,5522154.}
\end{figure*}

\begin{figure*}
 \includegraphics[width=\textwidth]{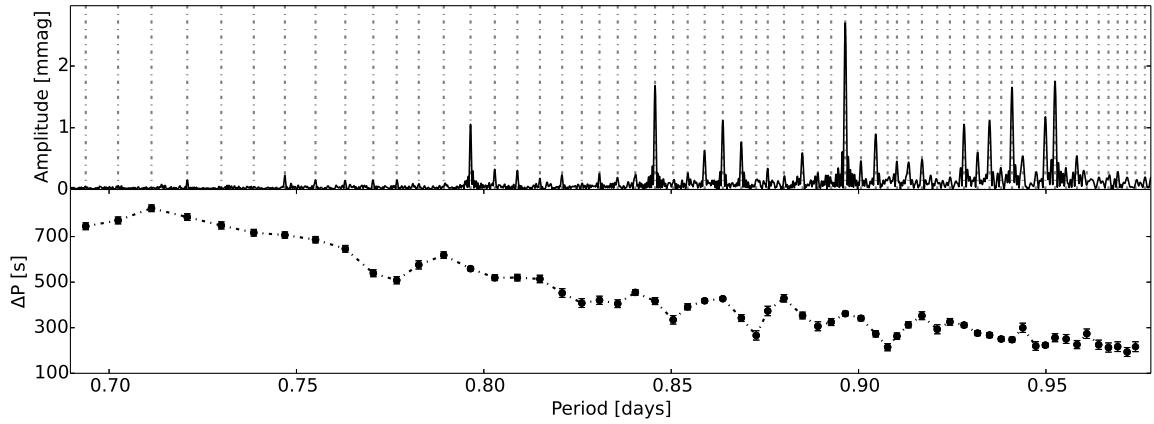}
 \caption{\label{fig:kic5708550}The period spacing patterns of KIC\,5708550.}
\end{figure*}

\begin{figure*}
 \includegraphics[width=\textwidth]{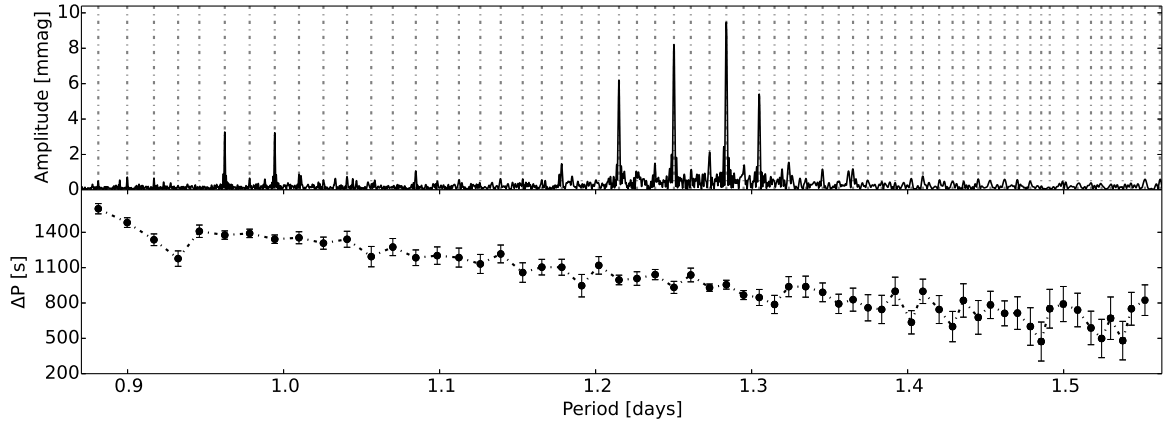}
 \caption{\label{fig:kic5788623}The period spacing patterns of KIC\,5788623.}
\end{figure*}

\begin{figure*}
 \includegraphics[width=\textwidth]{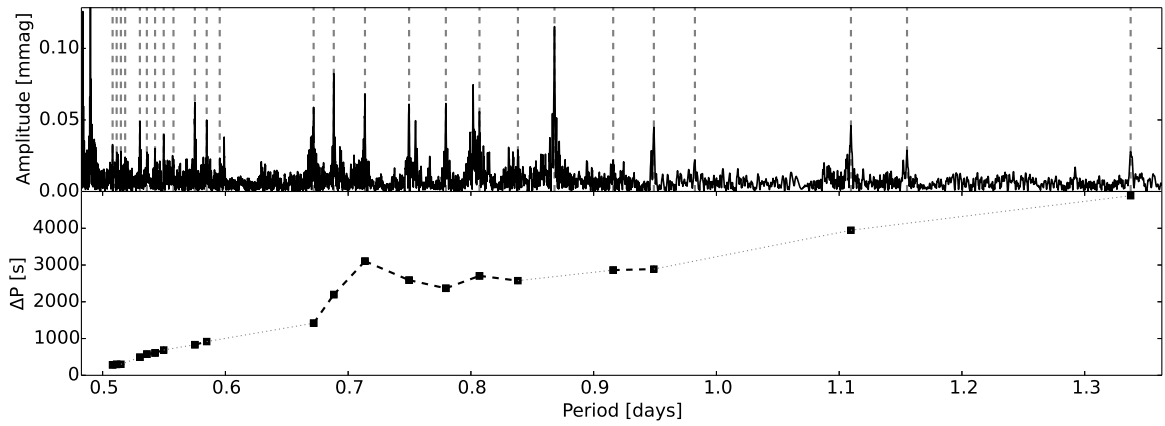}
 \caption{\label{fig:kic6185513}The period spacing patterns of KIC\,6185513.}
\end{figure*}

\begin{figure*}
 \includegraphics[width=\textwidth]{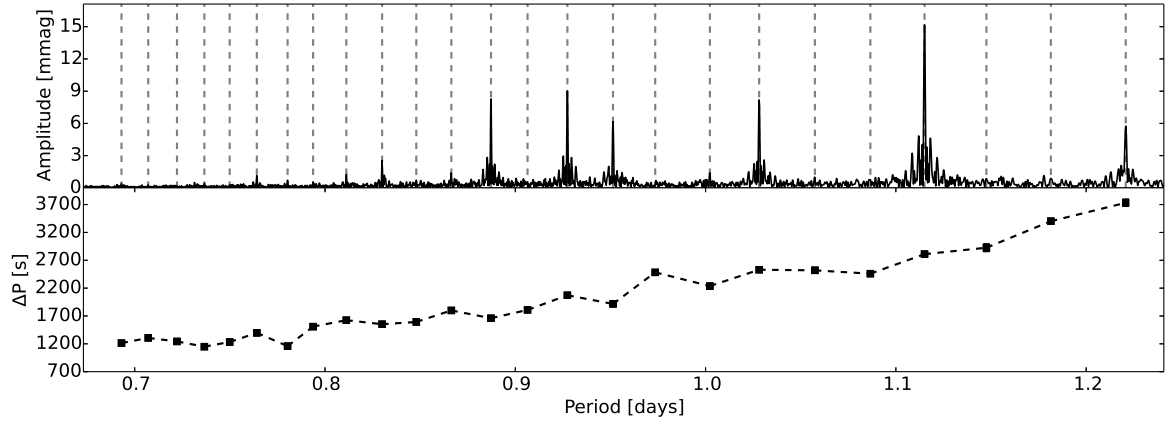}
 \caption{\label{fig:kic6425437}The period spacing patterns of KIC\,6425437.}
\end{figure*}

\begin{figure*}
 \includegraphics[width=\textwidth]{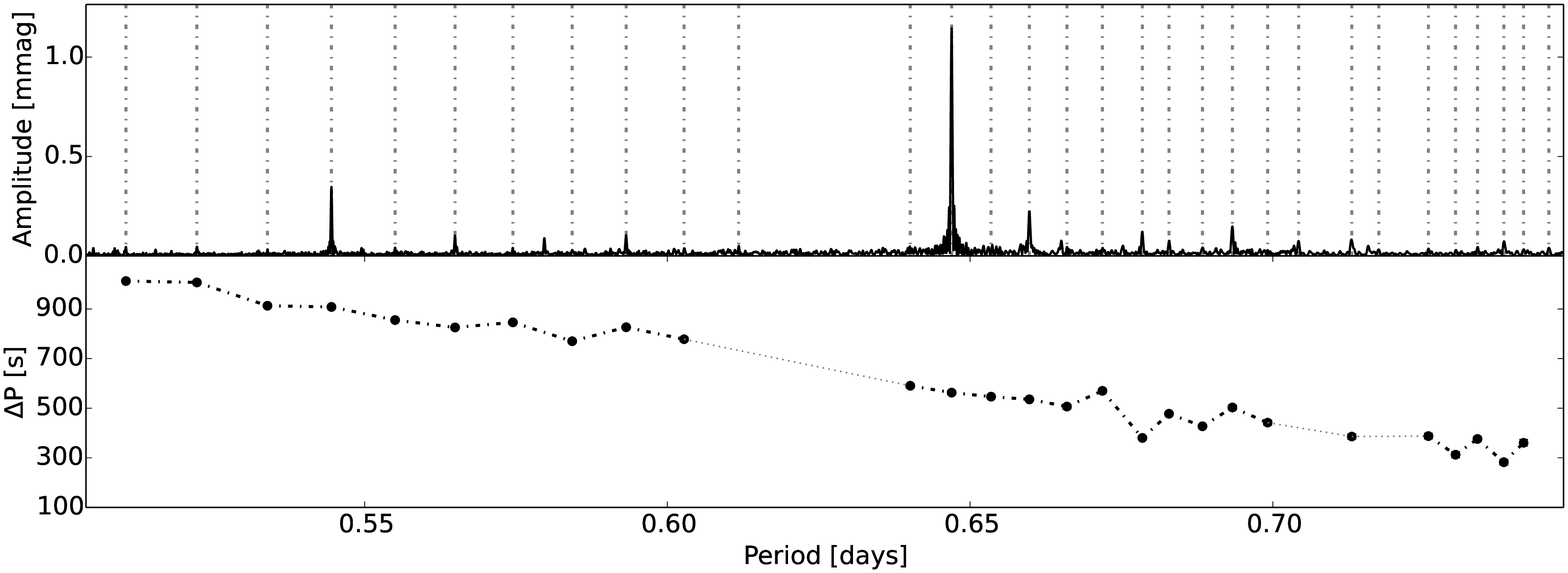}
 \caption{\label{fig:kic6468146}The period spacing patterns of KIC\,6468146.}
\end{figure*}

\begin{figure*}
 \includegraphics[width=\textwidth]{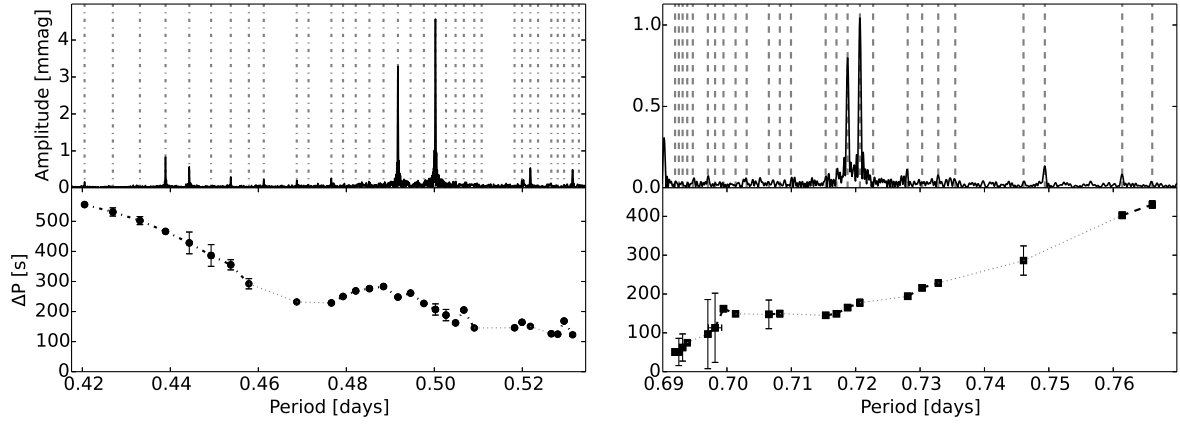}
 \caption{\label{fig:kic6468987}The period spacing patterns of KIC\,6468987.}
\end{figure*}

\begin{figure*}
 \includegraphics[width=\textwidth]{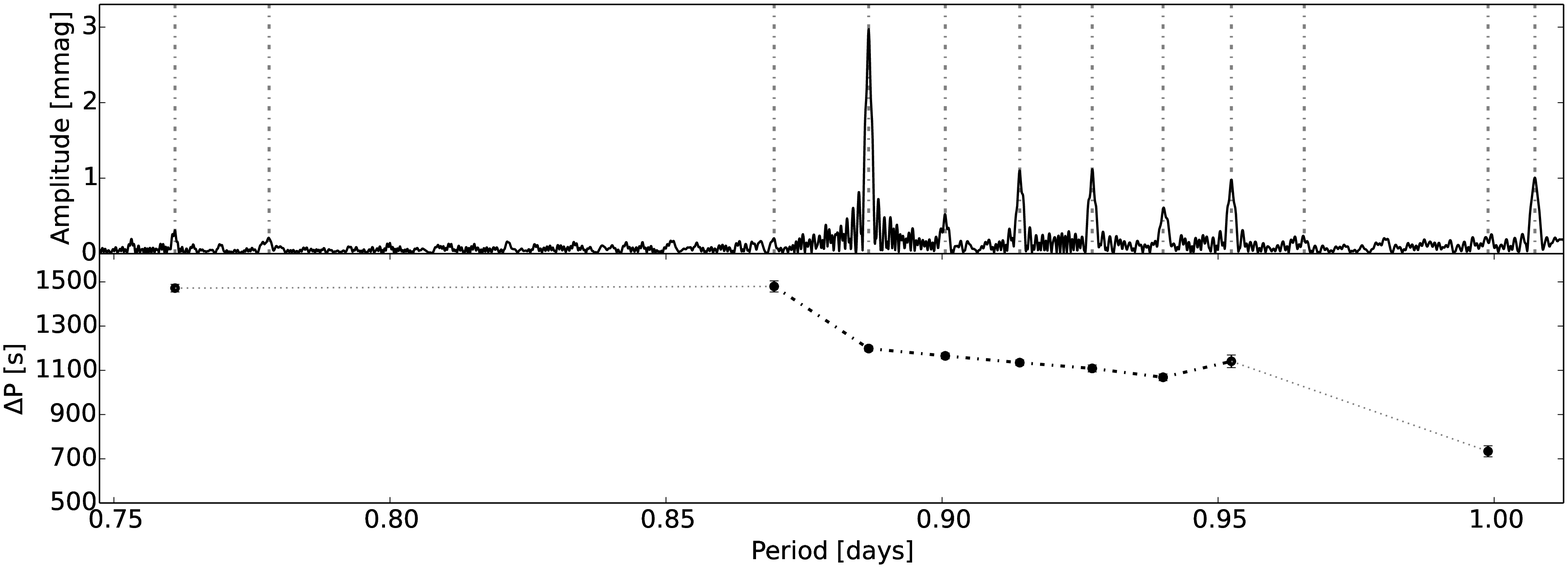}
 \caption{\label{fig:kic6678174}The period spacing patterns of KIC\,6678174.}
\end{figure*}

\begin{figure*}
 \includegraphics[width=\textwidth]{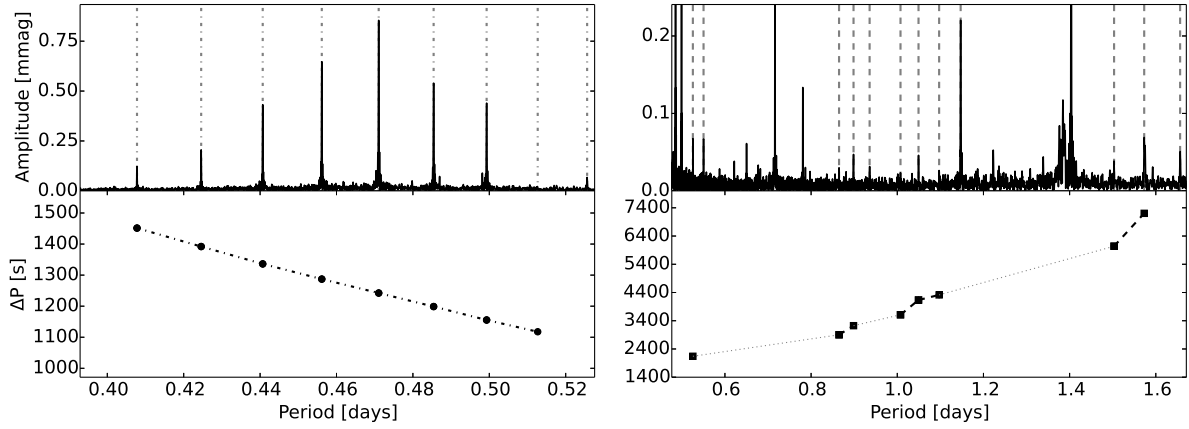}
 \caption{\label{fig:kic6778063}The period spacing patterns of KIC\,6778063.}
\end{figure*}

\begin{figure*}
 \includegraphics[width=\textwidth]{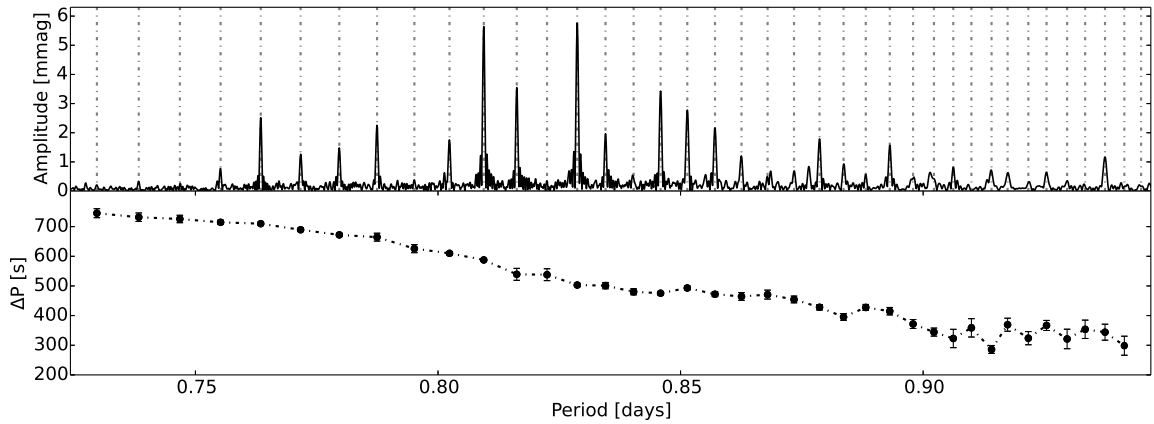}
 \caption{\label{fig:kic6935014}The period spacing patterns of KIC\,6935014.}
\end{figure*}

\begin{figure*}
 \includegraphics[width=\textwidth]{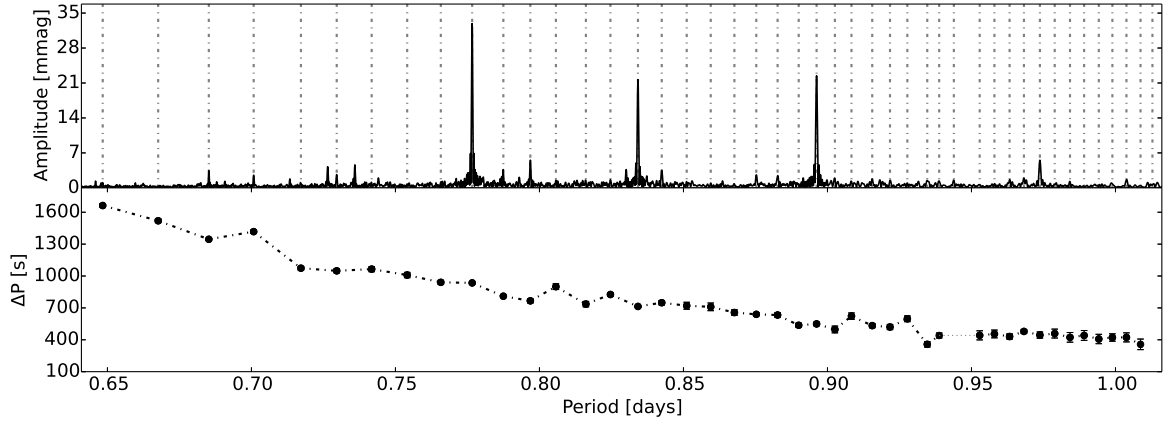}
 \caption{\label{fig:kic6953103}The period spacing patterns of KIC\,6953103.}
\end{figure*}

\begin{figure*}
 \includegraphics[width=\textwidth]{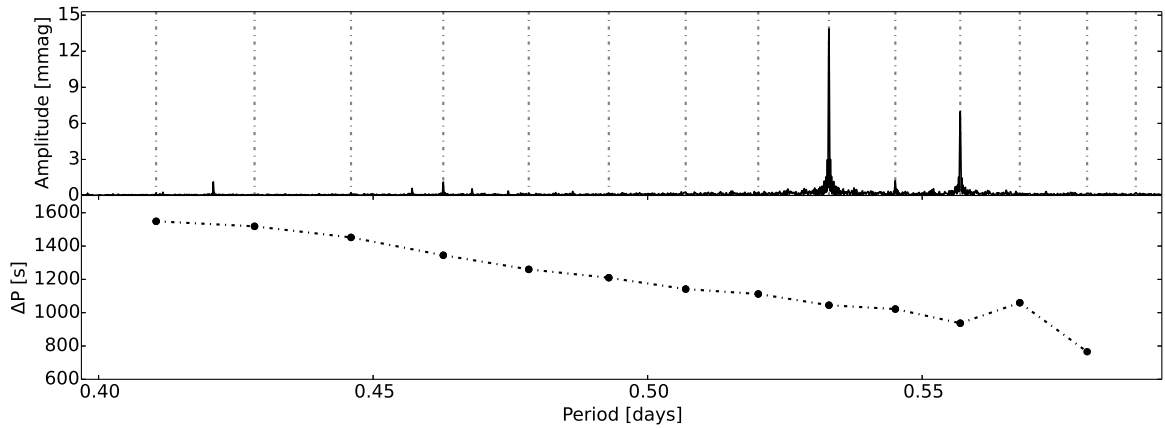}
 \caption{\label{fig:kic7023122}The period spacing patterns of KIC\,7023122.}
\end{figure*}
\clearpage
\begin{figure*}
 \includegraphics[width=\textwidth]{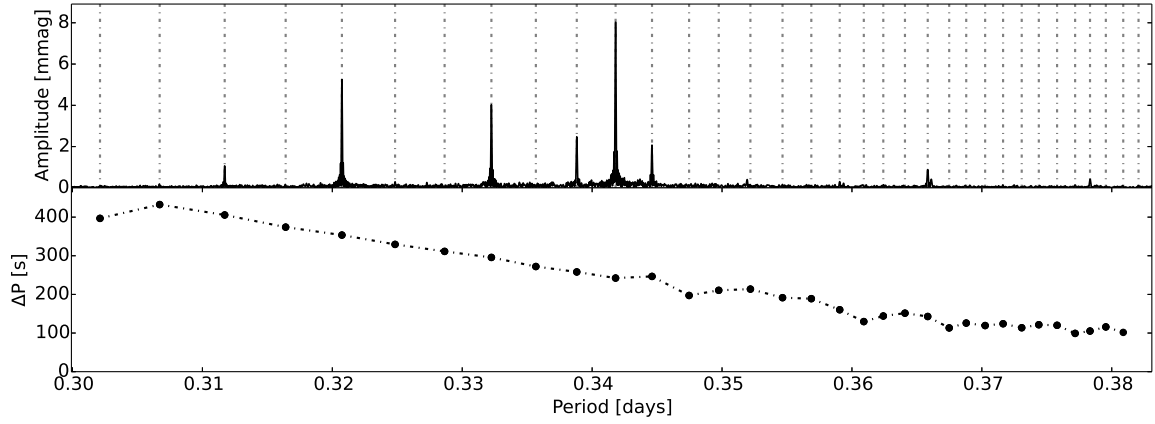}
 \caption{\label{fig:kic7365537}The period spacing patterns of KIC\,7365537.}
\end{figure*}

\begin{figure*}
 \includegraphics[width=\textwidth]{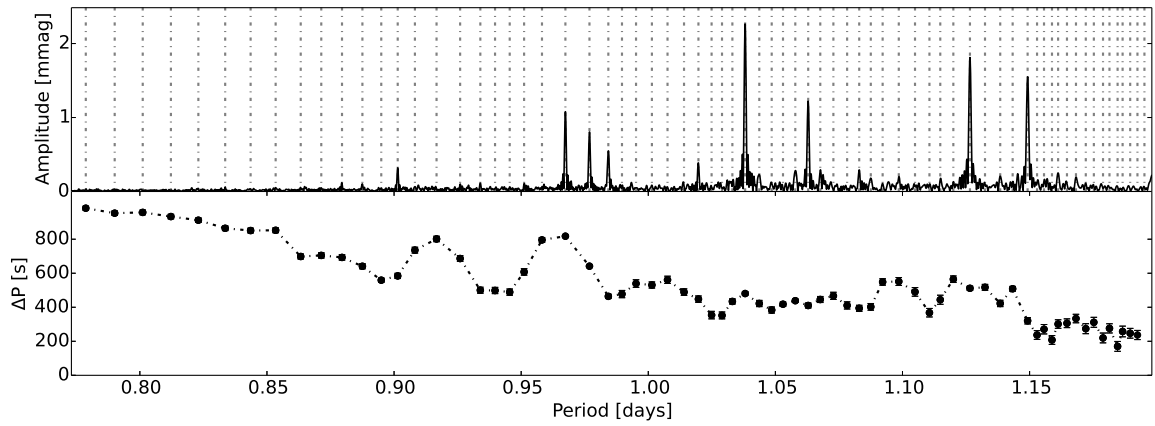}
 \caption{\label{fig:kic7380501}The period spacing patterns of KIC\,7380501.}
\end{figure*}

\begin{figure*}
 \includegraphics[width=\textwidth]{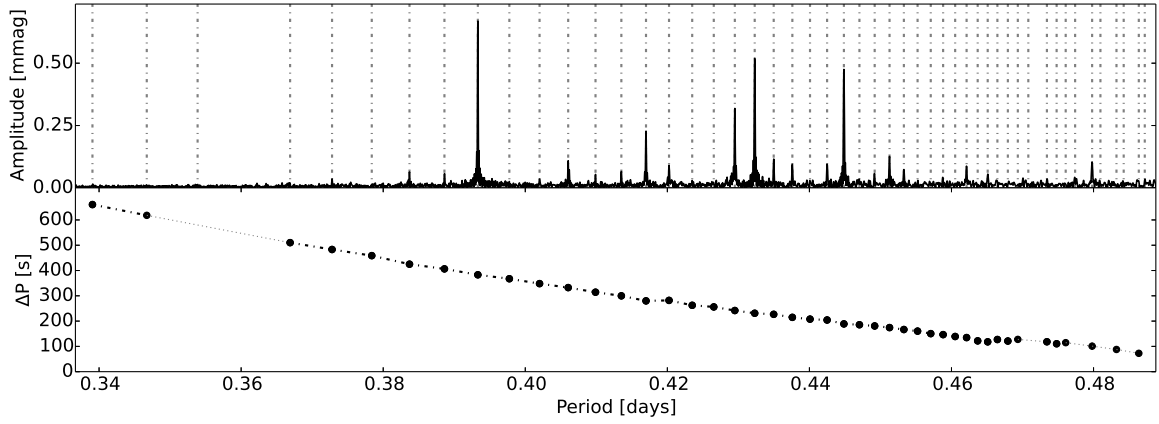}
 \caption{\label{fig:kic7434470}The period spacing patterns of KIC\,7434470.}
\end{figure*}

\begin{figure*}
 \includegraphics[width=\textwidth]{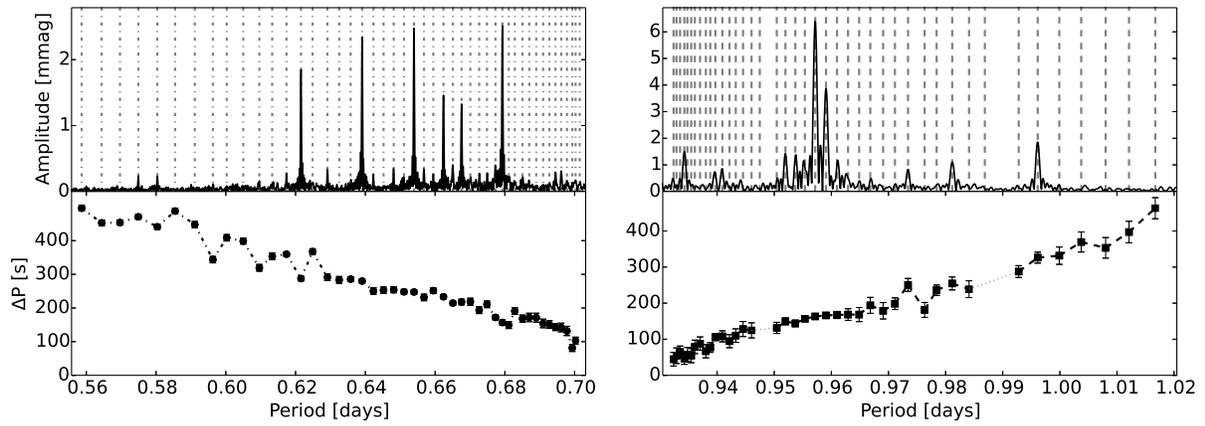}
 \caption{\label{fig:kic7583663}The period spacing patterns of KIC\,7583663.}
\end{figure*}

\begin{figure*}
 \includegraphics[width=\textwidth]{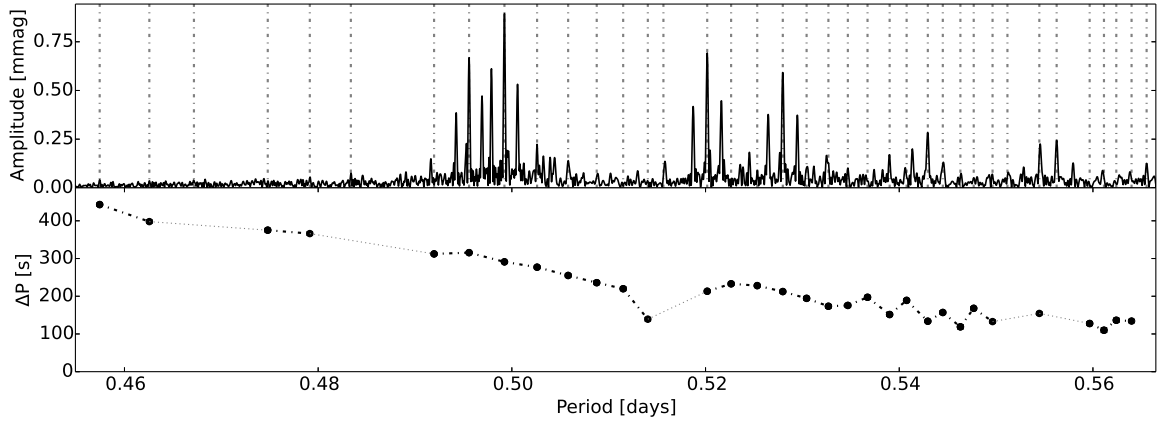}
 \caption{\label{fig:kic7746984}The period spacing patterns of KIC\,7746984.}
\end{figure*}

\begin{figure*}
 \includegraphics[width=\textwidth]{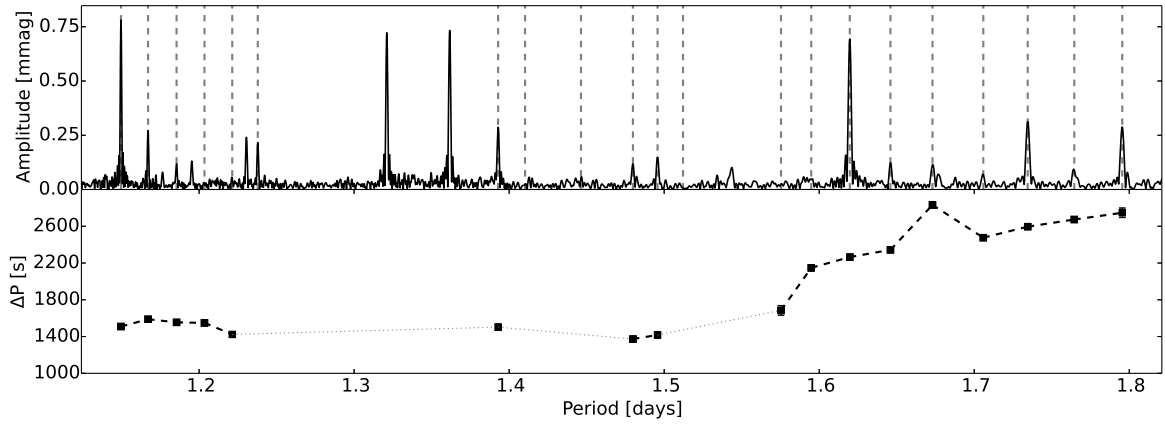}
 \caption{\label{fig:kic7867348}The period spacing patterns of KIC\,7867348.}
\end{figure*}

\begin{figure*}
 \includegraphics[width=\textwidth]{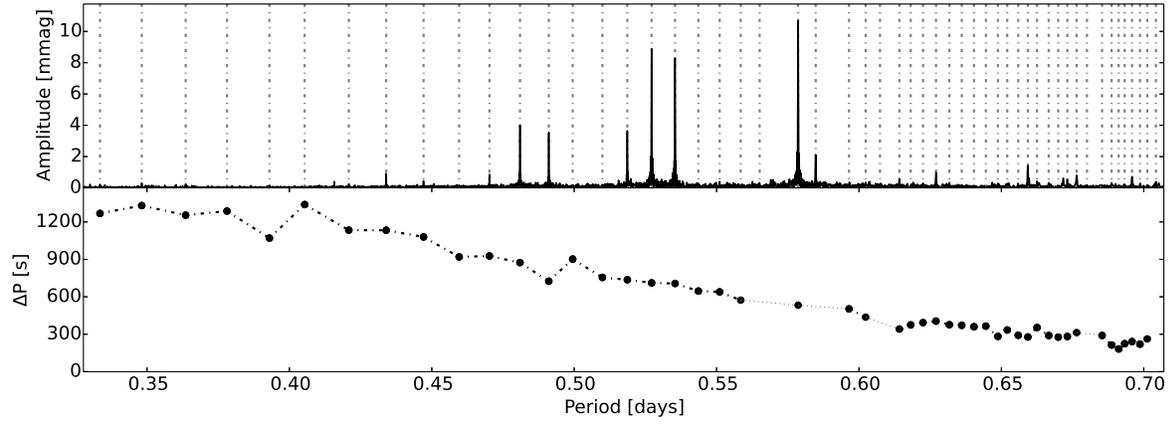}
 \caption{\label{fig:kic7939065}The period spacing patterns of KIC\,7939065.}
\end{figure*}

\begin{figure*}
 \includegraphics[width=\textwidth]{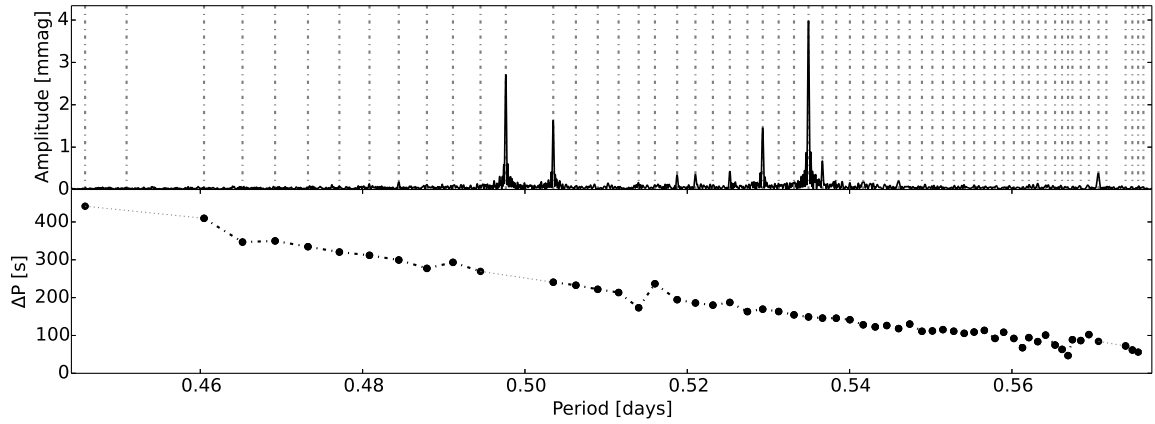}
 \caption{\label{fig:kic8364249}The period spacing patterns of KIC\,8364249.}
\end{figure*}

\begin{figure*}
 \includegraphics[width=\textwidth]{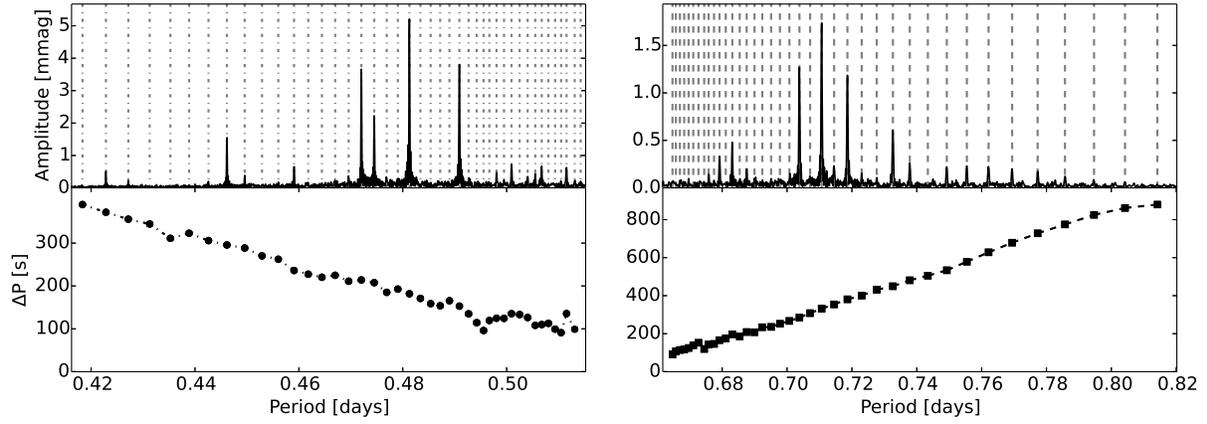}
 \caption{\label{fig:kic8375138b}The period spacing patterns of KIC\,8375138.}
\end{figure*}

\begin{figure*}
 \includegraphics[width=\textwidth]{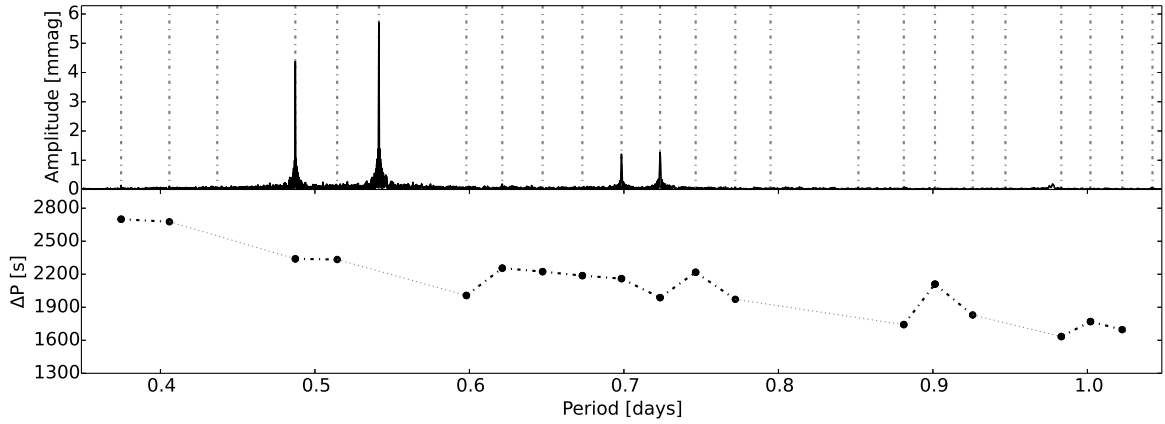}
 \caption{\label{fig:kic8645874}The period spacing patterns of KIC\,8645874.}
\end{figure*}

% \begin{figure*}
%  \includegraphics[width=\textwidth]{KIC8693972.eps}
%  \caption{\label{fig:kic8693972}The period spacing patterns of KIC\,8693972.}
% \end{figure*}

\begin{figure*}
 \includegraphics[width=\textwidth]{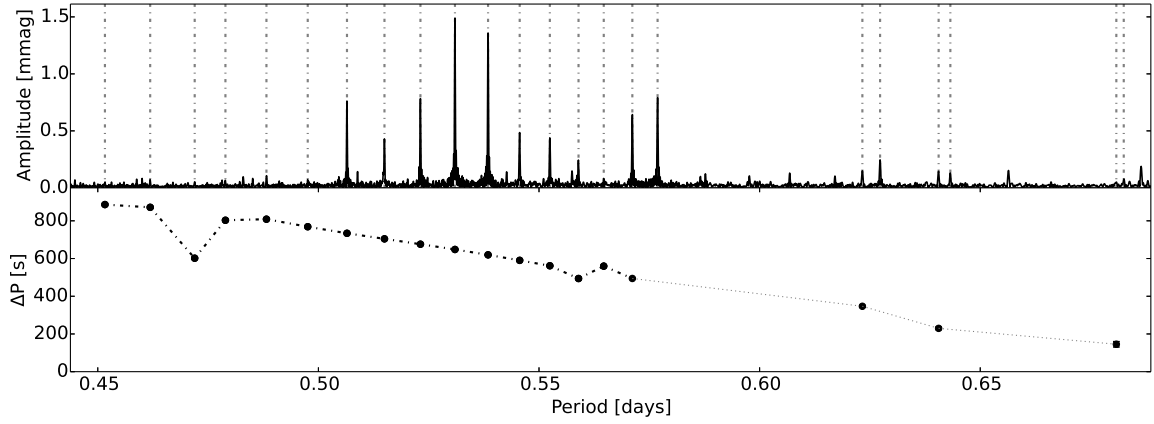}
 \caption{\label{fig:kic8836473}The period spacing patterns of KIC\,8836473.}
\end{figure*}

\begin{figure*}
 \includegraphics[width=\textwidth]{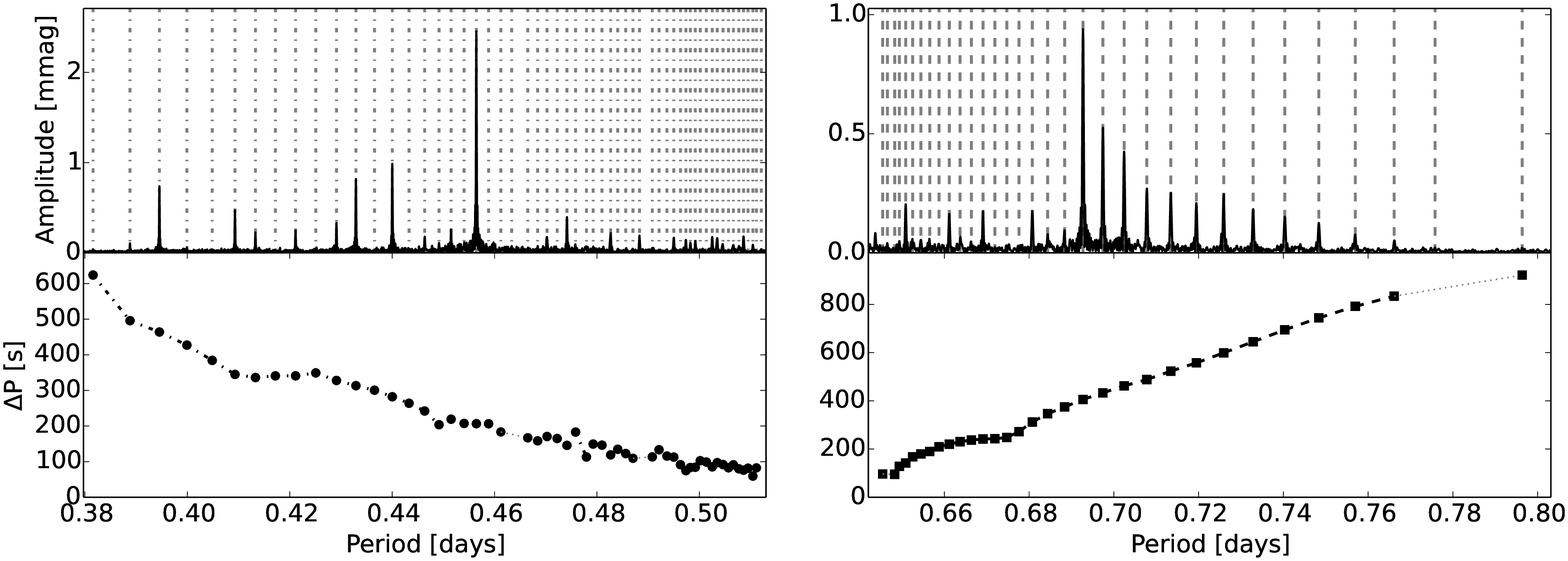}
 \caption{\label{fig:kic9210943}The period spacing patterns of KIC\,9210943.}
\end{figure*}

\begin{figure*}
 \includegraphics[width=\textwidth]{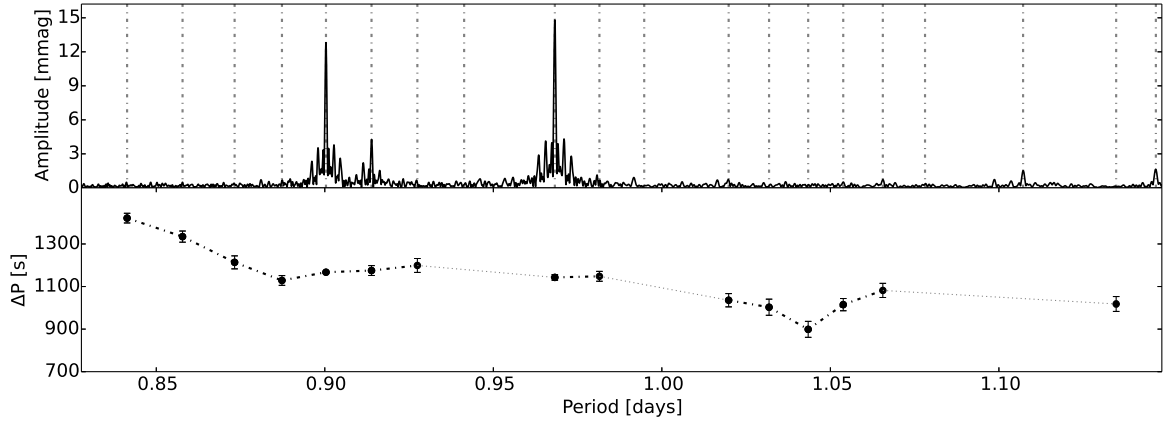}
 \caption{\label{fig:kic9419694}The period spacing patterns of KIC\,9419694.}
\end{figure*}

\begin{figure*}
 \includegraphics[width=\textwidth]{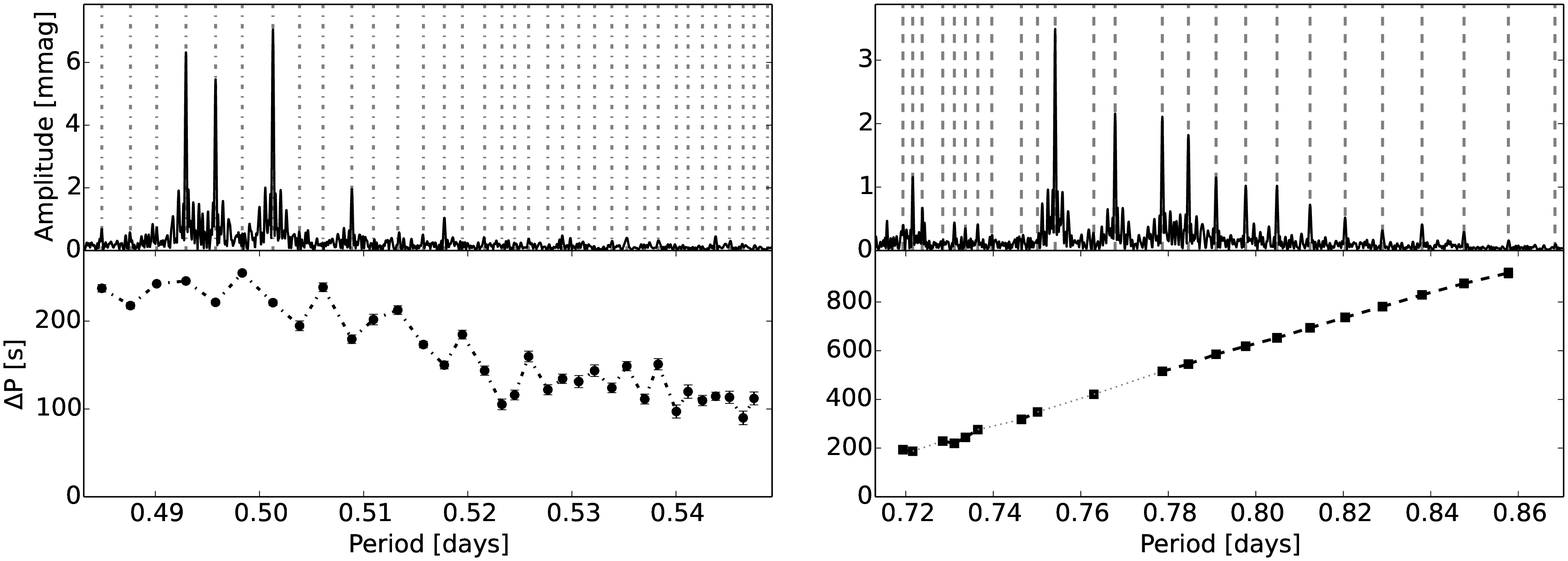}
 \caption{\label{fig:kic9480469}The period spacing patterns of KIC\,9480469.}
\end{figure*}

\begin{figure*}
 \includegraphics[width=\textwidth]{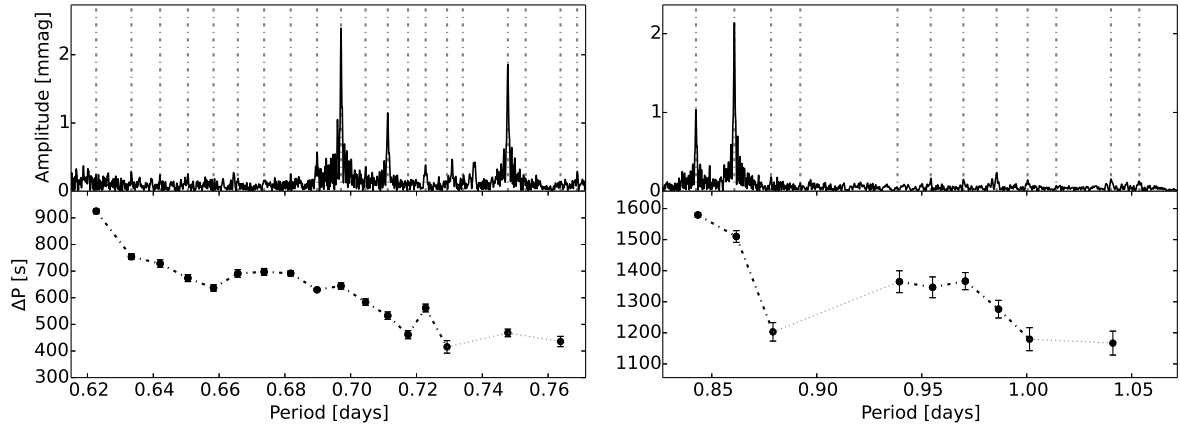}
 \caption{\label{fig:kic9595743}The period spacing patterns of KIC\,9595743.}
\end{figure*}

\begin{figure*}
 \includegraphics[width=\textwidth]{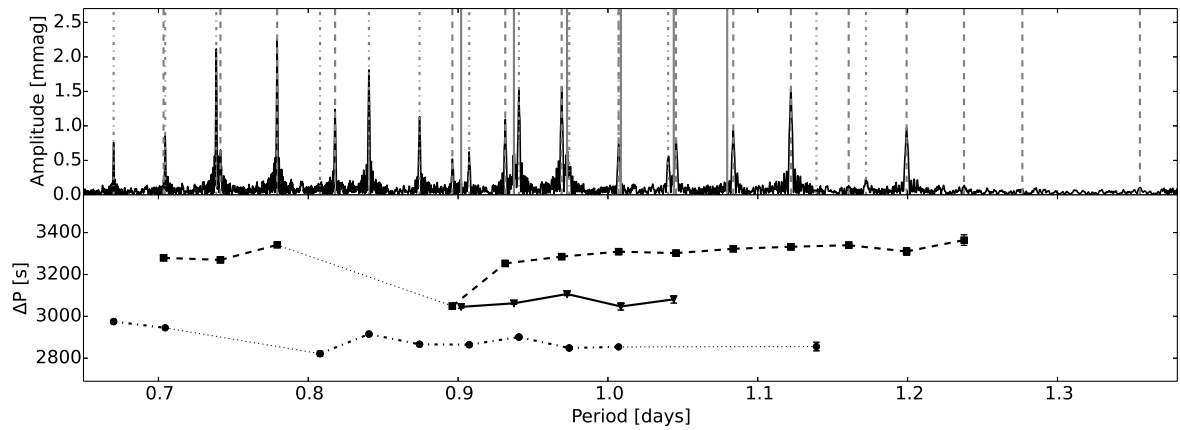}
 \caption{\label{fig:kic9751996b}The period spacing patterns of KIC\,9751996.}
\end{figure*}

\begin{figure*}
 \includegraphics[width=\textwidth]{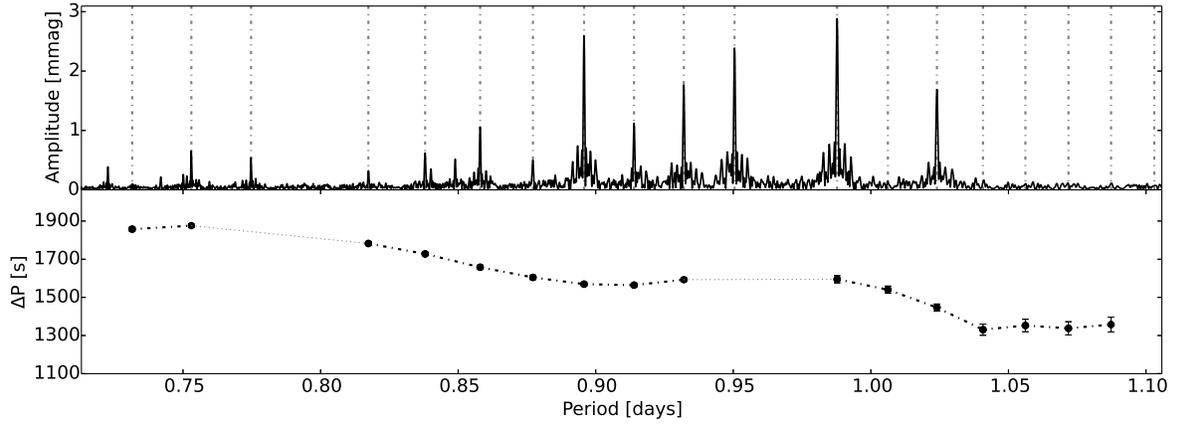}
 \caption{\label{fig:kic10224094}The period spacing patterns of KIC\,10224094.}
\end{figure*}

\begin{figure*}
 \includegraphics[width=\textwidth]{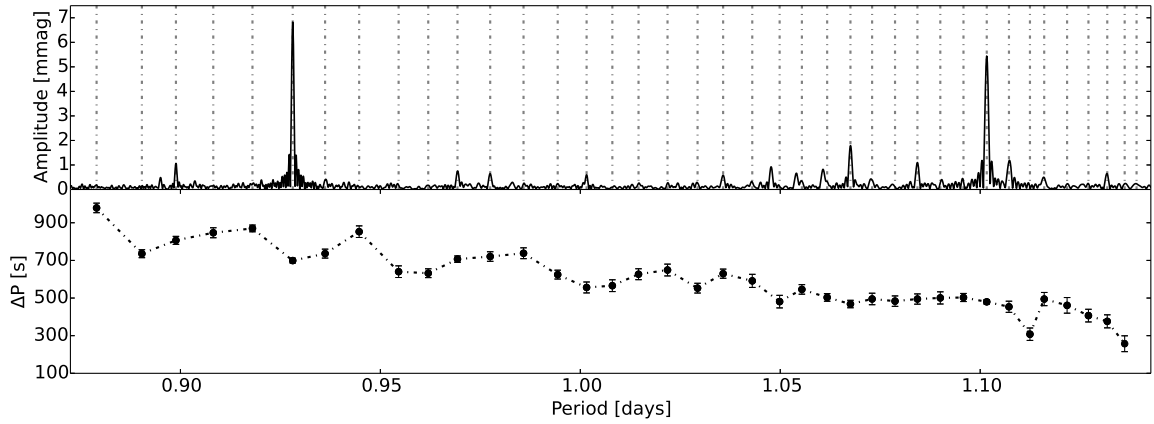}
 \caption{\label{fig:kic10256787}The period spacing patterns of KIC\,10256787.}
\end{figure*}
\clearpage
\begin{figure*}
 \includegraphics[width=\textwidth]{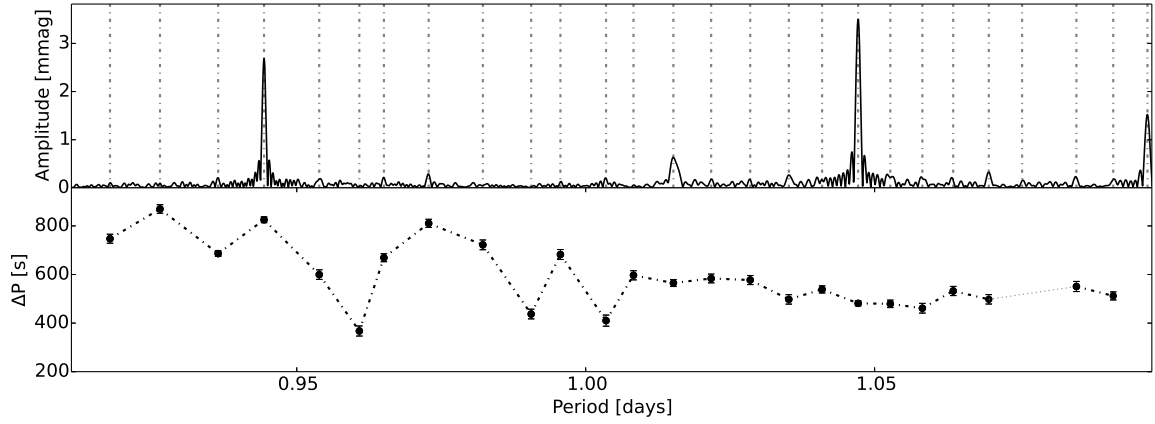}
 \caption{\label{fig:kic10467146}The period spacing patterns of KIC\,10467146.}
\end{figure*}

\begin{figure*}
 \includegraphics[width=\textwidth]{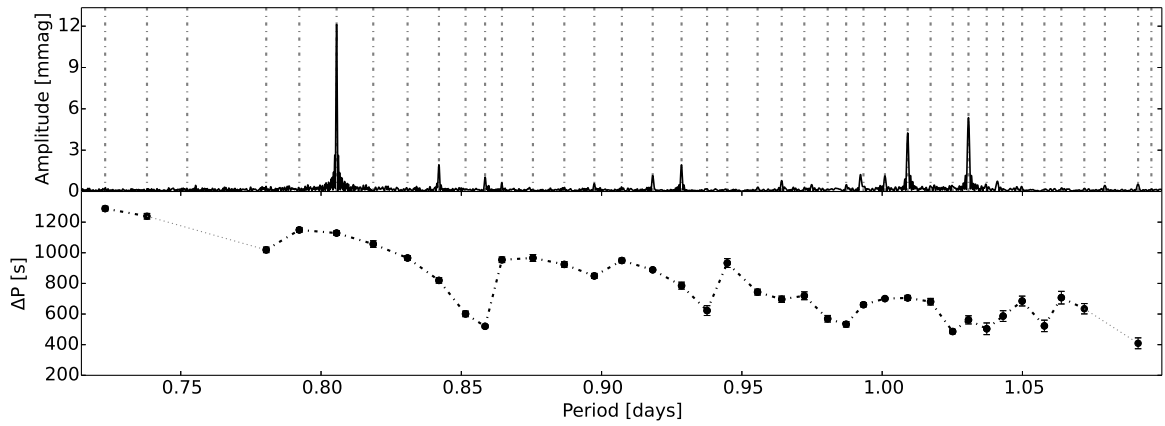}
 \caption{\label{fig:kic11080103}The period spacing patterns of KIC\,11080103.}
\end{figure*}

\begin{figure*}
 \includegraphics[width=\textwidth]{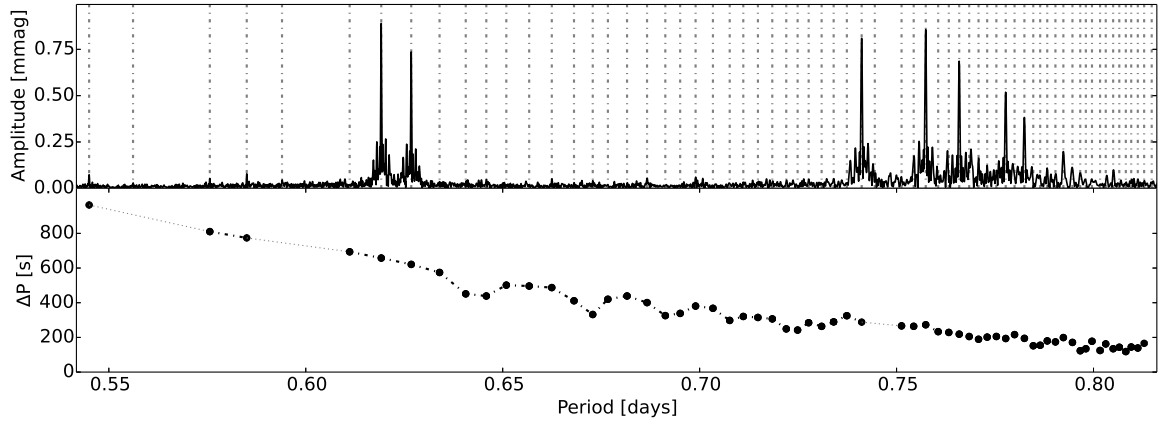}
 \caption{\label{fig:kic11099031}The period spacing patterns of KIC\,11099031.}
\end{figure*}

\begin{figure*}
 \includegraphics[width=\textwidth]{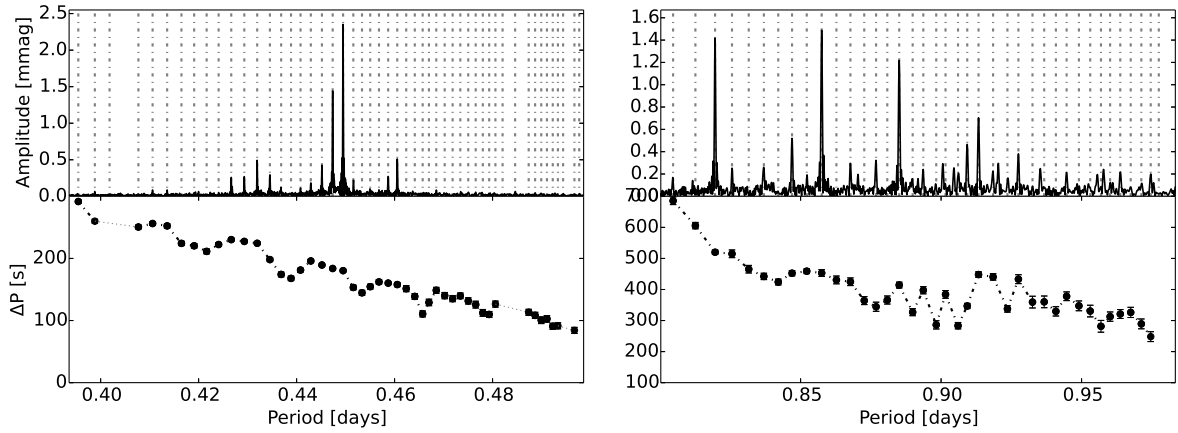}
 \caption{\label{fig:kic11294808}The period spacing patterns of KIC\,11294808.}
\end{figure*}

\begin{figure*}
 \includegraphics[width=\textwidth]{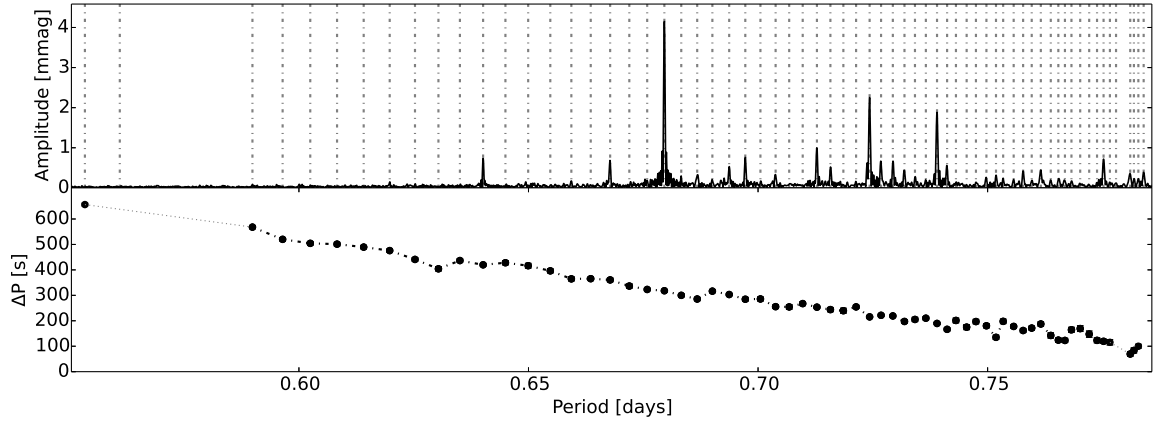}
 \caption{\label{fig:kic11456474}The period spacing patterns of KIC\,11456474.}
\end{figure*}

\begin{figure*}
 \includegraphics[width=\textwidth]{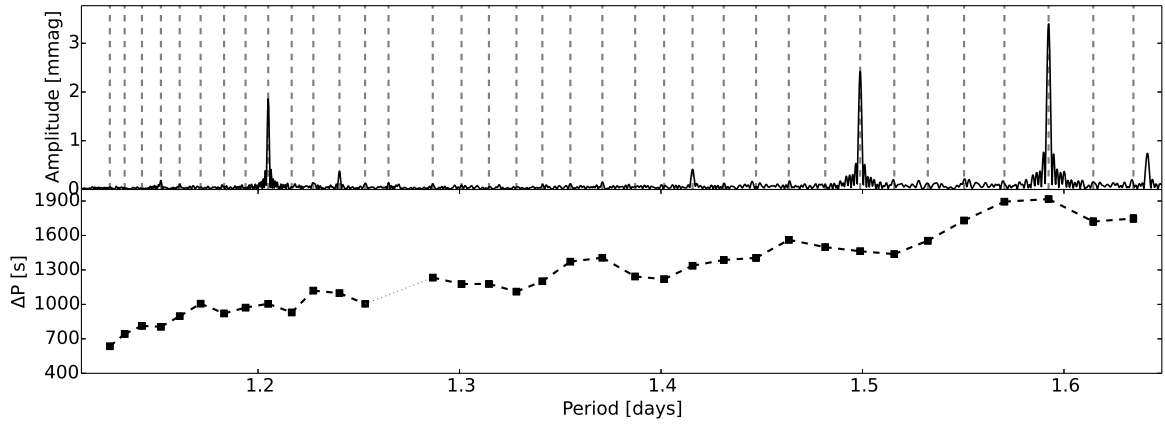}
 \caption{\label{fig:kic11668783}The period spacing patterns of KIC\,11668783.}
\end{figure*}

\begin{figure*}
 \includegraphics[width=\textwidth]{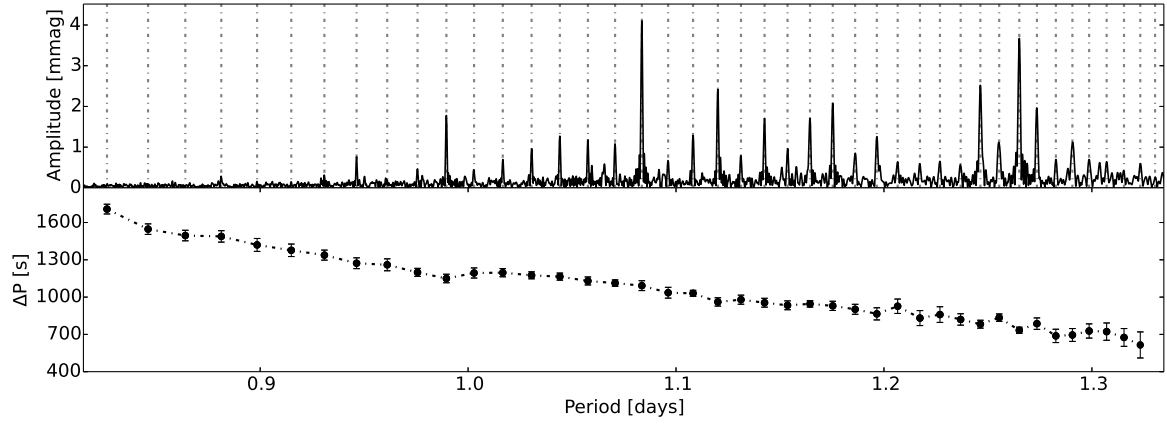}
 \caption{\label{fig:kic11721304}The period spacing patterns of KIC\,11721304.}
\end{figure*}

\begin{figure*}
 \includegraphics[width=\textwidth]{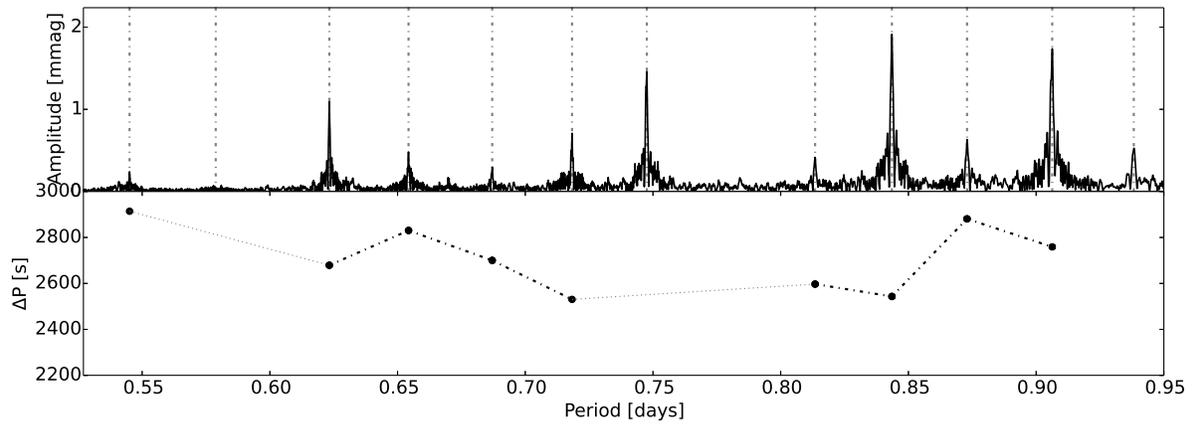}
 \caption{\label{fig:kic11754232}The period spacing patterns of KIC\,11754232.}
\end{figure*}

\begin{figure*}
 \includegraphics[width=\textwidth]{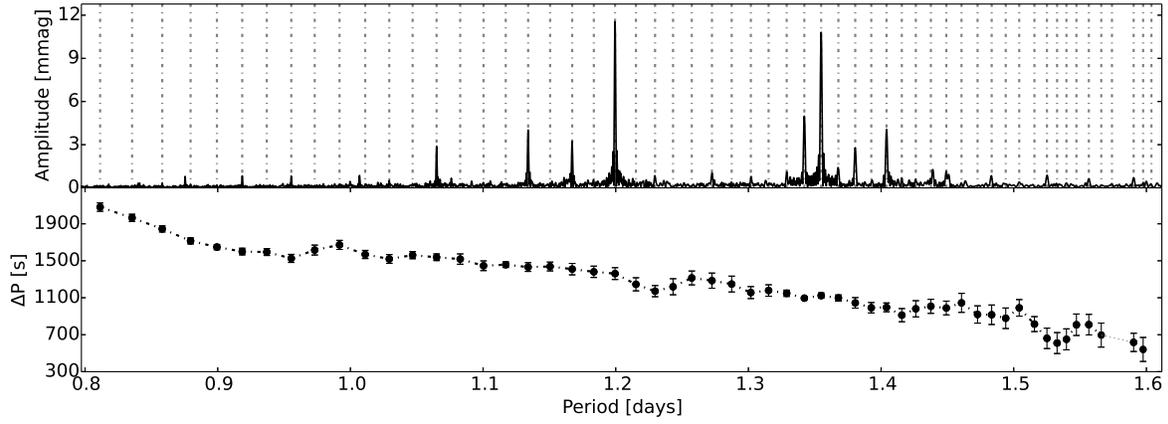}
 \caption{\label{fig:kic11826272}The period spacing patterns of KIC\,11826272.}
\end{figure*}

\begin{figure*}
 \includegraphics[width=\textwidth]{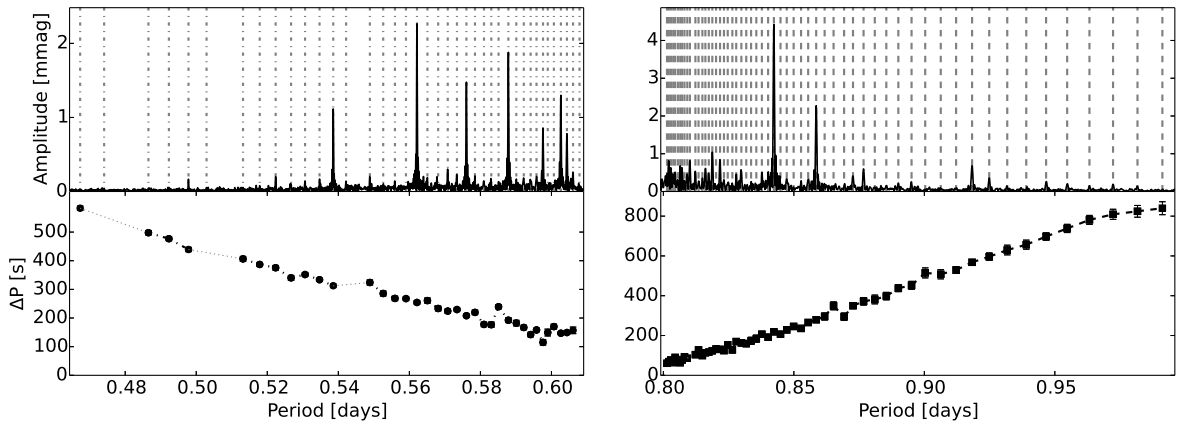}
 \caption{\label{fig:kic11907454}The period spacing patterns of KIC\,11907454.}
\end{figure*}

\begin{figure*}
 \includegraphics[width=\textwidth]{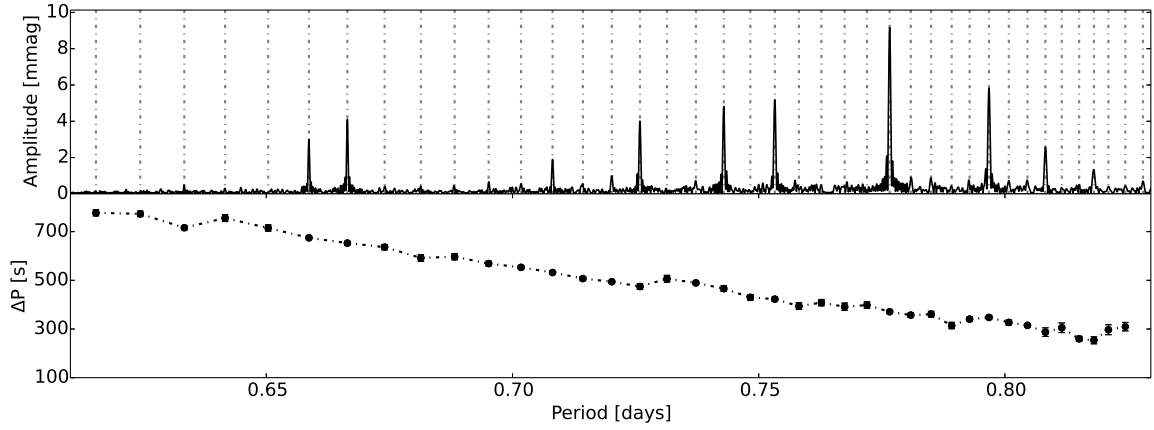}
 \caption{\label{fig:kic11917550}The period spacing patterns of KIC\,11917550.}
\end{figure*}

\begin{figure*}
 \includegraphics[width=\textwidth]{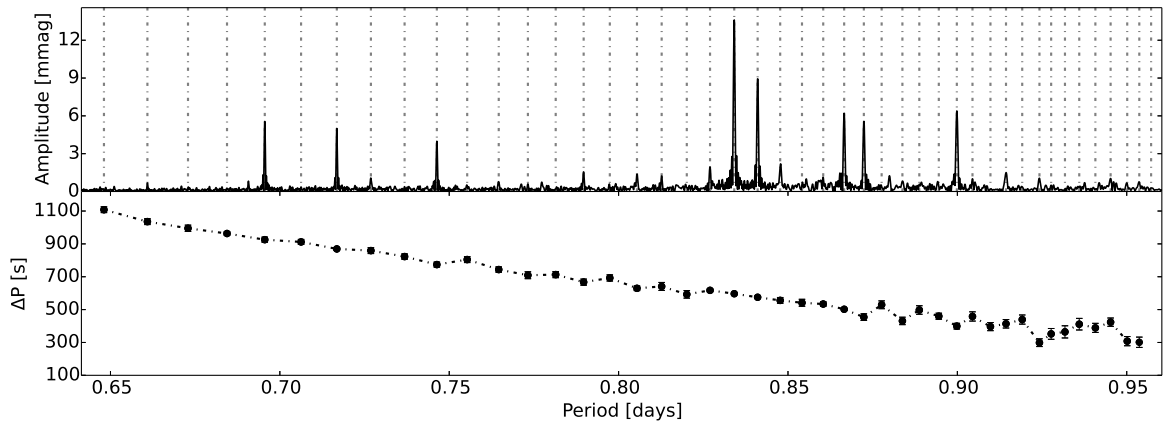}
 \caption{\label{fig:kic11920505}The period spacing patterns of KIC\,11920505.}
\end{figure*}

\begin{figure*}
 \includegraphics[width=\textwidth]{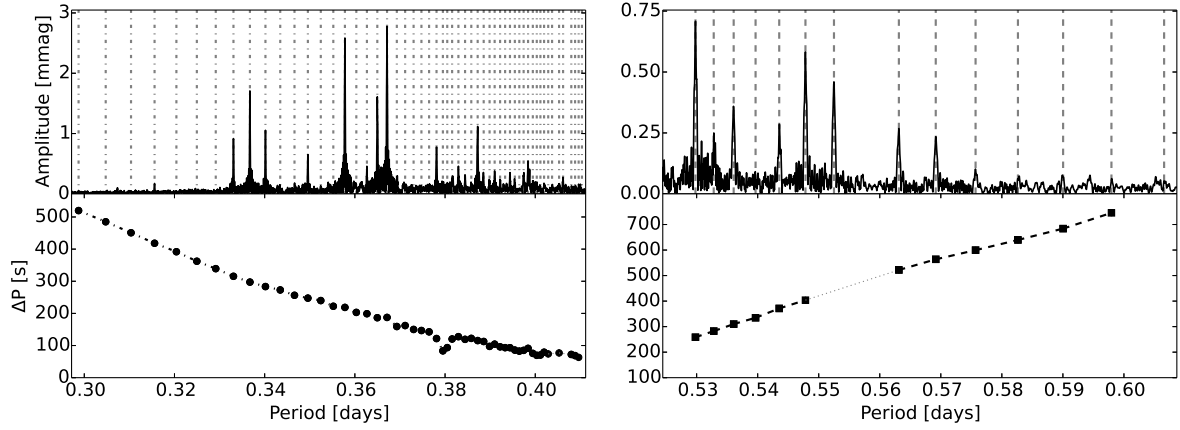}
 \caption{\label{fig:kic12066947}The period spacing patterns of KIC\,12066947.}
\end{figure*}

\begin{figure*}
 \includegraphics[width=\textwidth]{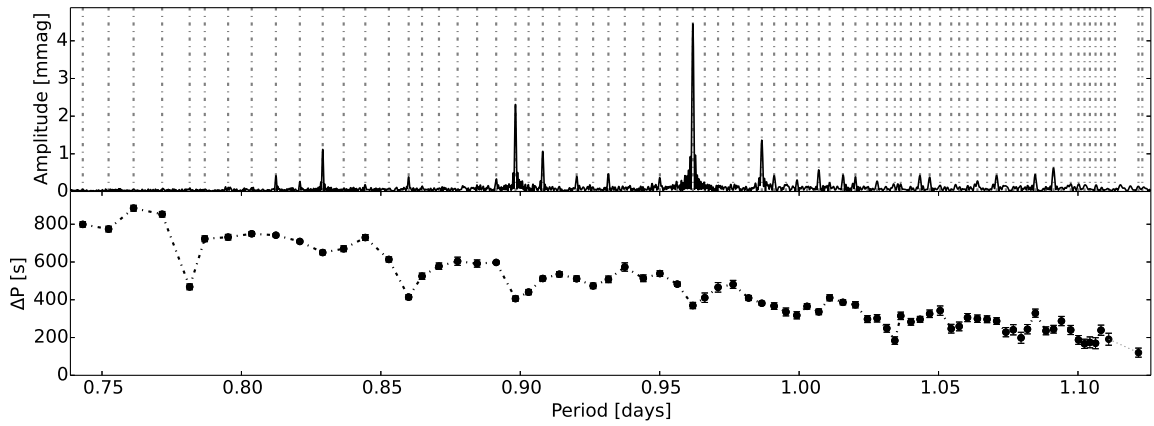}
 \caption{\label{fig:kic12458189}The period spacing patterns of KIC\,12458189.}
\end{figure*}

\end{document}